%% file: arxiv.tex
\documentclass[a4paper, 11pt]{article}
\usepackage[a4paper]{geometry}

\usepackage[longnamesfirst, round]{natbib}
\usepackage{url}

\usepackage[utf8]{inputenc}         
\usepackage{amsthm,amsmath,amsfonts,amssymb}                   
\usepackage{graphicx}                 
\usepackage{subfigure}                   

\usepackage{listings}                 
\usepackage{tikz}

\usepackage{algorithm2e}
\usepackage{mathtools}
\usepackage{enumerate}

\usepackage{authblk}
\usepackage{epstopdf}
\usepackage{bbm}

\def\eqd{\,{\buildrel d \over =}\,}

\begin{document}

\title{Does non-stationary spatial data always require non-stationary random fields?}
\author[1]{Geir-Arne Fuglstad\thanks{Corresponding author, fuglstad@math.ntnu.no}}
\author[1]{Daniel Simpson}
\author[2]{Finn Lindgren}
\author[1]{Håvard Rue}
\affil[1]{Department of Mathematical Sciences, NTNU, Norway}
\affil[2]{Department of Mathematical Sciences, University of Bath, UK}
\date{September 11, 2015}
\maketitle

\begin{abstract}
A stationary spatial model is an idealization and we expect that the true dependence
structures of physical phenomena are spatially varying, but how should we
handle this non-stationarity in practice? We study the challenges involved
in applying a flexible non-stationary model to
a dataset of annual precipitation in the conterminous US, where exploratory
data analysis shows strong evidence of a non-stationary covariance structure. 

The aim of this paper is to investigate the modelling pipeline once non-stationarity has been detected in spatial data.  We show that there is a real danger of over-fitting the model and that careful modelling is necessary in order to properly account for varying second-order structure.  In fact, the example shows that sometimes non-stationary Gaussian random fields are not necessary to model non-stationary spatial data.
\vspace{0.2cm}

\end{abstract}
\maketitle

\section{Introduction}
\label{sec:Introduction}

There are, in principle, two sources of non-stationarity present in any dataset: 
the non-stationarity in the mean and the non-stationarity in the covariance
structure. Classical geostatistical models based on stationary Gaussian random 
fields (GRFs) ignore the latter, but include the former through covariates 
that capture important structure in the mean. The focus of non-stationary 
spatial modelling is non-stationarity in the covariance structure.
However, it is impossible to separate the non-stationarity in the mean and the
non-stationarity in the covariance structure based on a single realization, and 
even with multiple realizations it is challenging. 

The Karhunen-Lo\`eve expansion states that under certain conditions a GRF
can be decomposed into an infinite linear combination of orthogonal
functions, which is weighted by independent Gaussian variables
with decreasing variances. For a single realization these orthogonal
functions will be confounded with the covariates in the mean, and the
mean structure and the covariance structure cannot be separated. This can give
apparent long range dependencies and global non-stationarity if spatially varying
covariates are missing in the mean. Such spurious global non-stationarity and
its impact on the local estimation of non-stationarity is an important topic
in the paper. 

A high degree of flexibility in the covariance structure combined with 
weak information about the covariance structure in the data makes 
overfitting by interpreting  ostensible patterns in the data as non-stationarity a 
critical issue. However, the most important point from an applied viewpoint is the
 computational
costs of running a more complex model versus the scientific gain. Non-stationarity in the mean is 
computationally cheap, whereas methods for non-stationarity in the covariance
structure are much more expensive. This raises two important questions: How much do we
gain by including non-stationarity in the covariance structure?  What sort of non-stationarity, if any, is most appropriate for the problem at hand?

The computational cost of non-stationary models usually comes from
a high number of highly dependent parameters that makes it expensive to run 
MCMC methods or likelihood optimizations, but the challenges with non-stationary 
models are not only computational. Directly specifying non-stationary covariance
functions is difficult and we need other ways of constructing models. 
Additionally, we need to choose where to put the non-stationarity. Should we
have non-stationarity in the range, the anisotropy, the marginal variance, the smoothness or the nugget effect? And how do we combine it all to 
a valid covariance structure?

\subsection{Non-stationarity}
Most of the early literature on non-stationary methods deals with data
from environmental monitoring stations where multiple realizations are available.
In this situation it is possible to calculate the empirical covariances 
between observed locations, possibly accounting for temporal dependence, and 
finding the required covariances through, for example, shrinkage towards a parametric 
model~\citep{Loader1989} or kernel smoothing~\citep{Oehlert1993}. It is also possible
to deal efficiently with a single realization with the moving window
approach of Haas~\citep{Haas1990b,Haas1990a, Haas1995}, but this 
method does not give valid global covariance structures. 

However, the 
most well-known method from this time period is the deformation method 
of~\citet{Sampson1992}, in which an underlying stationary process is made 
non-stationary by applying a spatial deformation. The original formulation has 
been extended to the Bayesian framework~\citep{Damian2001, Damian2003,Schmidt2003}, 
to a single realization~\citep{anderes2008estimating}, to covariates in
the covariance structure~\citep{Schmidt2011} and to higher dimensional
base spaces~\citep{Bornn2012}. 

Another major class of non-stationary methods is based on the process 
convolution method developed by Higdon~\citep{Higdon1998,higdon1999non}. In this 
method a spatially varying kernel is convolved with a white noise process to
create a non-stationary covariance structure. \citet{Paciorek2006} looked
at a specific case where it is possible to find a closed form expression for 
a Mat\'ern-like covariance function and \citet{RSSC:RSSC12027} used a kernel 
that depends on wind direction and strength to control the covariance structure. 
The process convolution methods have also been extended to dynamic
multivariate processes~\citep{Calder2007,Calder2008} and spatial
multivariate processes~\citep{Kleiber201276}.

It is possible to take a different approach to non-stationarity, where
instead of modelling infinite-dimensional Gaussian processes one uses 
a linear combination of basis functions and models the covariance 
matrix of the coefficients of the basis functions~\citep{Nychka2002,Nychka2014}. 
One such approach is the
fixed rank kriging method~\citep{Cressie2008}, which uses a linear
combination of a small number of basis functions and estimates the covariance 
matrix for the coefficients of the linear combination. This approach leads to a continuously
indexed spatial process with a non-stationary covariance structure. The predictive 
processes~\citep{Banerjee2008} corresponds to a specific choice of the 
basis functions and the covariance matrix, but does not give a
very flexible type of non-stationarity. All such methods are variations of the
same concept, but lead to different computational schemes with different
computational properties. The dimension of the finite-dimensional basis is
in all cases used to control the computational cost and the novelty of each method
lies in how the basis elements are selected and connected to each other, and the
computational methods used to exploit the structure.

An overview of the literature before around 2010 is given in \citet{Sampson2010}.
This overview also includes less well-known methods such as the piece-wise Gaussian process 
of~\citet{Kim2005}, processes based on weighted linear combination of stationary 
processes~\citep{Fuentes2001,Fuentes2002b,Fuentes2002a,Nott01122002}.

Recently, a new class of methods based on the SPDE-approach
introduced by~\citet{Lindgren2011} is emerging. This class of methods is based
on a representation of the spatial field as a solution of a stochastic partial
differential equation (SPDE) with spatially varying coefficients. The methodology
is closely connected with Gaussian Markov random fields (GMRFs)~\citep{Rue2005}
and is able to handle more observations
than is possible with the deformation method and the 
process convolution method. In a similar 
way as a spatial GMRF describes local behaviour for a discretely 
indexed process, an SPDE describes local behaviour for a continuously indexed 
process. This locality in the continuous description can 
be transferred to a 
GMRF approximation of the solution of the SPDE, and gives a GMRF with a spatial 
Markovian structure that can be exploited in computations. 

This type of methodology has been applied to global ozone 
data~\citep{Bolin2011} and to annual precipitation in Norway with covariates in 
the covariance structure~\citep{Rikke2013,ingebrigtsen2014estimation}. Additionally, \citet{sigrist2012} 
used similar type of modelling to handle a spatio-temporal process where
wind direction and strength enters in the covariance structure.

Despite all the work that has been done in non-stationary spatial modelling,
it is still an open field where no model stands out as the clear choice. 
However, we believe that modelling locally such as in the SPDE-based 
models is more attractive than modelling globally such as in the deformation
method and the process convolution method. Therefore, we choose to use an extension
of the model by~\citet{Fuglstad2014} that allows for both a spatially varying
correlation structure and a spatially varying marginal variance.
This method is closely connected to the already well-known deformation method 
of~\citet{Sampson1992} and the Mat\'ern-like process convolution 
of~\citet{Paciorek2006}, but is focused at the local behaviour and not the
global behaviour. 

In a similar way as in the model of~\citet{Paciorek2006} the global structure 
is defined through the combination of ellipses at each location that describe 
anisotropy. However, their model only combines the ellipses at  two 
locations and does not account for the local behaviour between locations. The new 
model incorporates the local anisotropy everywhere into the covariance for each 
pair of locations and is not the same as the model of~\citet{Paciorek2006}. 
The model works in a similar way as the deformation method. However,
instead of describing a global deformation, the ellipses augment the local
distances around each point and describe locally a change of distances
such that lengths are different in different directions, but does not, in general, 
lead to a deformation of \(\mathbb{R}^2\) to \(\mathbb{R}^2\). Such local modelling
tends to lead to a deformation in an ambient space of dimension higher than 2.
 The interest of this paper is 
to study the challenges and results of applying the method to a dataset of 
annual precipitation in the conterminous US.

\subsection{Annual precipitation in the conterminous US}
This case study of non-stationarity will use the measurements of monthly total
precipitation at different measurement stations in the conterminous US for 
the years 1895--1997 that are available at 
\url{http://www.image.ucar.edu/GSP/Data/US.monthly.met/}. This dataset
was chosen because it is publicly available in a form that is easily
downloaded and loaded into software, and because the large spatial scale 
of the dataset and the complexity of the physical process that
generates weather makes it intuitively feels like there must be 
non-stationarity in the dataset. The main focus of this study is to see
how much non-stationarity improves the spatial predictions compared to a
stationary model, but the approach could also be used to gain insight about 
the spatial structure of climate in the conterminous 
US~\citep{smith1996estimating}.

\begin{figure}
	\centering
	\includegraphics[width=11cm]{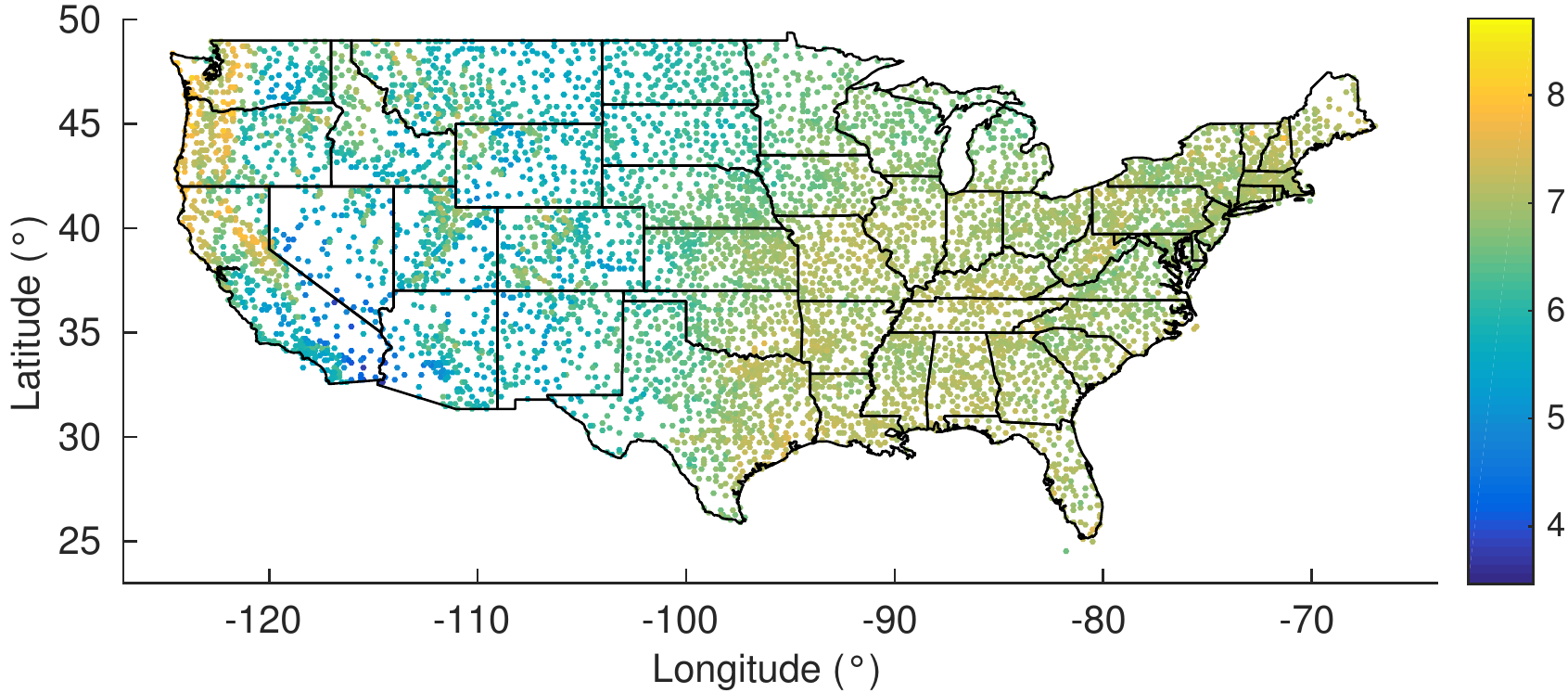}
	\caption{The logarithm of total yearly precipitation measured in millimetres at
			 7040 locations in the conterminous US for the year 1981.}
	\label{fig:observation}
\end{figure}

In total there are 11918 measurement stations in the dataset, but measurements are 
only available at a subset of the stations each month and the rest of the stations 
have in-filled data~\citep{Johns2003}. For each year, we aggregate the monthly
data at those stations which have measurements available at all months in that
year and produce a dataset of yearly total precipitation. This gives a different
number of locations for each year. We then take the logarithm of each observation
to create the transformed data that is used in this paper. 
Figure~\ref{fig:observation}
shows the transformed data at the 7040 stations available for 1981. 
The only covariate available in the dataset is the elevation at each station, and 
%since the focus of the paper is on the covariance structure, 
no work was done to find other covariates from alternate sources. 

If the focus were to model this data in the best possible way, it would, 
in general, be good to look for more covariates or consider alternatives such as 
spatially heterogeneous coefficients~\citep{fotheringham2003geographically,gelfand2003spatial} 
before using a full non-stationary model. Non-stationarity can be achieved both
through a flexible model for the mean structure and through a flexible model for 
the covariance structure. The difference between the two alternatives lies in the
interpretation of the model, and with focus on computations and predictions, a simple 
second-order structure and a flexible first-order structure could be an attractive option.
But it is our explicit intention to focus on non-stationarity in the second-order
structure and we keep the model for the mean simple.

We will assume that the transformed data can be treated as Gaussian, which is a 
reasonable assumption because we are modelling annual precipitation data. However,
it would not be a reasonable assumption, for example, for daily data, and it
would be necessary to consider not only how to deal with non-stationarity, but
also how to deal with the lack of Gaussianity.  
\citet{bolin2013non} compare the predictions made by
a stationary Gaussian model, a stationary Gaussian model for transformed data 
and two stationary
non-Gaussian models for monthly precipitation for two different months 
from the same dataset as in this paper. They apply the non-Gaussian model
of~\citet{bolin2014spatial}, but do not find clear evidence that one model
perform better than the others. The approach of~\citet{bolin2014spatial} is
based on the discretization of an SPDE in a similar way as the approach
in this paper, but using a non-Gaussian noise process, 
and a possible extension of the non-stationary model in this paper would be to 
non-Gaussian data.

\begin{figure}
	\centering
	\includegraphics[width=11cm]{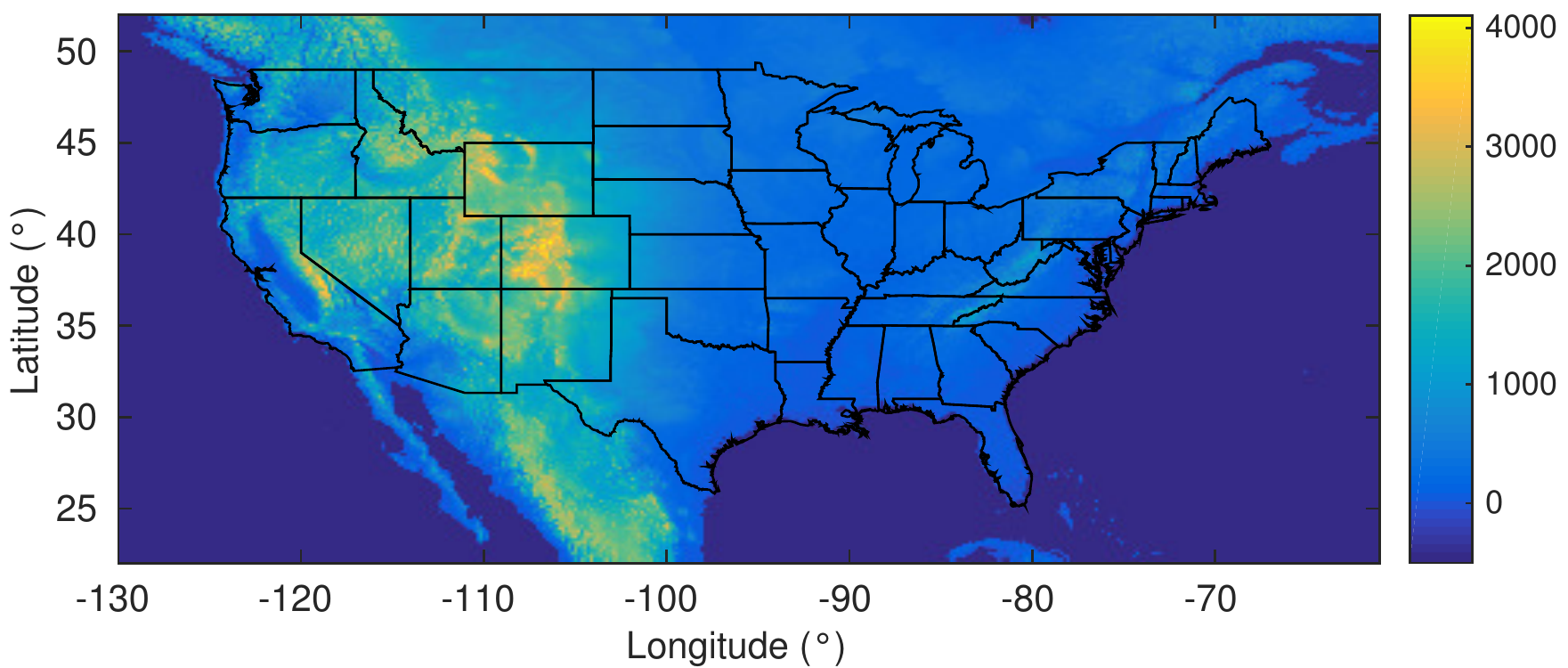}
	\caption{Elevation in the US measured in meters. Data from GLOBE data 
	         set~\citep{Globe1999}}
	\label{fig:elevation}
\end{figure}

The main motivation for focusing on 
the year 1981 is that~\citet{Paciorek2006} previously studied the annual
precipitation in the subregion of Colorado for this year. They did not see major
improvements over a stationary model and our preliminary analysis showed that
there was little non-stationarity left in the subregion after introducing a 
joint mean and elevation.
However, Colorado constitutes a small part of the conterminous US, and as shown
in Figure~\ref{fig:elevation} there are large differences in the topography of 
the western and the eastern part of the conterminous US. A large proportion of the 
western part is mountainous whereas in the eastern part a large proportion is 
mostly flat. This varied topography is a strong indication that 
the process cannot possibly be stationary.

\begin{figure}
	\centering
	\includegraphics[width=7cm]{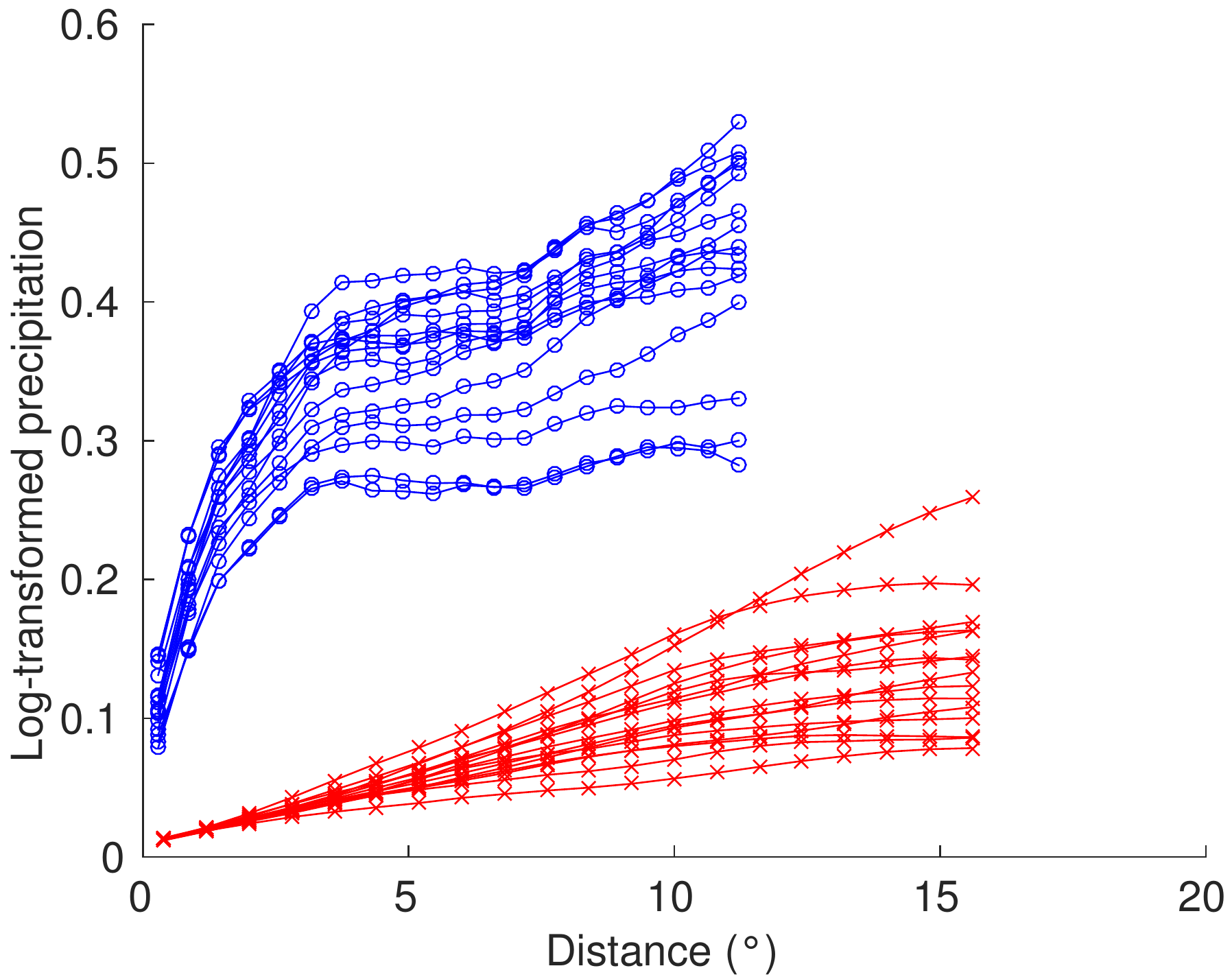}
	\caption{Estimated semi-variograms for the years 1971 to 1985 using the 
			 locations to the west of \(100\,^\circ\mathrm{W}\) 
			 coloured in blue and marked with circles and locations
			 to the east of \(100\,^\circ\mathrm{W}\) coloured in red and 
			 marked with crosses.}
	\label{fig:semiVar}
\end{figure}

To substantiate our claims of non-stationarity we explore the 
difference in the covariance structure in the western and eastern part through 
variograms. The data from years 1971--1985 is selected and divided into two regions: 
the area west of \(100\,^\circ \mathrm{W}\) and the area east of
\(100\, ^\circ \mathrm{W}\). For each year the variogram of each region 
is calculated. Figure~\ref{fig:semiVar} shows that there 
is no overlap between the variograms of the western region and the eastern 
region. There is significant variation within each region, but the overall
appearance clearly indicates different covariance structures within the regions.
Based on the evidence of non-stationarity seen in the variograms for the
full region, we want to know if a non-stationary model will improve the 
predictions. It has been observed by several authors~\citep{Schmidt2011,RSSC:RSSC12027}
and it has also been the experience of the authors that non-stationary models
do not lead to much difference in the predicted values, and that the differences
are found in the prediction variances. However, predictions should always have
associated error estimates and when we write improved predictions,
we are interested in whether the predictive distributions, summarized by the
predicted values and their associated prediction variances, better describe
the observed values.

There are two cases of interest: a single realization and multiple
realizations. In the former it is impossible to separate the non-stationarity
in the mean and in the covariance structure, and the non-stationary model might
be more accurately described as adaptive smoothing, but many spatial datasets 
are of this form and a non-stationary model might still perform better than
a stationary model. We will
investigate both of these cases and evaluate whether the non-stationary model
improves predictions and whether the computational costs are worth it. It
is clear that stationarity is not the truth, but that does not mean that it
does not necessarily constitutes a sufficient model for predictions.

\subsection{Overview}
The paper is divided into five sections. Section~\ref{sec:Theory} describes 
how we model the data. We discuss what type of non-stationarity is present in 
the model and how it is specified, how we parametrize the non-stationarity 
and how we perform computations with the non-stationary model.
Then in Section~\ref{sec:Application} a hierarchical model incorporating 
the non-stationary model is applied to annual precipitation in a single 
realization setting, and in Section~\ref{sec:detrend} the data is studied from 
a multiple realizations perspective. The differences between the estimated 
covariance structures and the prediction scores for the different models are 
discussed. The paper ends with discussion and concluding remarks in 
Section~\ref{sec:Discussion}.

\section{Modelling the data}
\label{sec:Theory}
Before analyzing the data we need to introduce the model that will
be used. Particularly, we need to say which types of non-stationarity will
be present in the model and how this non-stationarity will be modelled. 
A good spatial model should provide a useful 
way to do both the theoretical modelling and the associated computations. We
first discuss how non-stationarity is introduced and 
then how to parametrize the non-stationary.

\subsection{Modelling the non-stationarity}
\label{sec:GRF}
\begin{figure}
	\centering
	\includegraphics[width=10cm]{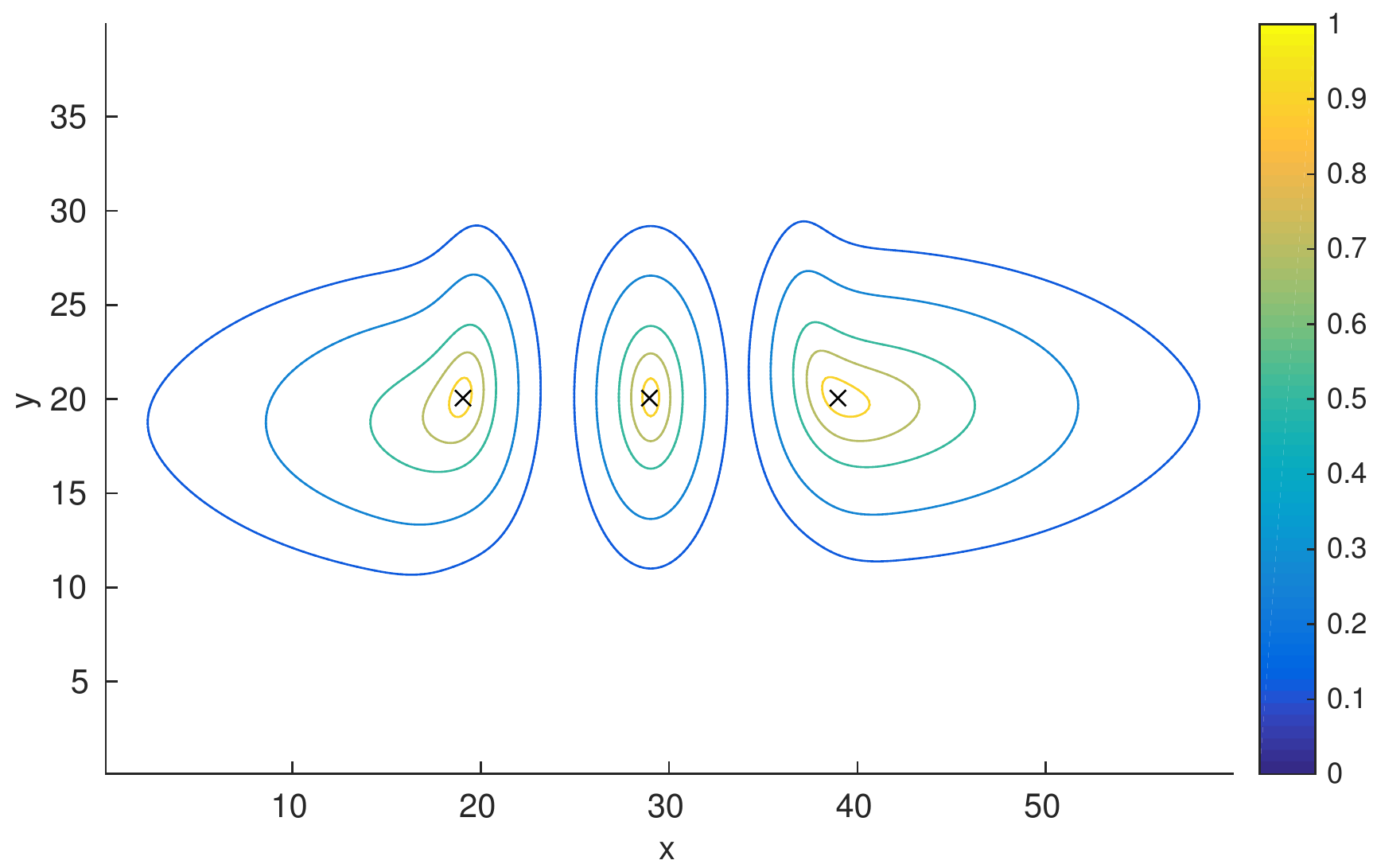}
	\caption{Example of a correlation function caused by varying local behaviour.
			 For each location marked with a black cross, the 0.9, 0.7, 0.5, 0.25 
			 and 0.12 level contours of the correlation function are shown.}
	\label{fig:CorrExample}
\end{figure}

It is difficult to specify a global covariance function when one only
has intuition about local behaviour. Consider the situation in 
Figure~\ref{fig:CorrExample}. The left hand side and the right hand side have locally 
large ``range'' in the horizontal direction and somewhat shorter ``range'' in 
the vertical direction, and the middle area has locally much shorter ``range'' in 
the horizontal direction, but slightly longer in the vertical direction. We write
``range'' with quotation marks because the concept of a global range does not
have a well-defined meaning in non-stationary modelling. Instead we will think of range as a 
local feature and use the word to mean what happens to dependency in a small region
around each point. From 
the figure one can see that for the point in the middle, the chosen contours look 
more or less unaffected by the two other regions since they are fully contained 
in the middle region, but that for the point on the left hand side and the 
point on the right hand side, there is much skewness introduced by the 
transition into a different region. 

It would be hard to specify a fitting global correlation
function for this situation. However, if one instead starts with an isotropic 
process and then stretches the left hand side and the right hand side in the 
\(x\)-direction, the task is much easier. This is a flexible way to create
interesting covariance structures and is the core of the deformation 
method~\citep{Sampson1992}, but can be challenging since one has to create
a valid \emph{global} deformation. We present instead a model where the
modelling can be done 
\emph{locally} without worrying about the \emph{global} structure. We let 
the local structure automatically specify a valid global structure. In 
this example one would only specify that locally the range is longer in 
the horizontal direction in the left hand side and the right hand side, and 
then let this implicitly define the global structure without directly 
modelling a global deformation.

In the SPDE-based approach the correlation between two spatial locations is 
determined implicitly by the behaviour between the spatial locations. If 
there are mountains, the model could specify that locally the distances 
are longer than they appear on the map and the correlation will 
decrease more quickly when crossing those areas, and if there are plains, the 
model could specify that distances are shorter than they appear on the map 
and the
correlation will decrease more slowly in those areas. 
A major advantage of this approach is that the local specification naturally
leads to a spatial GMRF with good computational properties. It is possible
to approximate the local continuous description with a local discrete
description. The result is a spatial GMRF with a very
sparse precision matrix

% In the SPDE-based approach we do not try to model the global behaviour
% directly, but rather how distances and variances behave locally. This
% implies that specifying a specific correlation between two locations
% is hard, but this is not what we are trying to do. If we were just given
% two locations and asked to give the correlation between the annual precipitation
% at those two locations without any information about what happens in-between, there 
% would be no way to give that value. The correlation should be determined 
% by what happens between the locations. That is whether there are plains
% or lakes where one might believe that the correlation decreases slowly or a mountain 
% where one might believe that the correlation decreases quickly. A great advantage
% about this 
% type of local specification is that it naturally leads to a spatial GMRF that 
% has good computational properties. We can specify a spatial GMRF where each location 
% is conditionally dependent only on locations close to itself and give it a sparse 
% precision matrix.

The starting point for the non-stationary SPDE-based model 
is the stationary SPDE introduced in~\citet{Lindgren2011},
\begin{equation}
	(\kappa^2-\nabla\cdot\nabla)u(\vec{s}) = \sigma \mathcal{W}(\vec{s}), \qquad \vec{s}\in\mathbb{R}^2,
	\label{eq:LindgrenSPDE}
\end{equation}
where \(\kappa>0\) and \(\sigma>0\) are constants, 
\(\nabla = (\frac{\partial}{\partial x}, \frac{\partial}{\partial y})^\mathrm{T}\) and \(\mathcal{W}\) 
is a standard Gaussian white noise process. The SPDE describes the GRF \(u\) as a 
smoothed version of the Gaussian white noise on the right hand side of the
equation. \citet{Whittle1954,whittle1963stochastic} showed
that any stationary solution of this SPDE has the Mat\'{e}rn covariance function
\begin{equation}
	r(\vec{s}_1, \vec{s}_2) = \frac{\sigma^2}{4\pi \kappa^2} (\kappa \lvert\lvert \vec{s}_2-\vec{s}_1 \rvert\rvert)K_{1}(\kappa \lvert\lvert \vec{s}_2-\vec{s}_1 \rvert\rvert),
	\label{eq:MaternCov}
\end{equation} 
where \(K_1\) is the modified Bessel function of second kind, order 1.
This covariance function is a member of the commonly-used Mat\'ern family of 
covariance functions, and one can see from Equation~\eqref{eq:MaternCov} 
that one can first use \(\kappa\) to select the range and then \(\sigma\) 
to achieve the desired marginal variance. In some methods for non-stationarity
it is possible to spatially vary the smoothness, but this is not a feature that is
available in the non-stationary model presented here. However, with the 
flexibility present in the rest of the non-stationarity it is not clear if the 
smoothness would be jointly identifiable.

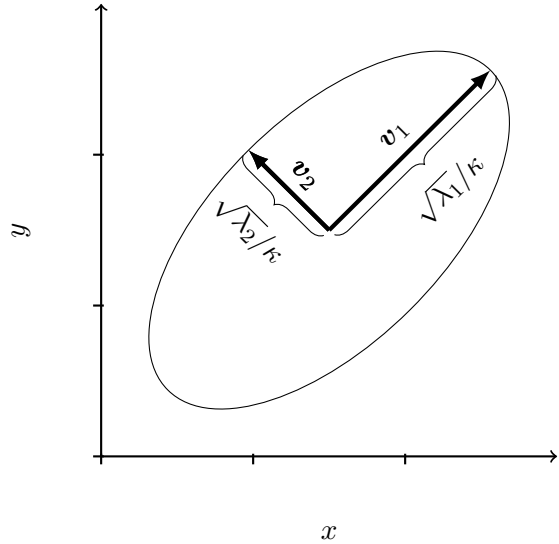
\begin{figure}
	\centering
	\input{Tikz/statCovariance.tex}
	\caption{Iso-correlation curve for the 0.6 level, where \((\lambda_1, \vec{v}_1)\) and \((\lambda_2, \vec{v}_2)\) are the eigenpairs of \(\mathbf{H}\).}
	\label{fig:CovFunction}
\end{figure}

The next step is to generate a GRF with an anisotropic Mat\'{e}rn covariance function. The
cause of the isotropy in SPDE~\eqref{eq:LindgrenSPDE} is that the Laplacian, 
\(\Delta = \nabla\cdot\nabla\) is invariant to a change of coordinates that 
involves rotation and translation. To change this a \(2\times 2\)
matrix \(\mathbf{H}>0\) is introduced into the operator to give the SPDE
\begin{equation}
	(\kappa^2-\nabla\cdot\mathbf{H}\nabla)u(\vec{s}) = \sigma \mathcal{W}(\vec{s}).
	\label{eq:AniSPDE}
\end{equation}
This choice is closely related to the change of coordinates 
\(\tilde{\vec{s}}=\mathbf{H}^{1/2}\vec{s}\)~\citep[Section 3]{Fuglstad2014} and gives the covariance function
\begin{equation}
	r(\vec{s}_1, \vec{s}_2) = \frac{\sigma^2}{4\pi\kappa^2\sqrt{\det(\mathbf{H})}} (\kappa \lvert\lvert \mathbf{H}^{-1/2}(\vec{s}_2-\vec{s}_1) \rvert\rvert)K_{1}(\kappa \lvert\lvert \mathbf{H}^{-1/2}(\vec{s}_2-\vec{s}_1) \rvert\rvert).
	\label{eq:aniMatern}
\end{equation}
Compared to Equation~\eqref{eq:MaternCov} there is a change in the marginal variance and
a directionality is introduced through a distance measure different than the 
standard Euclidean distance. Figure~\ref{fig:CovFunction} shows how the eigenpairs of \(\mathbf{H}\)
and the value of \(\kappa\) act together to control range. One can see that the
construction leads to elliptic iso-covariance curves.
In what follows \(\sigma\) is assumed to be equal to 1 since the marginal variance can
be controlled by varying \(\kappa^2\) and \(\mathbf{H}\) together.

The final step is to construct a non-stationary GRF where the local behaviour at each location
is governed by SPDE~\eqref{eq:AniSPDE} with \(\sigma = 1\) and the values of \(\kappa^2\) and
\(\mathbf{H}\) varying over the domain. The intention is to create a GRF by 
chaining together processes with different local covariance structures. The SPDE becomes
\begin{equation}
	(\kappa^2(\vec{s})-\nabla\cdot\mathbf{H}(\vec{s})\nabla)u(\vec{s}) = \mathcal{W}(\vec{s}).
	\label{eq:FinalSPDE}
\end{equation}
For technical reasons concerned with the discretization of the SPDE, \(\kappa^2\) 
is required to be continuous and \(\mathbf{H}\) is required to be continuously 
differentiable. This does not present any problems and is easily achieved by
using continuously differentiable basis functions for \(\kappa^2\) and
\(\mathbf{H}\). The restricted form where \(\kappa^2\) is constant was
investigated in~\citet{Fuglstad2014}, but this restricted form only 
allows for varying local anisotropy without control over the marginal
variances. This extended model allows for spatially varying ``range'', 
anisotropy and marginal variance.

The non-stationary covariance structure is fully described by 
SPDE~\eqref{eq:FinalSPDE}, but before the model can be used in practice 
the description must be brought into a form 
which is useful for computations. This can be done by discretizing the SPDE 
using a finite-dimensional basis expansion where the distribution
of the coefficients is a sparse GMRF that possesses approximately the same covariance 
structure as \(u\). See \ref{app:Details} for more details.
This kind of construction alleviates one of the largest problems 
with GMRFs, namely that they are hard to specify in a spatially
coherent manner. The computational benefits of 
spatial GMRFs are well known, but a GMRF needs to be constructed through its
conditional distributions and it is notoriously hard
to do this for non-stationary models. But with this approach
it is possible to model the problem with an SPDE and then do computations
with the computational benefits of a spatial GMRF.

\subsection{Parametrizing the non-stationarity}
Before we can turn the theoretical and computational description of the 
non-stationary model into a statistical model, we need to describe the
non-stationarity through parameters. This means both decomposing the model
into parameters and connecting the parameters together through a penalty.

The first step is to decompose the function 
\(\mathbf{H}\), which must give positive definite \(2 \times 2\) matrices
at each location, into simpler functions. One usual way to do this is to 
use two strictly positive functions
\(\lambda_1\) and \(\lambda_2\) for the eigenvalues and a function \(\phi\) for the
angle between the \(x\)-axis and the eigenvector associated with \(\lambda_1\).
However, with a slight re-parametrization
\(\mathbf{H}\) can be written as the sum of an isotropic effect, described by a constant
times the identity matrix, plus an additional anisotropic effect, described by
direction and magnitude.

Express \(\mathbf{H}\) through
the  scalar functions \(\gamma\), \(v_x\) and \(v_y\) by
\[
	\mathbf{H}(\vec{s}) = \gamma(\vec{s})\mathbf{I}_2 + \begin{bmatrix} v_x(\vec{s}) \\ v_y(\vec{s}) \end{bmatrix} \begin{bmatrix} v_x(\vec{s}) & v_y(\vec{s}) \end{bmatrix},
\]
where \(\gamma\) is required to be strictly positive. 
The eigendecomposition of this matrix has eigenvalue
\(\lambda_1(\vec{s}) = \gamma(\vec{s})+v_x(\vec{s})^2+v_y(\vec{s})^2\) with eigenvector
\(\vec{v}_1(\vec{s}) = (v_x(\vec{s}), v_y(\vec{s}))\)
and eigenvalue \(\lambda_2(\vec{s}) = \gamma(\vec{s})\) with eigenvector \(\vec{v}_2(\vec{s}) =(-v_y(\vec{s}), v_x(\vec{s}))\). From Figure~\ref{fig:CovFunction} this means that for
a stationary model, \(\gamma\) affects the length of the shortest semi-axis 
of the iso-correlation curves and \(\vec{v}\) specifies the direction of
and how much larger the longest semi-axis is. The above decomposition through
\(\gamma\), \(v_x\) and \(v_y\) is general and is valid for every symmetric 
positive-definite \(2\times 2\) matrix.

Since we want flexible covariance structures, some
representation of the functions \(\kappa^2\), \(\gamma\), \(v_x\) and \(v_y\) is
needed. To ensure positivity of \(\kappa^2\) and \(\gamma\), they are first 
transformed into \(\log(\kappa^2)\) and \(\log(\gamma)\). Each of these functions
will be expanded in a basis, and requires a penalty that imposes regularity and 
makes sure the function
is not allowed to vary too much. 
The choice was made to give \(\log(\kappa^2)\), \(\log(\gamma)\), \(v_x\) and \(v_y\)
spline-like penalties. The steps that follow are the same for each function. 
Therefore, they are only shown for \(\log(\kappa^2)\).

The function \(\log(\kappa^2)\) is given a penalty according to the distribution 
generated from the SPDE
\begin{equation}
	-\Delta \log(\kappa^2(\vec{s})) = \mathcal{W}_\kappa(\vec{s})/\sqrt{\tau_\kappa}, \qquad \vec{s}\in\mathcal{D},
	\label{eq:Prior}
\end{equation}
where \(\tau_\kappa > 0\) is the parameter controlling the penalty, with the 
Neumann boundary condition of zero derivatives at the edges. This extra 
requirement is used to restrict
the resulting distribution so it is only invariant to the addition of
a constant function, and the penalty parameter
is used to control how much \(\log(\kappa^2)\) can vary from a constant function.
The penalty defined through SPDE~\eqref{eq:Prior} is in this paper called a 
two-dimensional second-order random walk due to its similarity 
to a one-dimensional second-order random walk~\citep{Lindgren2008}.

The first step of making the above penalty applicable for the 
computational model is to expand \(\log(\kappa^2)\) in a basis through a linear
combination of basis functions,
\[
	\log(\kappa^2(\vec{s})) = \sum_{i=1}^{k}\sum_{j=1}^{l} \alpha_{ij} f_{ij}(\vec{s}),
\]
where \(\{\alpha_{ij}\}\) are the parameters and
\(\{f_{ij}\}\) are real-valued basis functions. For convenience, the basis
is chosen in such a way that all basis functions satisfy the boundary conditions
specified in SPDE~\eqref{eq:Prior}. If this is done, one immediately satisfies
the boundary condition. The remaining tasks are then
to decide which basis functions to use and what the resulting penalties on the
parameters are.

Due to a desire to make
\(\mathbf{H}\) continuously differentiable and a desire to have ``local'' basis
functions, the basis functions are chosen to be based on 2-dimensional, 
second-order B-splines (piecewise-quadratic functions). The basis is constructed
as a tensor product of two 1-dimensional B-spline bases constrained to 
satisfy the boundary condition.

The penalty is based on the distribution defined by SPDE~\eqref{eq:Prior}, 
so the final step is to determine a Gaussian distribution for the parameters
such that the distribution of \(\log(\kappa^2)\) is close to a solution of
SPDE~\eqref{eq:Prior}. The approach taken is based on a least-squares 
formulation of the solution
and is described in \ref{app:Prior}. Let \(\vec{\alpha}\) 
be the \(\{\alpha_{ij}\}\) parameters stacked row-wise, then the
result is that \(\alpha\) should be given
a zero-mean Gaussian distribution with precision matrix 
\(\tau_\kappa\mathbf{Q}_{\mathrm{RW2}}\). This matrix has rank \((kl-1)\),
due to the Neumann boundary conditions, and
the distribution is invariant to the addition of a vector of only the same values, 
but for convenience the 
penalty will still be written as \(\vec{\alpha}\sim\mathcal{N}_{kl}(\vec{0}, \mathbf{Q}_{\mathrm{RW2}}^{-1}/\tau_\kappa)\).

\subsection{Hierarchical model}
\label{sec:FullModel}
Observations \(y_1, y_2, \ldots, y_N\) are made at locations 
\(\vec{s}_1, \vec{s}_2, \ldots, \vec{s}_N\). The observed
value at each location is assumed to be the sum of a fixed
effect due to covariates, a spatial ``smooth'' effect and a random effect. The
covariates at location \(\vec{s}_i\) are described by the \(p\)-dimensional
row vector \(\vec{x}(\vec{s}_i)^\mathrm{T}\) and the spatial field is
denoted by \(u\). This gives the observation equation
\[
	y_i = \vec{x}(\vec{s}_i)^\mathrm{T}\vec{\beta} + u(\vec{s}_i) + \epsilon_i,
\]
where \(\vec{\beta}\) is a \(p\)-variate random vector for the 
coefficients of the covariates and 
\(\epsilon_i \sim \mathcal{N}(0, 1/\tau_{\mathrm{noise}})\) is the random
effect for observation \(i\), for \(i = 1, 2, \ldots, N\). 

The \(u\) is modelled and parametrized as described in the previous
sections and the GMRF approximation is used for computations. In this GMRF 
approximation the domain is divided into a regular grid consisting
of rectangular cells and each element of the GMRF approximation
describes the average value on one of these cells. So
\(u(\vec{s}_i)\) is replaced with the approximation
\(\vec{e}(\vec{s}_i)^\mathrm{T}\vec{u}\), where 
\(\vec{e}(\vec{s}_i)^\mathrm{T}\) is the \(mn\)-dimensional
row vector selecting the element of \(\vec{u}\) which corresponds
to the cell which contains location \(\vec{s}_i\). In total, this gives
\begin{equation}
	\vec{y} = \mathbf{X}\vec{\beta}+\mathbf{E}\vec{u} + \vec{\epsilon},
	\label{eq:DataEquation}
\end{equation}
where \(\vec{y} = (y_1, y_2, \ldots, y_N)\), the matrix \(\mathbf{X}\) 
has \(\vec{x}(\vec{s}_1)^\mathrm{T},\ldots,\vec{x}(\vec{s}_N)^\mathrm{T}\)
as rows and the matrix \(\mathbf{E}\) has 
\(\vec{e}(\vec{s}_1)^\mathrm{T},\ldots,\vec{e}(\vec{s}_N)^\mathrm{T}\) as 
rows. In this equation the spatial effect is approximated with a discrete model,
but the covariate has not been gridded and is at a higher resolution than the
grid.

The
model for the observations can also be written in the form
\[
	\vec{y}|\vec{\beta},\vec{u},\log(\tau_\mathrm{noise}) \sim \mathcal{N}_N(\mathbf{X}\vec{\beta} + \mathbf{E}\vec{u}, \mathbf{I}_N/\tau_\mathrm{noise} ).
\]
The parameter \(\tau_{\mathrm{noise}}\) 
acts as the precision of a joint effect 
from measurement noise and small scale spatial variation~\citep{Diggle2007}.
We make the underlying model for the \(p\)-dimensional random variable
\(\vec{\beta}\) proper by introducing a weak Gaussian penalty,
\[
	\vec{\beta} \sim \mathrm{N}_p(\vec{0}, \mathbf{I}_p/\tau_{\beta}).
\]
The penalty can be made stronger, but we do not believe it will have a strong effect
on the estimates for this dataset with only an intercept and one covariate.

To describe the full hierarchical model, we introduce symbols 
to denote the parameters that control the spatial field \(u\). Denote the parameters that
control \(\log(\kappa^2)\), \(\log(\gamma)\), \(v_x\) and \(v_y\) by 
\(\vec{\alpha}_1\), \(\vec{\alpha}_2\), \(\vec{\alpha}_3\) and
\(\vec{\alpha}_4\), respectively. Further, denote the corresponding 
penalty parameters for each function by \(\tau_1\), \(\tau_2\), \(\tau_3\) and 
\(\tau_4\). With this notation the full model becomes
\begin{align*}
	\text{Stage 1: } & \vec{y} | \vec{\beta}, \vec{u}, \log(\tau_{\mathrm{noise}}) \sim \mathcal{N}_N(\mathbf{X}\vec{\beta}+\mathbf{E}\vec{u}, \mathbf{I}_N/\tau_{\mathrm{noise}}) \\
	\text{Stage 2: } &  \vec{u}|\vec{\alpha}_1, \vec{\alpha}_2, \vec{\alpha}_3, \vec{\alpha}_4 \sim \mathcal{N}_{nm}(\vec{0}, \mathbf{Q}^{-1}), \, \, \, \, \vec{\beta} \sim \mathcal{N}_p(\vec{0}, \mathbf{I}_p/\tau_{\beta})\\
	\text{Stage 3: } & \vec{\alpha}_i| \tau_i \sim \mathcal{N}_{kl}(\vec{0}, \mathbf{Q}_{\mathrm{RW2}}^{-1}/\tau_i) \,\, \text{for}\,\, i = 1,2,3,4,
\end{align*}
where \(\tau_1\), \(\tau_2\), \(\tau_3\), \(\tau_4\) and \(\tau_\beta\) are 
penalty parameters that must be pre-selected.

An important model choice when constructing the GMRF approximation of the spatial process
is the selection of the resolution of the approximation. The approximation does not allow for variation of the spatial field within a grid cell 
and the spatial resolution must be chosen high enough to capture variations on the scale
at which observations were made. The variation at sub-grid scale cannot be 
captured by the approximation and will be captured by the nugget effect. 

\subsection{Penalized likelihood and inference}
The two things of main interest to us in this case study are the covariance 
parameters \(\vec{\theta} = (\vec{\alpha}_1, \vec{\alpha}_2, \vec{\alpha}_3, \vec{\alpha}_4, \log(\tau_{\mathrm{noise}}))\) and the
predictive distributions for unmeasured locations. To estimate the covariance
parameters, we need the integrated likelihood
where the latent field consisting of the coefficients of the fixed effects and 
the spatial effect are integrated out. This integration can be done explicitly 
because the spatial field by construction is Gaussian and the parameters of the
fixed effects are Gaussian due to the choice of a Gaussian penalty. 

First, collect the fixed effect and the spatial effect in
\(\vec{z} = (\vec{u}^\mathrm{T}, \vec{\beta}^\mathrm{T})\). The
model given the value of \(\vec{\theta}\) can then be written as
\[
	\vec{z}|\vec{\theta}  \sim \mathcal{N}_{mn+p}(\vec{0}, \mathbf{Q}_{z}^{-1})
\]
and
\[
	\vec{y} | \vec{z}, \vec{\theta} \sim \mathcal{N}_{N}(\mathbf{S}\vec{z}, \mathbf{I}_N/\tau_{\mathrm{noise}}),
\]
where
\begin{align*}
	\mathbf{S} = \begin{bmatrix} \mathbf{E} & \mathbf{X} \end{bmatrix} \quad \text{and} \quad \mathbf{Q}_z = \begin{bmatrix} \mathbf{Q} & \mathbf{0} \\ \mathbf{0} & \tau_\beta \mathbf{I}_p \end{bmatrix}.
\end{align*}
We then use the fact that both these distributions are Gaussian to 
integrate out \(\vec{z}\) from the likelihood, as shown in 
\ref{app:CondDist}. This gives the full penalized log-likelihood
\begin{align}
\notag \log(\pi(\vec{\theta}|\vec{y})) &= \mathrm{Const} -\frac{1}{2}\sum_{i=1}^4\vec{\alpha}_i^\mathrm{T}\mathbf{Q}_{\mathrm{RW2}}\vec{\alpha}_i\cdot\tau_i +\frac{1}{2}\log(\det(\mathbf{Q}_z)) +\frac{N}{2}\log(\tau_\mathrm{noise})+ \\
	&-\frac{1}{2}\log(\det(\mathbf{Q}_\mathrm{C}))-\frac{1}{2}\vec{\mu}_\mathrm{C}^\mathrm{T}\mathbf{Q}_z\vec{\mu}_\mathrm{C}-\frac{\tau_\mathrm{noise}}{2}(\vec{y}-\mathbf{S}\vec{\mu}_\mathrm{C})^\mathrm{T}(\vec{y}-\mathbf{S}\vec{\mu}_\mathrm{C}),
	\label{eq:fullPost}
\end{align}
where \(\mathbf{Q}_\mathrm{C} = \mathbf{Q}_z+\mathbf{S}^\mathrm{T}\mathbf{S}\cdot\tau_\mathrm{noise}\)
and \(\vec{\mu}_\mathrm{C} = \mathbf{Q}_\mathrm{C}^{-1}\mathbf{S}^\mathrm{T}\vec{y}\cdot\tau_\mathrm{noise}\).

The first step of the inference scheme is to estimate the covariance parameters 
\(\vec{\theta}\) with the value \({\skew{3}\hat{\vec{\theta}}}\)
that maximizes Equation~\eqref{eq:fullPost}. This value is then used to calculate
predictions and prediction standard deviations at new locations
\(\vec{y}^*\) by using the predictive distribution 
\(\vec{y}^*|{\skew{3}\hat{\vec{\theta}}}, \vec{y}\). However, the 
penalty parameters that control the penalty of the covariance parameters are 
difficult to estimate. The profile likelihoods are hard to calculate and there is
not enough information on such a low stage of the hierarchical model to estimate
them together with the covariance parameters.
Thus they have to be pre-selected, based on intuition about how much the
covariance structure should be allowed to vary, or chosen with a 
cross-validation procedure based on a scoring rule for the predictions. 

During implementation of the inference scheme it became apparent that an 
analytic expression for the gradient was needed for the optimization to converge. 
Its form is given in \ref{app:Gradient}, and its value can be computed for 
less cost than a finite difference approximation of the gradient for the number 
of parameters used in the application in this paper. The calculations require the 
use of techniques for calculating only parts of the inverse of a sparse precision 
matrix~\citep{col27}.

\section{Non-stationarity in a single realization}
\label{sec:Application}
\subsection{Adaptive smoothing framework}
We begin by considering the common situation in spatial statistics where only 
a single realization is available. In this situation it is theoretically 
impossible to separate non-stationarity in the mean and in the covariance structure, 
and the non-stationary model is better described as adaptive 
smoothing. The non-stationary model allows the degree of smoothing to vary 
over space, and areas with long range will have high smoothing and areas with 
short range will have low smoothing. In practice, parts of the non-stationarity in 
the mean structure will be captured in the covariance structure,
but this is not necessarily a problem and might lead to better predictions.
The main interest is finding out whether the complex non-stationary model 
improves predictions at unobserved locations and at whether the computational
costs are worth it.

We select the
year 1981 which has 7040 measurement stations and want to predict the annual precipitation in the entire conterminous US with associated prediction standard
deviations. Two covariates are used in the mean structure: 
an intercept and a linear effect of elevation. This means that 
the design matrix, \(\mathbf{X}\), in Equation~\eqref{eq:DataEquation} has two 
columns.
The first column contains only ones, and corresponds to the intercept, and the second
column contains elevations measured in kilometres. There should be strong information
about the two covariates and a weak penalty is applied to the coefficients of the 
fixed effects,
\(\vec{\beta} \sim \mathcal{N}_2(\vec{0}, \mathbf{I}_2\cdot 10^{4})\).

\subsection{Stationary model}
\label{sec:StatModel}
The spatial effect is constructed on a rectangular domain with 
longitudes from \(130.15\,^\circ\mathrm{W}\) to \(60.85\,^\circ\mathrm{W}\)
and latitudes from \(21.65\,^\circ \mathrm{N}\) to \(51.35\,^\circ \mathrm{N}\). 
This is larger than the actual size of the conterminous US as can be seen
in Figure~\ref{fig:observation}, and is chosen to reduce
boundary effects. The domain is discrectized into a \(400 \times 200\) grid
and the parameters \(\log(\kappa^2)\), \(\log(\gamma)\), \(v_x\), \(v_y\) and
\(\log(\tau_{\mathrm{noise}})\) are estimated. In this case the second order 
random walk penalty is not used as no basis (except a constant) is needed for 
the functions. The estimated values with associated approximate standard deviations
are shown in Table~\ref{tab:StatPar}. The approximate standard deviations 
are calculated from the observed information matrix.

\begin{table}
	\centering
	\caption{Estimated values of the parameters and associated approximate standard deviations
			 for the stationary model.}
	\begin{tabular}{lll}
		\textbf{Parameter}			& \textbf{Estimate}	& \textbf{Standard deviation} \\
		\(\log(\kappa^2)\)			& \(-1.75\)			& \(0.15\)	\\
		\(\log(\gamma)\)			& \(-0.272\)		& \(0.042\)	\\
		\(v_x\)						& \(0.477\)			& \(0.053\) \\
		\(v_y\)						& \(-0.313\)		& \(0.057\)	\\
		\(\log(\tau_{\mathrm{noise}})\)	& \(4.266\)			& \(0.030\)
	\end{tabular}
	\label{tab:StatPar}
\end{table}

\begin{figure}
	\centering
	\includegraphics[width=10cm]{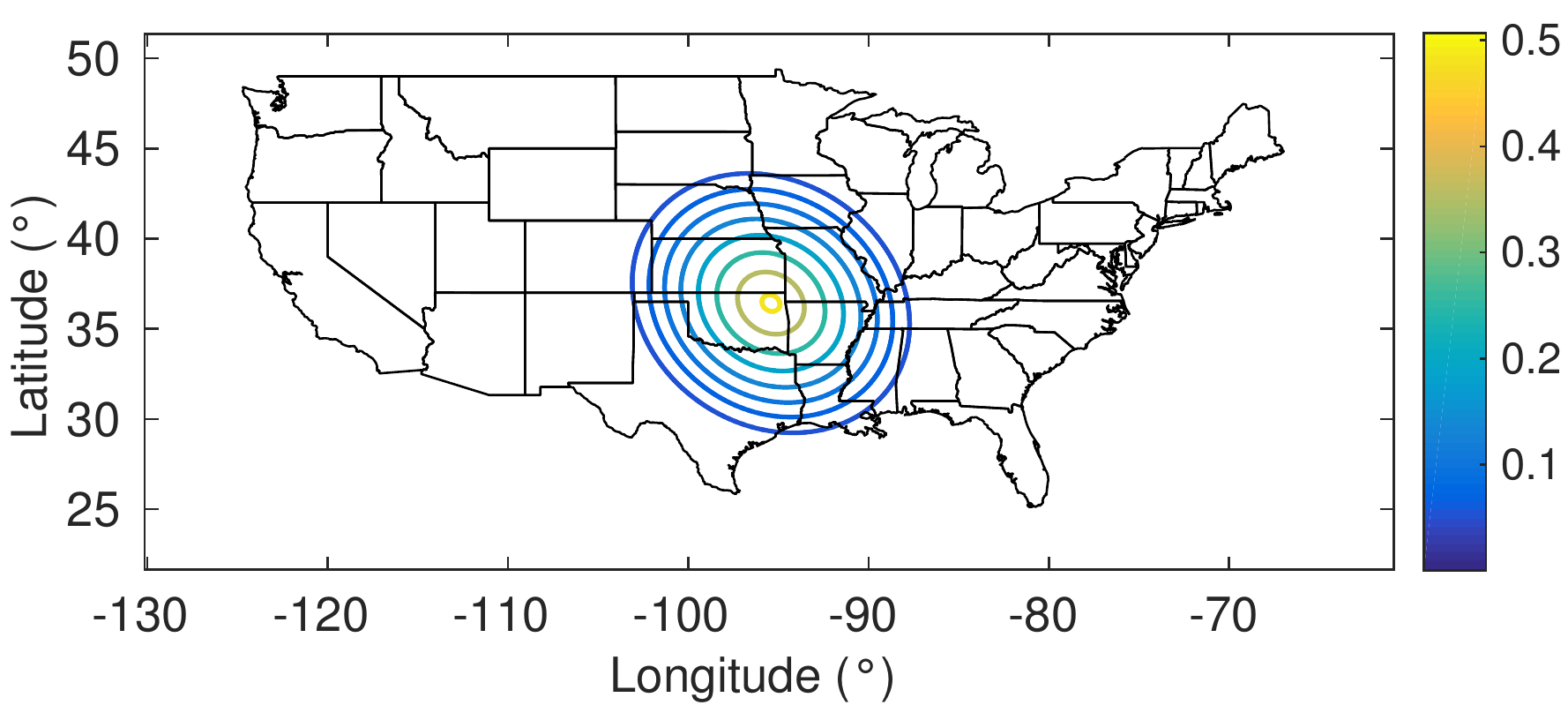}
	\caption{The 0.95, 0.70, 0.50, 0.36, 0.26, 0.19, 0.14 and 0.1 level correlation contours of the estimated covariance function for the stationary model.}
	\label{fig:statCov}
\end{figure}

From Section~\ref{sec:GRF} one can see that the estimated model implies a covariance
function approximately equal to the Mat\'{e}rn covariance function
\[
	r(\vec{s}_1,\vec{s}_2) = \hat{\sigma}^2 \left\lvert\left\lvert \left(\hat{\mathbf{H}}/\hat{\kappa}^2\right)^{-1/2}(\vec{s}_2-\vec{s}_1)\right\rvert\right\rvert K_1\left(\left\lvert\left\lvert \left(\hat{\mathbf{H}}/\hat{\kappa}^2\right)^{-1/2}(\vec{s}_2-\vec{s}_1)\right\rvert\right\rvert\right),
\]
where \(\hat{\sigma}^2 = 0.505\) and
\[
	\frac{\hat{\mathbf{H}}}{\hat{\kappa}^2} = \begin{bmatrix} 5.71 & -0.86 \\ -0.86 & 4.96 \end{bmatrix},
\] 
together with a nugget effect with precision \(\hat{\tau}_\mathrm{noise} = 71.2\).
Figure~\ref{fig:statCov} shows contours of the estimated covariance function with 
respect to a chosen location. One can see that the model gives high dependence within
a typical-sized state, whereas there is little dependence between the centres of
different typically-sized states.

\begin{figure}
	\centering
	\subfigure[Prediction for the stationary model]{
		\includegraphics[width=5.5cm]{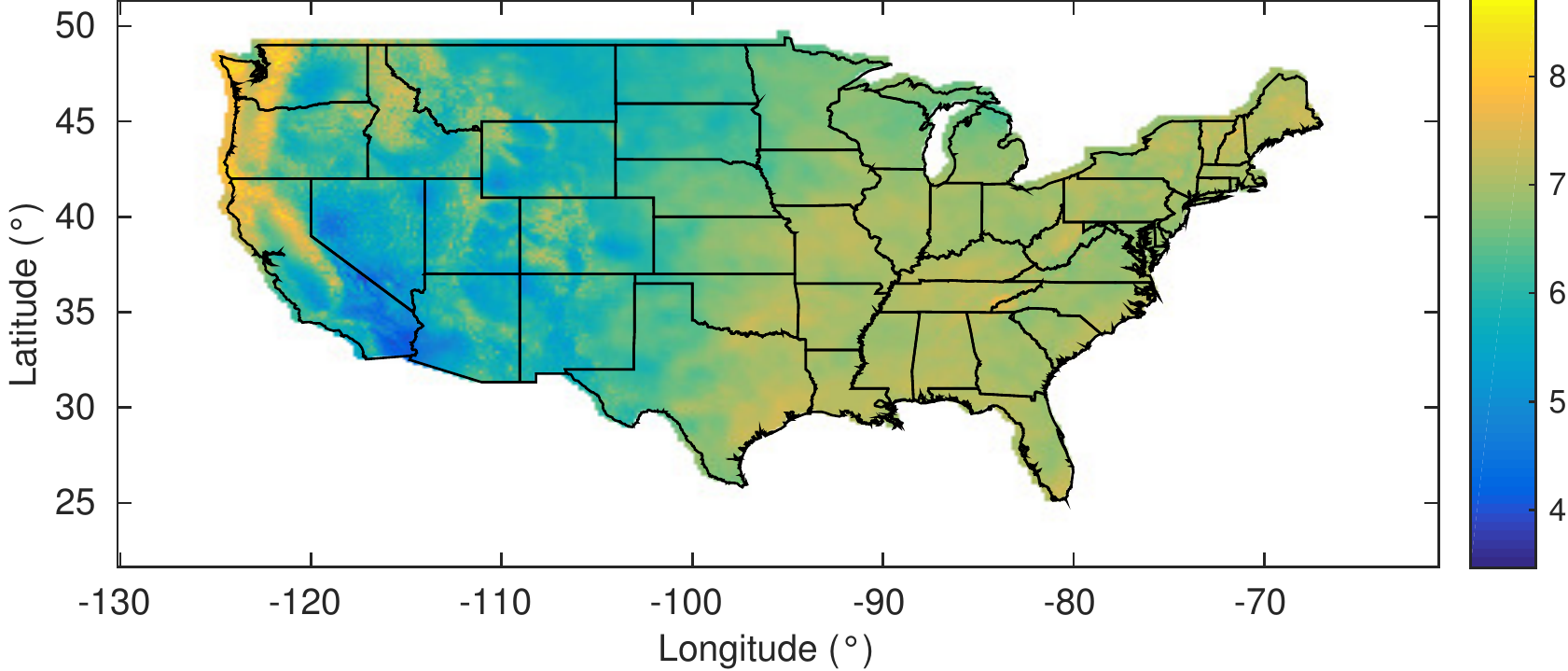}
		\label{fig:StatPred:Pred}
	}	
	\subfigure[Prediction for the non-stationary model]{
		\includegraphics[width=5.5cm]{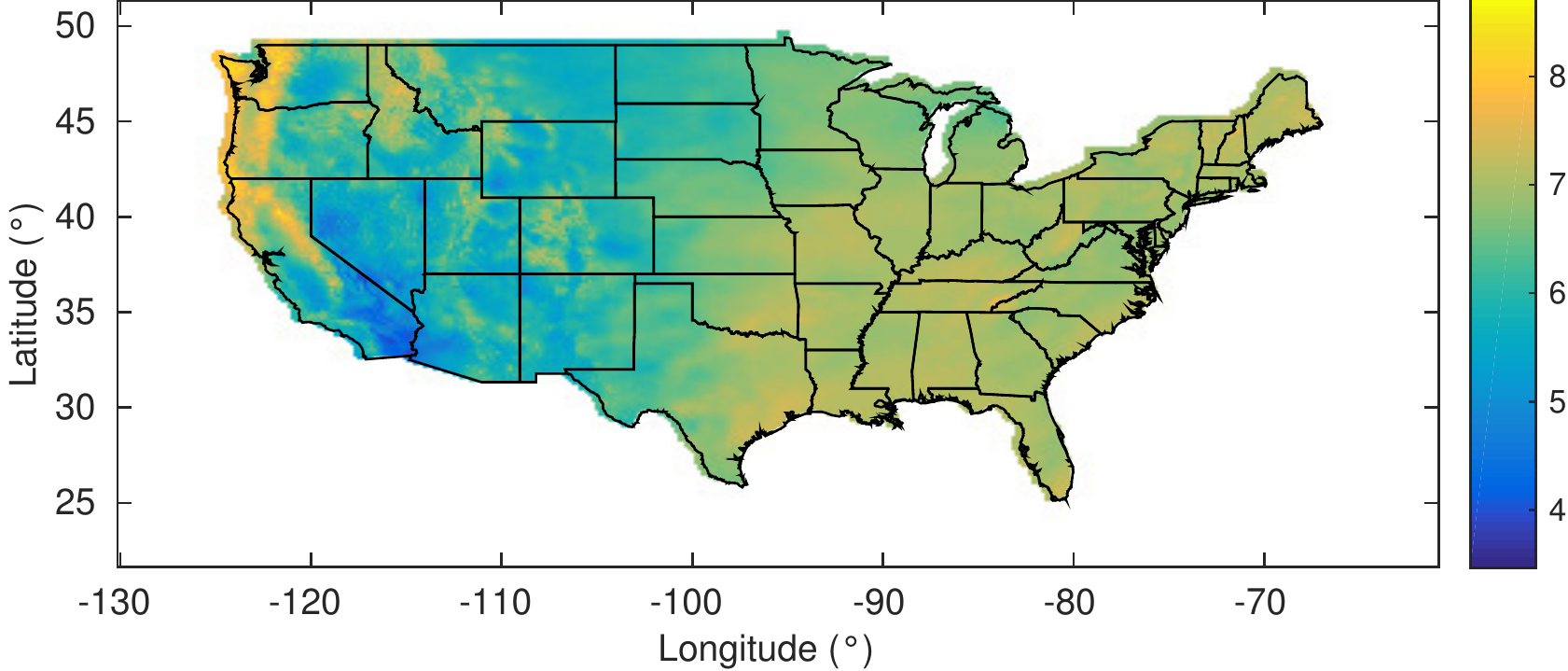}
		\label{fig:NonStatPred:Pred}
	}\\
		\subfigure[Prediction standard deviations for the stationary model]{
		\includegraphics[width=5.5cm]{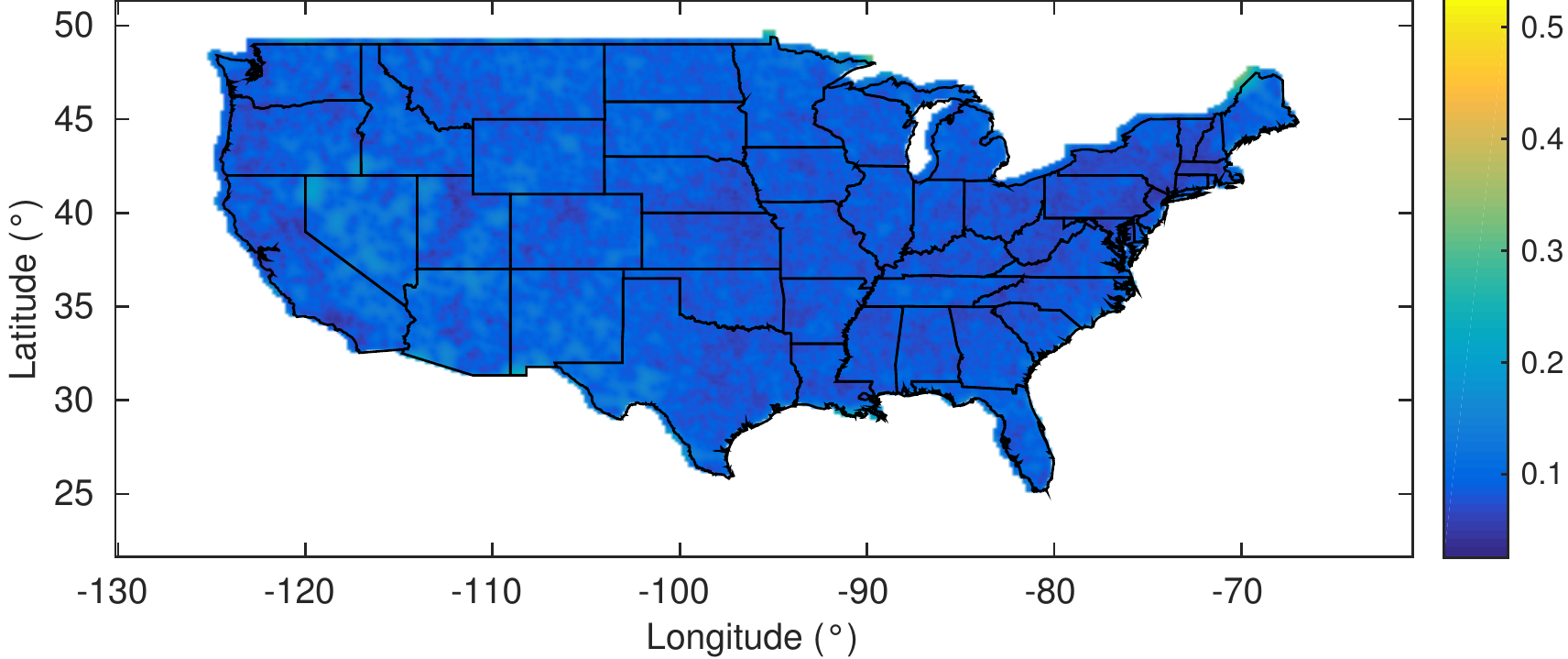}
		\label{fig:StatPred:StdDev}
	}
	\subfigure[Prediction standard deviations for the non-stationary model]{
		\includegraphics[width=5.5cm]{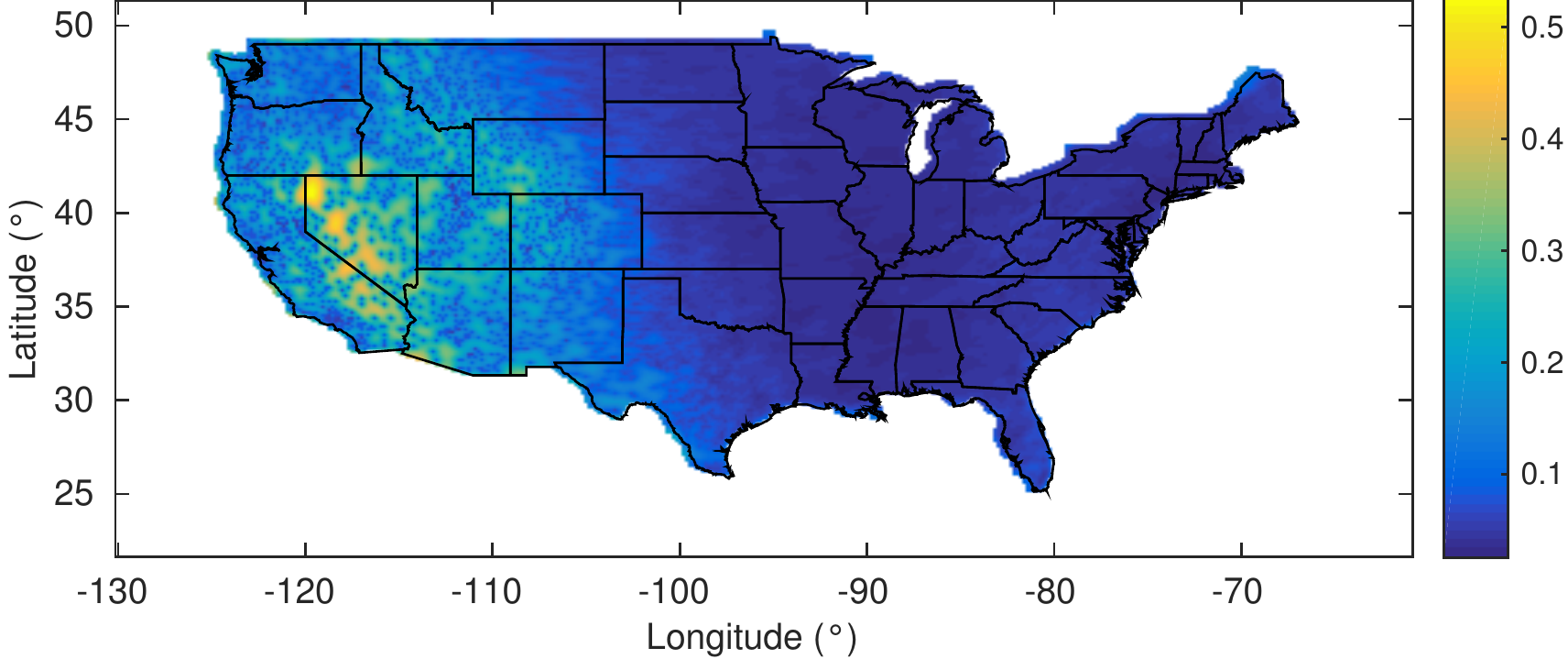}
		\label{fig:NonStatPred:StdDev}
	}

	\caption{Predicted values and prediction standard deviations for the
			 stationary model and the non-stationary model.}
	\label{fig:StatPred}
\end{figure}

Next, the parameter values are used together with the observed
logarithms of annual precipitations to predict the logarithm of
annual precipitation at the centre of each cell in the discretization.
The elevation covariate
for each location is selected from bilinear interpolation from the closest 
points in the high resolution elevation data set GLOBE~\citep{Globe1999}. 
The predictions and prediction standard deviations are shown in 
Figures~\ref{fig:StatPred:Pred} and \ref{fig:StatPred:StdDev}. 
Since there only are observations within the
conterminous US and this is the area of interest, 
the locations outside are coloured white. 

\subsection{Non-stationary model}
\label{sec:SmoothSel}
The parameters \(\tau_1\), \(\tau_2\), \(\tau_3\) and \(\tau_4\), that appear in 
the penalty for the functions \(\log(\kappa^2)\), \(\log(\gamma)\), \(v_x\)
and \(v_y\), respectively, have to be chosen before the rest of the inference
is started. The parameters are chosen with 5-fold cross-validation based on 
the log-predictive density. The data is randomly divided into five parts 
and in turn one part is used as test data and the other four parts are
used as training data. For each choice of \(\tau_1\), \(\tau_2\), \(\tau_3\) and
\(\tau_4\) the cross-validation error is calculated by
\[
	\mathrm{CV}(\tau_1, \tau_2, \tau_3, \tau_4) = -\frac{1}{5}\sum_{i=1}^{5} \log(\pi(\vec{y}_i^*|\vec{y}_i, {\skew{3}\hat{\vec{\theta}}}_i),
\]
where \(\vec{y}_i^*\) is the test data and \({\skew{3}\hat{\vec{\theta}}}_i\)
is the estimated covariance parameters based
on the training data \(\vec{y}_i\) using the selected \(\tau\)-values. 
The cross validation is done over \(\log(\tau_i)\in\{2,4,6,8\}\) for 
\(i = 1, 2, 3, 4\). We selected four values for each parameter to have
a balance between the need to test strong and weak penalties and 
to make the problem computationally feasible. Controlling the penalty
on non-stationarity is important, but appropriate penalty values are
not easily deduced from the model. Therefore, different values were tested
to determine values of \(\tau_i\) that corresponds to a weak penalty and
a strong penalty and then four points were chosen linearly on log-scale since
\(\tau_i\) acts as a scale parameter.
We use the same domain size as for the stationary model, but
reduce the grid size to  \(200 \times 100\) with \(8 \times 4\) basis functions 
for each function. The choice that gave the smallest cross-validation error was
\(\log(\tau_1) = 2\), \(\log(\tau_2) = 4\), \(\log(\tau_3) = 2\) 
and \(\log(\tau_4) = 8\).

After the penalty parameters are selected, the grid size is increased to
\(400 \times 200\) and each of the four functions in the SPDE is given a 
\(16 \times 8\) basis functions. Together with the precision
parameter of the random effect this gives a total of 513 parameters. These parameters
are estimated together based on the integrated likelihood. Note that 
there are 
not 513 ``free'' parameters as they are connected together in four different
penalties enforcing slowly changing functions. This means that an increase in the 
number of parameters increases the resolutions of the functions, but not directly
the degrees of freedom in the model.

\begin{figure}
	\centering
	\subfigure[Marginal standard deviations]{
		\includegraphics[width=7cm]{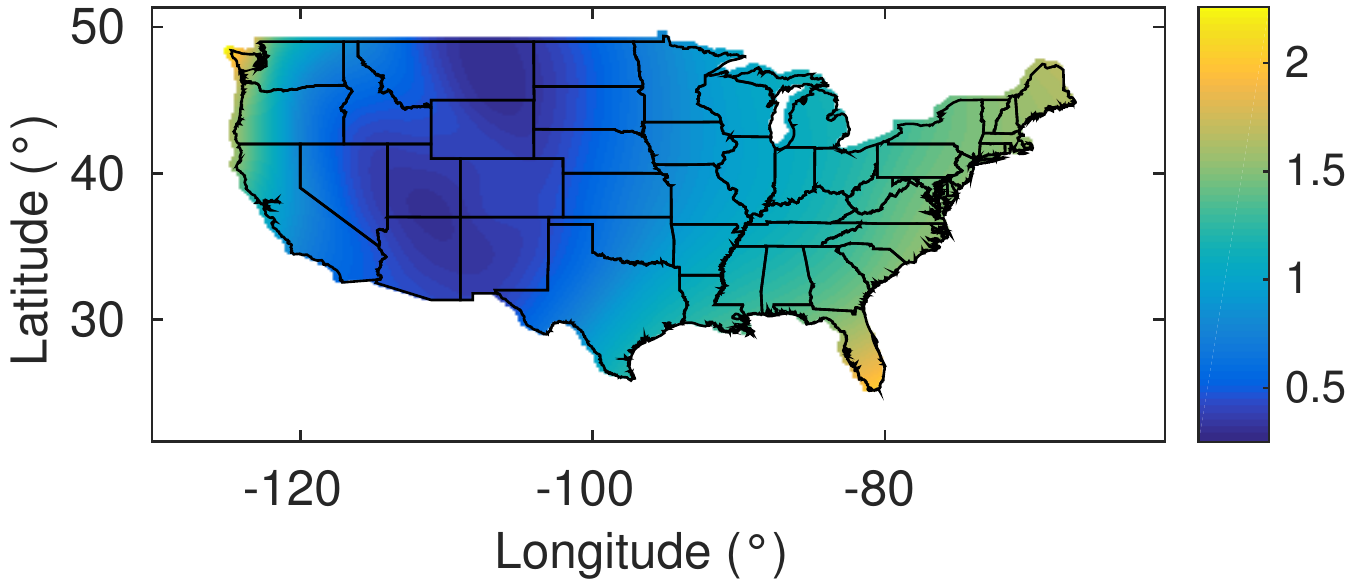}
		\label{fig:NonStatPar:TrueStdDev}
	}
	\subfigure[Correlation structure]{
		\includegraphics[width=7cm]{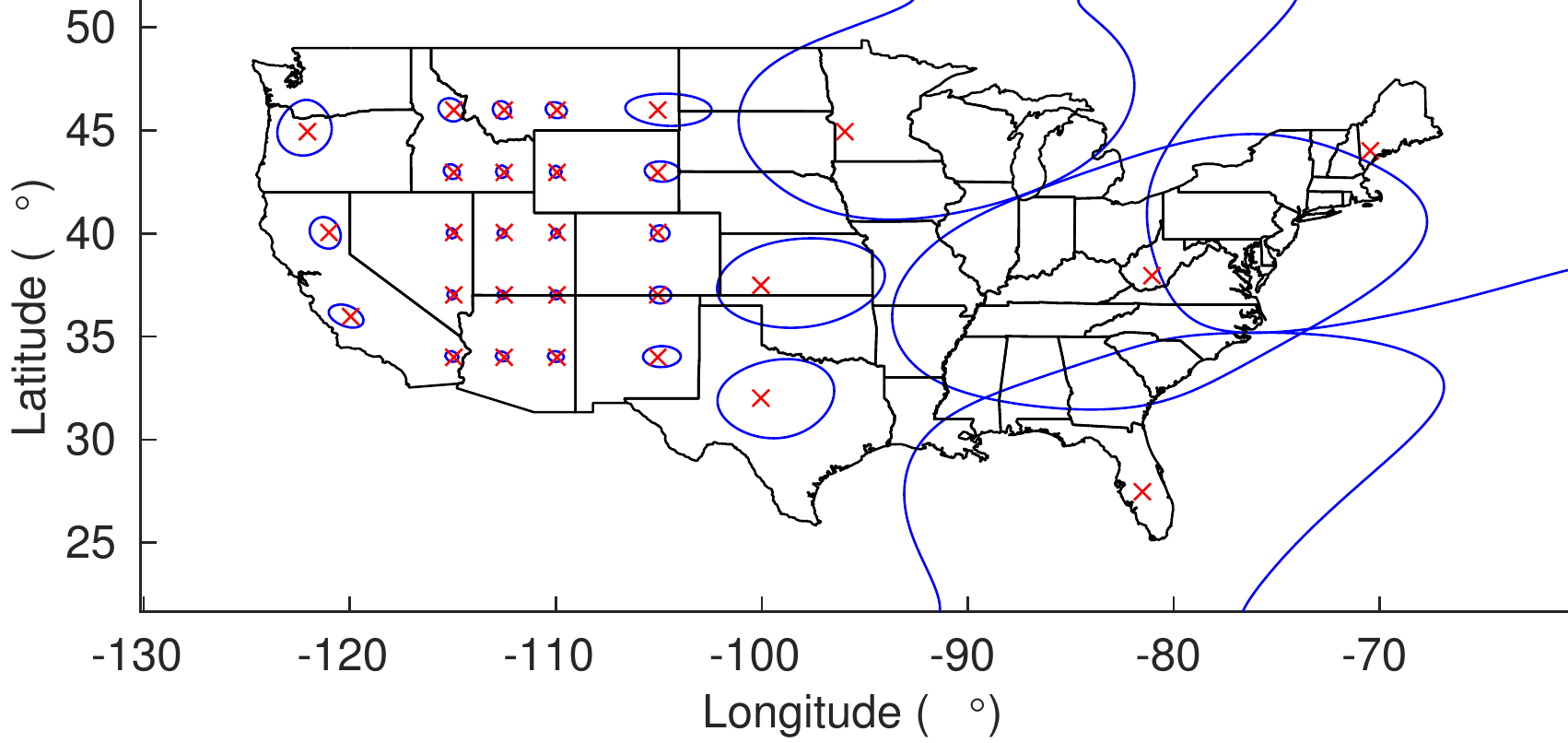}
		\label{fig:NonStatPar:TrueCorr}
	}
	\caption{Estimated covariance structure of the spatial field.
			 \subref{fig:NonStatPar:TrueStdDev} Marginal standard deviations
			 \subref{fig:NonStatPar:TrueCorr} Contours of 0.7 correlation
			 for selected locations marked with red crosses
			 }
	\label{fig:NonStatPar}
\end{figure}

The nugget effect is estimated to have a precision of 
\(\hat{\tau}_\mathrm{noise} = 107.4\). The estimates of \(\kappa^2\) and \(\mathbf{H}\)
are not shown since the exact values themselves are not interesting. We calculate
instead the marginal standard deviations for all locations and 0.7 level
correlation contours for selected locations in 
Figure~\ref{fig:NonStatPar:TrueStdDev} and Figure~\ref{fig:NonStatPar:TrueCorr},
respectively.
From these figures one can see that the estimated covariance structure is different
from the estimated covariance structure for the stationary model shown in 
Figure~\ref{fig:statCov}. In the non-stationary model we have a much longer range
in the eastern part and a much short range in the mountainous areas in the west.

The estimated covariance structure implies strong smoothing in the eastern
region and weak smoothing in the western region. This must be understood to 
say something about both how well the covariates describe the data at 
different locations and the underlying non-stationarity in the covariance structure
of the physical phenomenon. 
In this case there is a good fit for the elevation covariate in the mountainous areas 
in the western part, but it offers less information in the eastern part. From 
Figure~\ref{fig:observation} one can see that at around longitude 
\(97^\circ\, \mathrm{W}\) there is an increase in precipitation 
which cannot be explained by elevation, and thus is not captured by the
covariates. This jump must therefore be explained by the covariance structure, and
in this case it is explained by having the covariates fit well in the western region
and explaining the high values in the eastern region as being caused, randomly, by
a spatial process with a long range. 

In the same way as in Section~\ref{sec:StatModel} the logarithm of annual
precipitation is predicted at the centre of each cell in the discretization. 
This gives predictions for \(400 \times 200\) regularly distributed locations, 
where the value of the elevation covariate at each location is selected with 
bilinear interpolation from the closest points in the GLOBE~\citep{Globe1999} dataset.
The prediction and prediction standard deviations are shown in 
Figures~\ref{fig:NonStatPred:Pred} and \ref{fig:NonStatPred:StdDev}. As for the 
stationary model, the values outside
the conterminous US are coloured white. One can see that the overall look of the 
predictions is similar to the predictions from the stationary model, 
but that the prediction standard deviations differ. The prediction standard 
deviations vary strongly over the spatial domain because of the extreme 
differences in spatial range for the estimated non-stationary model.

\subsection{Evaluation of predictions}
\label{sec:ComparisonOneYear}
The predictions of the stationary model and the non-stationary model are
compared with the continuous rank probability score (CRPS)~\citep{Gneiting2005} 
and the logarithmic
scoring rule. CRPS is defined for a univariate distribution as
\[
	\mathrm{crps}(F, y) = \int_{-\infty}^{\infty}\! (F(y)-\mathbbm{1}(y\leq t))^2\, \mathrm{d}t,
\]
where \(F\) is the distribution function of interest, \(y\) is an observation and
\(\mathbbm{1}\) is the indicator function. This gives a measure of how well
a single observation fits a distribution. The total score is calculated
as the average CRPS for the test data,
\[
	\mathrm{CRPS} = \frac{1}{N}\sum_{i=1}^{N}\mathrm{crps}(F_k, y_k),
\]
where \(\{y_k\}\) is the test data and \(\{F_k\}\) are the corresponding marginal
predictive
distributions given the estimated covariance parameters and the training data. 
The logarithmic scoring rule is based on the joint predictive distribution of 
the test data \(\vec{y}^*\) given the estimated covariance parameters 
\(\hat{\vec{\theta}}\) and the training data \(\vec{y}\),
\[
	\mathrm{LogScore} = -\log\pi\left(\vec{y}^*|\hat{\vec{\theta}},\vec{y}\right).
\]

\begin{figure}
	\centering
	\subfigure[Log-predictive score]{
		\includegraphics[height=3.3cm]{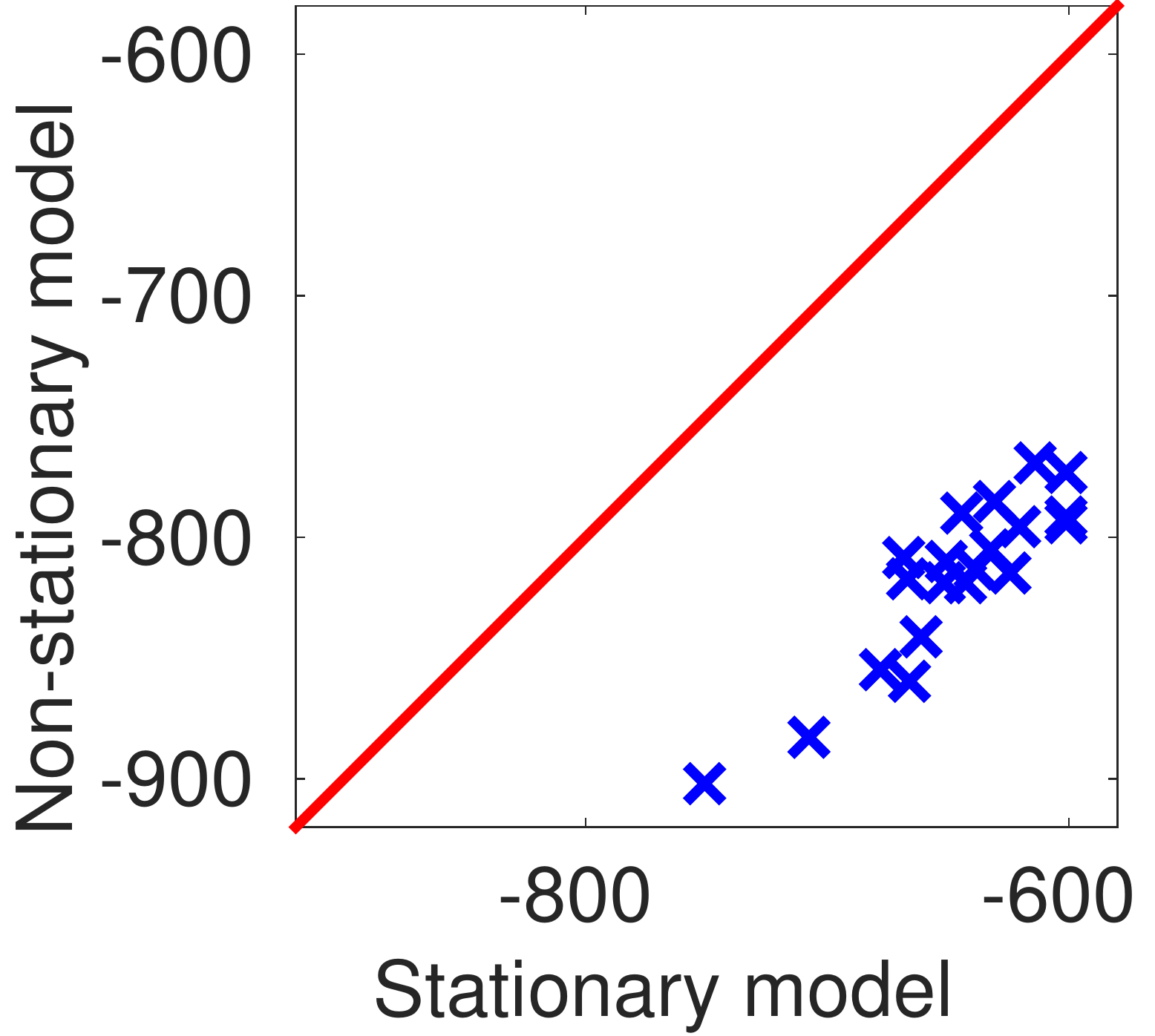}
		\label{fig:Comp:LPD}
	}
	\subfigure[CRPS]{
		\includegraphics[height=3.3cm]{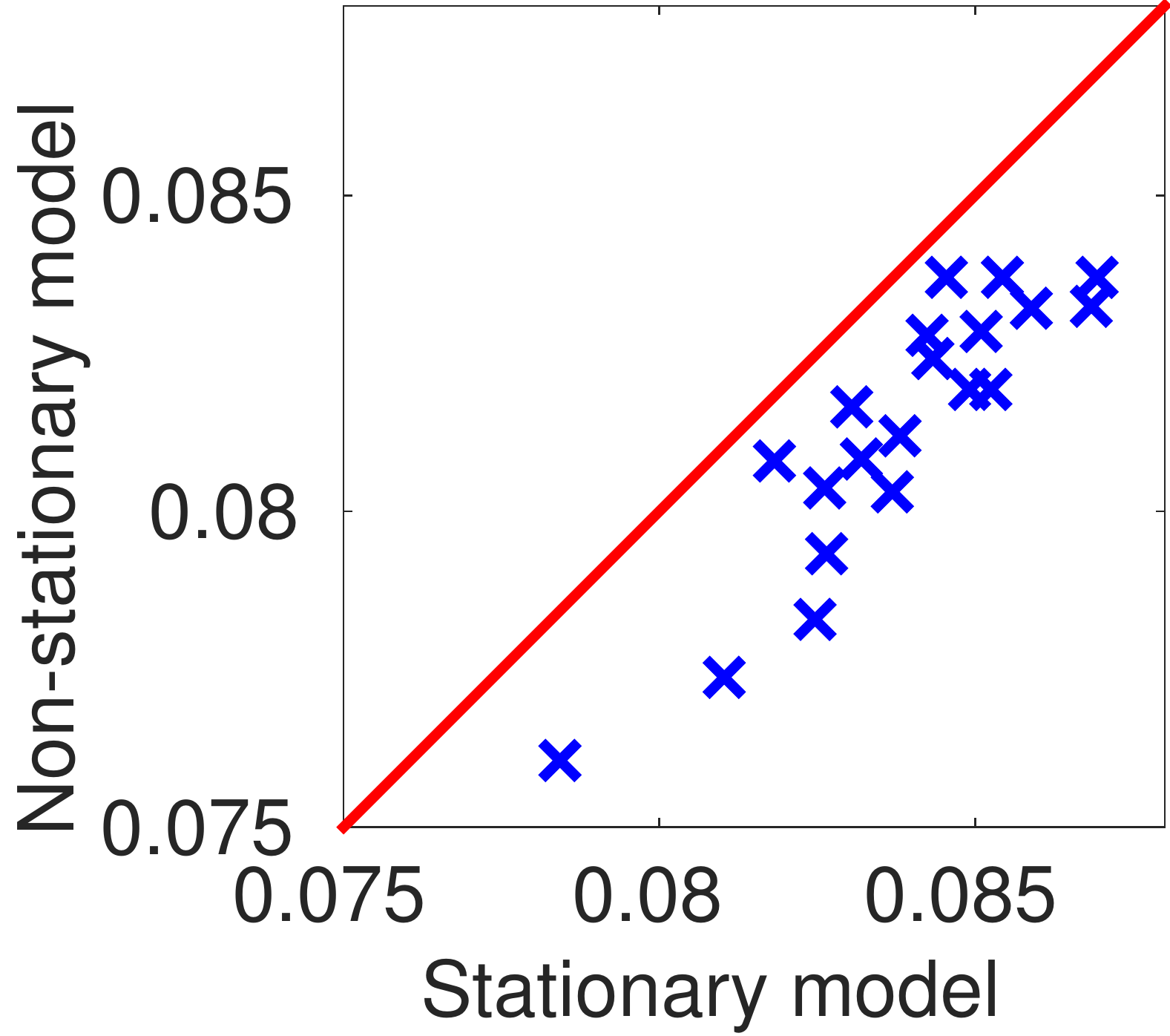}
		\label{fig:Comp:CRPS}
	}
	\subfigure[RMSE]{
		\includegraphics[height=3.3cm]{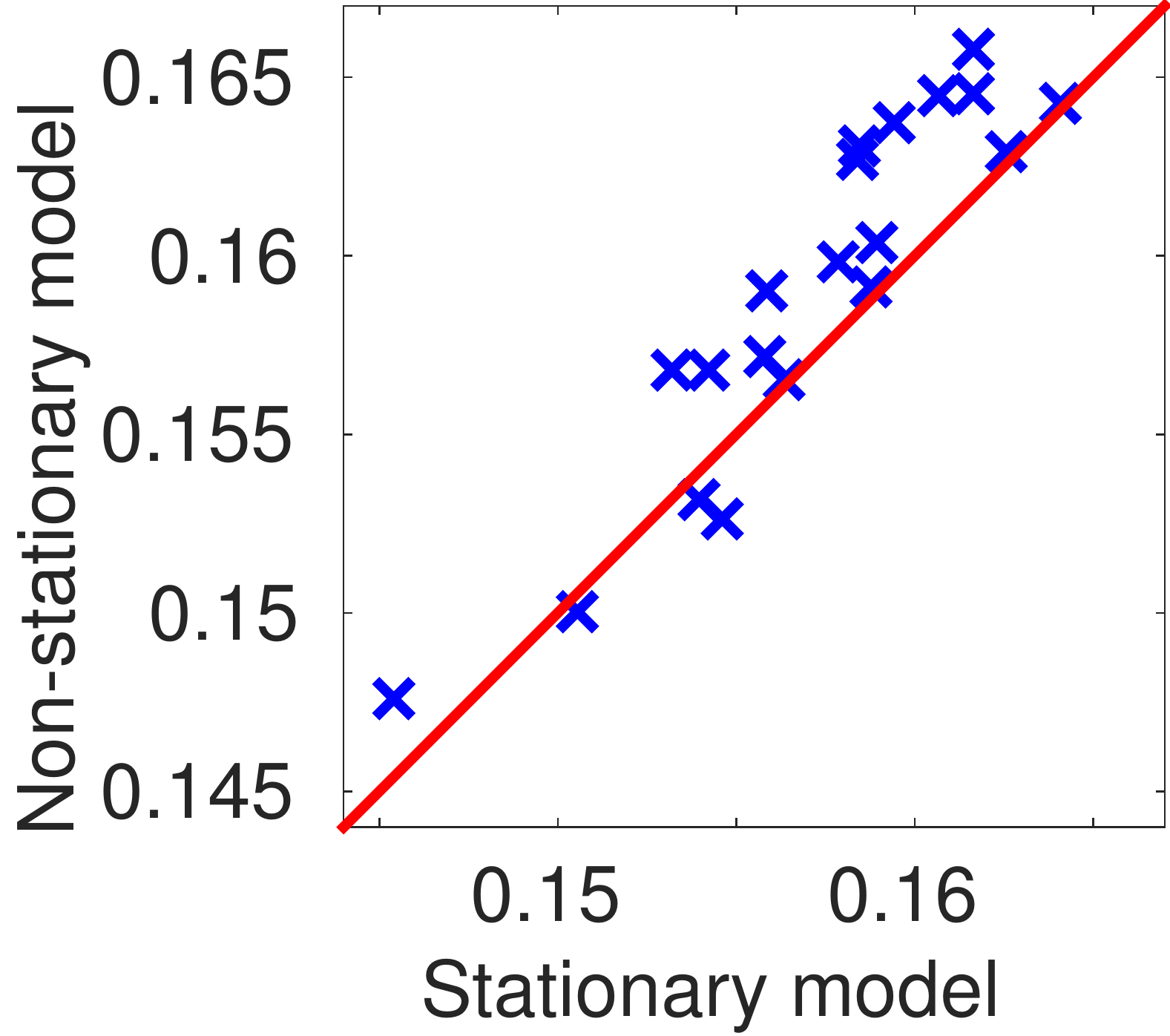}
		\label{fig:Comp:RMSE}
	}
	\caption{Scatter plots of prediction scores from the stationary and
			 the non-stationary model. 20\% of the locations are randomly chosen
			 to be held out and the remaining 80\% are used to estimate parameters
			 and predict the 20\% held out data. This was repeated 20 times. For values
			 below the line the non-stationary model is better, and conversely for
			 values above the line.}
	\label{fig:Comp}
\end{figure}

The comparison of the models is done using holdout sets where 
each holdout set consists of 20\% of the locations chosen randomly. The remaining 
80\% of the locations are used to estimate the parameters and to predict the 
values at the locations in the holdout set. This procedure is repeated 20 times. 
For each repetition the CRPS, the logarithmic score and the root mean square
error (RMSE) are calculated. From Figure~\ref{fig:Comp} one can see that measured 
by both log-predictive score and CRPS the non-stationary model gives better
predictions, but that the RMSE does not show any improvement. 

However, the RMSE is based only on the point predictions and does not
incorporate the prediction variances. The log-predictive score and the
CRPS are more interesting since they say something about how well the predictive
distributions fit.  
The difference in log-predictive score is large and indicates
that the non-stationary model is better, but the difference in CRPS is small
and indicates only a small improvement. The likely cause for this is 
that the log-predictive score evaluates the joint predictive distributions and 
there are difference which are not showing in the univariate predictive 
distributions. 

The full cross-validation procedure for selecting 
the penalty parameters is expensive and takes weeks and must be evaluated against
the potential gain in any application.
The results shows that the choice of scoring rule has a strong influence
on the conclusion of whether the non-stationary model was worth it. The CRPS does
not show evidence that all the extra computation time was worth it, but according
to the log-predictive score there is a large improvement.

\subsection{Criticism}
The log-predictive score and CRPS are better for the non-stationary model for each
hold-out set, but the covariance structure shown in Figure~\ref{fig:NonStatPar}
is troubling. The range was estimated long and the marginal variances were
estimated high in the eastern part because this was the ``best'' way to explain 
the changes observed, but we do not truly believe the estimates. The long 
estimated range 
means that most of the eastern part is highly correlated and the high marginal 
variance means that next year there might be a large change in the level in the 
eastern part. Whereas the low marginal variance in the west means that there will
be far less changes in the spatial field there the next year. This is clearly 
wrong since the data for different years do not show huge changes,
which are compatible with the estimated standard deviations of the spatial 
field, in the level of precipitation between years in the eastern region. 

It is well-known that the range and the marginal variance of the stationary
Mat\'ern model are not identifiable from a fixed-size observation 
window~\citep{Zhang2004}, and the situation is not
likely to improve for a complex model with spatially varying marginal variances
and covariance structure, but what we are seeing is the result of forcing the
model to include mean structure in the covariance structure.
Based on data from multiple years it is clear that  
the difference in level between the western and eastern region is actually caused by
a change in the mean. Further, the short range in the west
is also problematic because it means that few of surrounding data points are being used
to predict values in this part of the domain. This could mean that the spatial
effect is weak in this region, but the estimated covariance structure gives 
evidence that we need to investigate the cause more thoroughly.

This makes an important point regarding the worth of the non-stationary model. 
Whether we have improved the CRPS and the log-predictive
score is not the only question worth asking. We have gained understanding about 
issues in the estimated covariance structure that we need to investigate to
understand where the non-stationarity is coming from and whether it is 
correctly captured in the model. In this case we have 
gained something more than an improvement in prediction scores. 
We have identified two potential issues with the model: the wrongly specified 
mean, which we knew about, and the weak spatial effect in the western 
region, which we need investigate.

\section{Non-stationarity in multiple realizations}
\label{sec:detrend}
\subsection{Non-stationary modelling framework}
If we use multiple realizations, the non-stationarity in the mean and the 
non-stationarity in the covariance structure are separable if the spatio-temporal
process can be assumed to be stationary in time. Modelling the mean structure and the covariance structure
separately goes beyond adaptive smoothing 
and is a situation where the term non-stationary modelling is accurate.
The goal in this section is to separate out the non-stationarity in the mean
and to investigate the two issues we discovered in the analysis of a single year in 
the previous section: over-smoothing in the eastern region and under-smoothing 
in the western region.

We repeat the analysis using data from the years 
1971--1985, and we want to see how much the predictions improve and how 
the estimated non-stationary changes with a better model for the mean. 
Ideally, one could fit a full spatio-temporal model to these years, 
but since the focus is on the spatial non-stationarity we will assume that the 
15 years are independent realizations of the same spatial process. Since we are
using precipitation data aggregated to yearly data, the temporal dependence is weak
and this is a reasonable simplification.

\subsection{De-trending}
The first step in the analysis is to de-trend the dataset. Each year
has a different number of observations and some observations are at different
locations, which means that there will be different missing locations for
each year. The de-trending is done with a simple model that assumes 
that each year is an independent realization of a stationary spatial field and
 is observed with measurement noises with the same variance. 
The model is estimated based on the observations, and 
the values at locations of interest at each year is filled in based 
on the posterior marginal conditional means. Then we take the average of the 
fitted values over the 15 year period as an estimate of the true mean.

The simple model is fitted using the \textsf{R} package \textsf{INLA},
which is based on the INLA method of~\citep{Rue2009}. The model used is
\[
	y(\vec{s}_i, t) = \mu + x(\vec{s}_i)\beta+ u_t(\vec{s}_i) + a_t+\epsilon_{i,t}, \quad i=1,2,\ldots, N_t \quad t = 1971,1972,\ldots, 1985,
\]
where \(\mu\) is the joint mean for all observations, \(x(\vec{s}_i)\) is
the elevation at location \(\vec{s}_i\) and \(\beta\) is the
associated coefficient for the covariate, \(u_t\) for \(t=1971, 1972,\ldots,1985\) 
are independent realizations of the spatial effect for each year, \(a_t\) is an 
AR(1) process supposed to capture temporal changes in the joint mean between years,
and \(\epsilon_{i,t}\) are independent Gaussian measurement errors. The
spatial effect is approximately Mat\'ern with smoothness parameter
\(\nu = 1\). The model is estimated and used to predict the values at all locations 
of interest in all 15 years. The estimate of the true mean 
\(\hat{\mu}(\vec{s})\), at location \(\vec{s}\), is found by taking 
the average over the estimated value at each year.

\begin{figure}
	\centering
	\includegraphics[width=11cm]{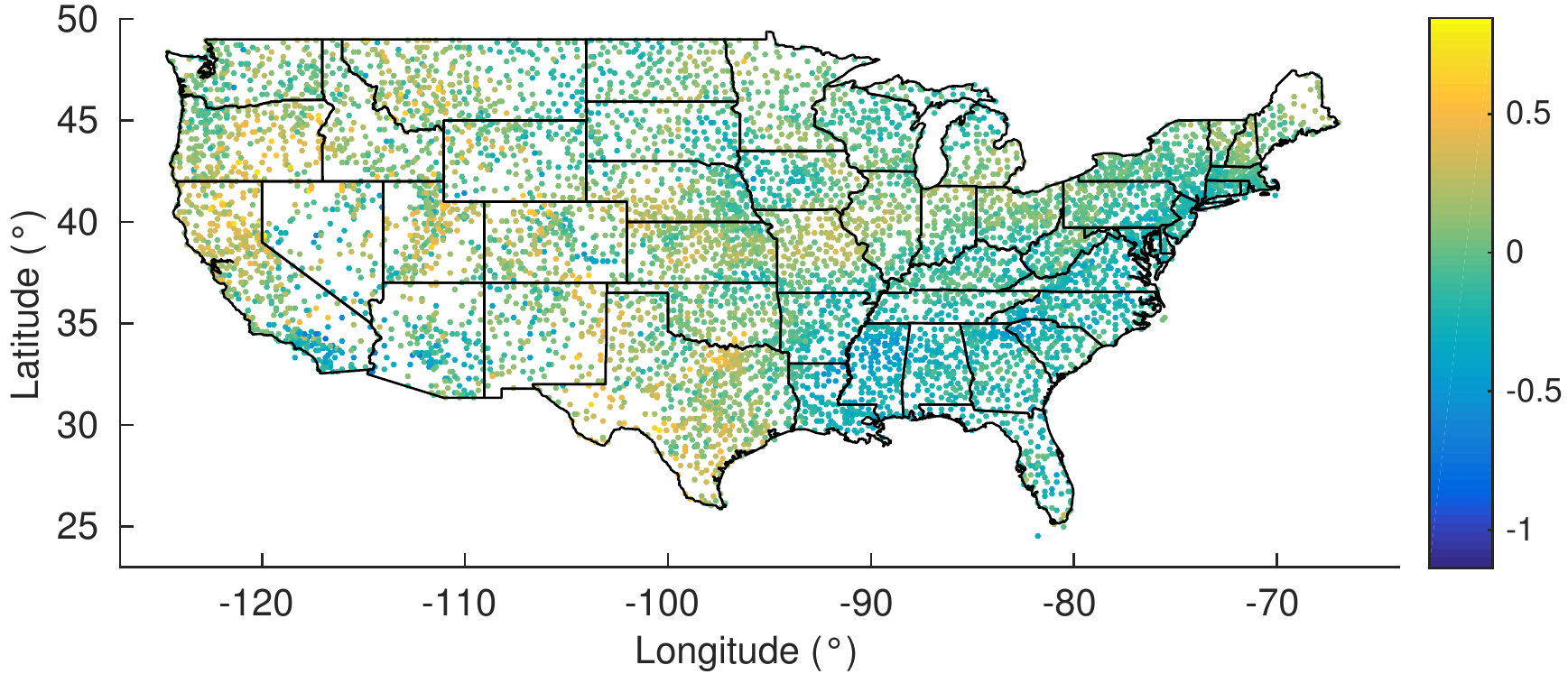}
	\caption{De-trended observations of log-transformed total annual precipitation measured in millimeter for 1981.}
	\label{fig:deTrendObs}
\end{figure}

In the rest of the section we focus on the residuals 
\(y(\vec{s}_i, t) - \hat{\mu}(\vec{s}_i)\). This means that the
estimate of the mean is assumed to be without uncertainty. The intention is 
to remove most of the non-stationarity in the mean and then 
evaluate whether there is remaining non-stationarity in the covariance structure of
the de-trended data that benefits from being modelled with a non-stationary model. 
However, if the mean structure is not estimated well with the de-trending procedure,
there could be significant non-stationarity left in the mean structure. Any such
residual structure will make the estimated covariance structure biased, but the 
de-trended data for 1981 shown in Figure~\ref{fig:deTrendObs} shows much less clear
evidence for a spatially varying mean than the original data in 
Figure~\ref{fig:observation}. For example, the de-trending has removed the obvious shift
in the level of the precipitation between the western and eastern sides.

\subsection{Fitting the non-stationary model}
\label{sec:underEast}
We fit a stationary model (STAT1) and a non-stationary model (NSTAT1) as 
in Section~\ref{sec:Application}, but without covariates and with the 
assumption that there are 15 independent replications of the residuals. 
Each year has observations at potentially different locations, but this does 
not pose any problems in the SPDE-based model since the entire field is 
modelled explicitly through the values on each cell in the discretization. 
The observations are mapped to statements about the values on the grid cells 
in each year and the inference proceeds in a similar way as for the adaptive
smoothing application that used only the year 1981. 

The penalty parameters \(\tau_1\), \(\tau_2\), \(\tau_3\) and \(\tau_4\) 
should be changed, but with 15 realizations the cross-validation becomes 
far more computationally expensive. Therefore, we performed an exploratory
analysis where the fits for low, medium and high smoothing were compared, and
we decided to use \(\log(\tau_1) = 10\), \(\log(\tau_2) = 10\), 
\(\log(\tau_3) = 10\) and \(\log(\tau_4) = 10\). This might not lead to 
the highest possible decrease in the prediction scores, but at this point 
the main interest lies in the qualitative changes in the estimated structure. 
And, it would, potentially, be a waste of time to put in the
required effort before we are certain that there are not major components
missing in the model.

\begin{figure}
	\centering
	\subfigure[Prediction for STAT1]{
		\includegraphics[width=5.5cm]{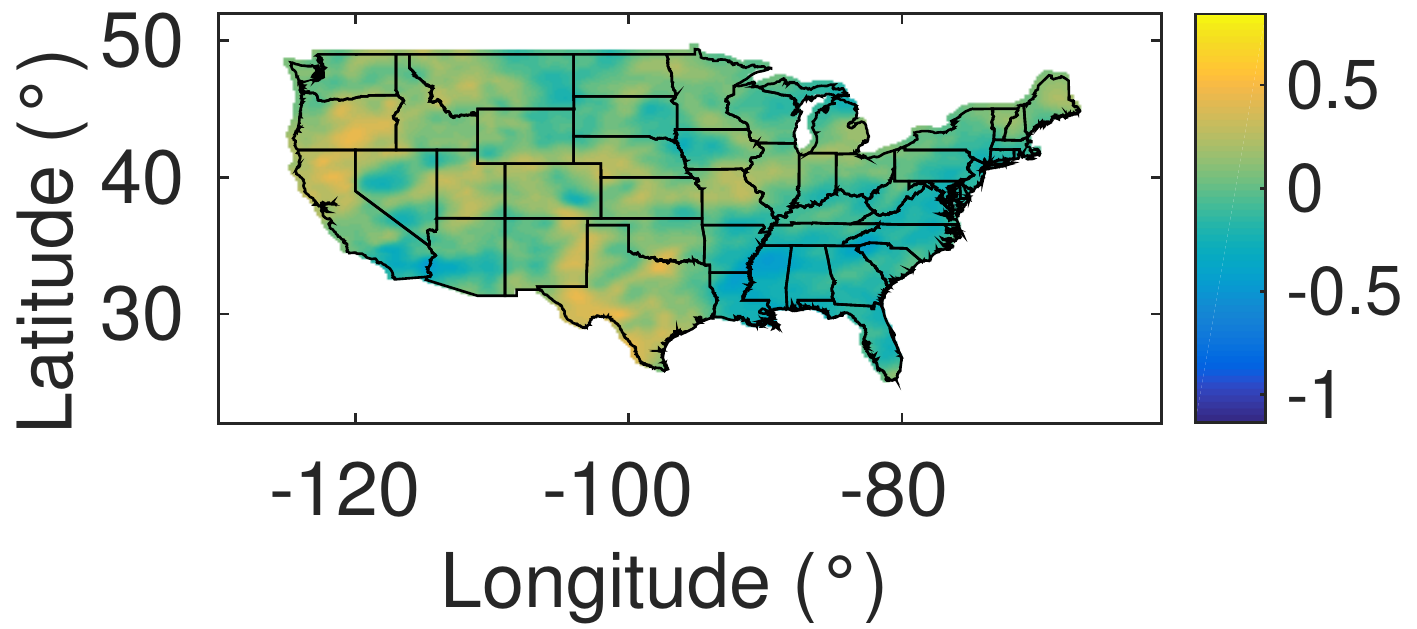}
		\label{fig:sub:deTrendPredRes1}
	}
	\subfigure[Prediction for NSTAT1]{
		\includegraphics[width=5.5cm]{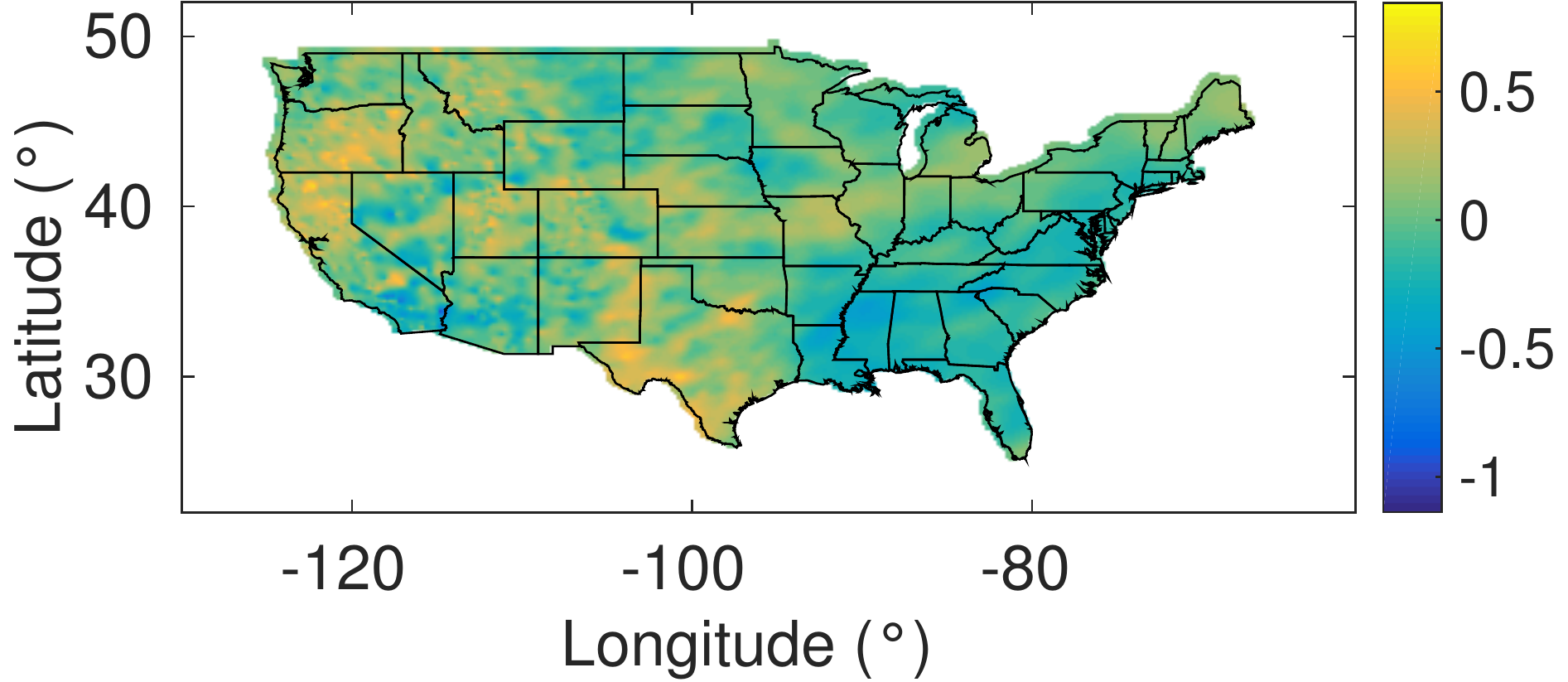}
		\label{fig:sub:deTrendPredRes1NS}
	}\\
	\subfigure[Prediction standard deviations for STAT1]{
		\includegraphics[width=5.5cm]{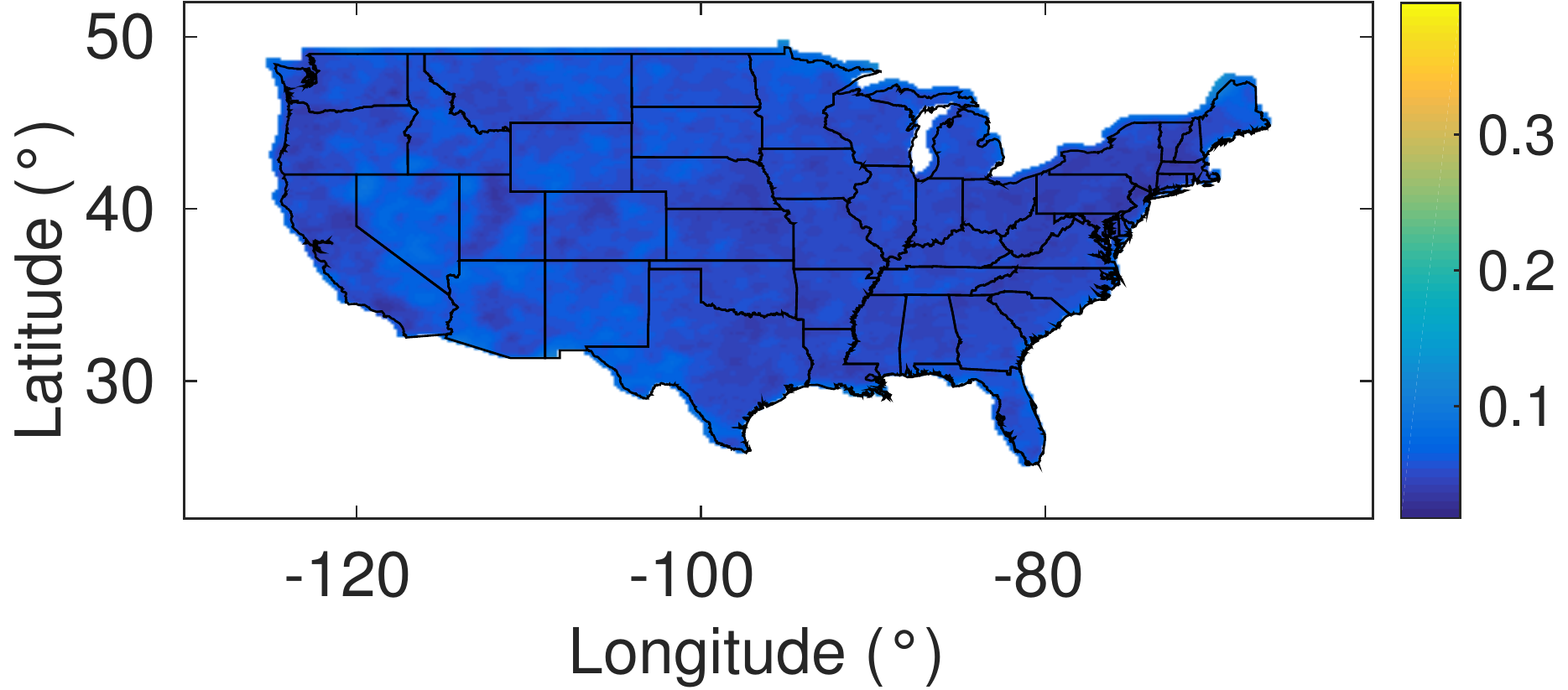}
		\label{fig:sub:deTrendPredStdDev1}
	}
	\subfigure[Prediction standard deviations for NSTAT1]{
		\includegraphics[width=5.5cm]{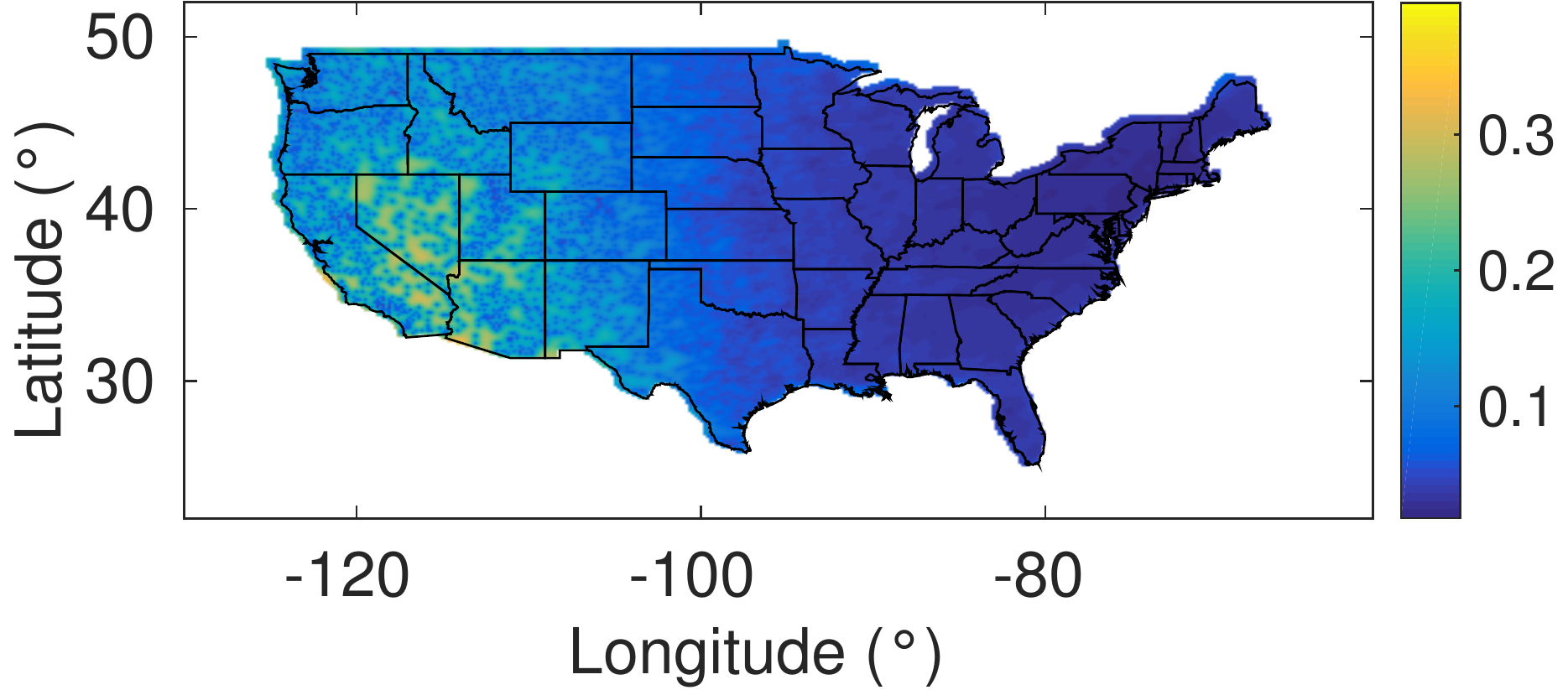}
		\label{fig:sub:deTrendPredStdDev1NS}
	}
	\caption{Prediction for de-trended data for year 1981 based
			 on the 15 year period 1971--1985. \subref{fig:sub:deTrendPredRes1}
			 shows the prediction for STAT1,
			 \subref{fig:sub:deTrendPredRes1NS} shows the prediction for NSTAT1, 
			 \subref{fig:sub:deTrendPredStdDev1} shows the
			 prediction standard deviations for STAT1 and
			 \subref{fig:sub:deTrendPredStdDev1NS} shows the prediction 
			 standard deviations for NSTAT1.}
	\label{fig:deTrendPred1}
\end{figure}

The parameters were estimated in the same way as in 
Section~\ref{sec:Application}, and the maximum penalized likelihood estimates
for non-stationarity were used to give the predictions shown in 
Figure~\ref{fig:deTrendPred1}. The figure shows both the predictions and 
the prediction standard deviations for STAT1 and NSTAT1. There are several
interesting features in these plots. First, the predicted values are similar 
for the two models and the main difference is found in the prediction standard
deviations. Second, the prediction standard deviations for the western region 
is troubling for NSTAT1. The range appears to be too short 
and the spatial effect appears to be close to independent measurement noise 
in this area. This is not consistent with Figure~\ref{fig:deTrendObs}, which
appears to have a spatial effect in this region as well. 

\begin{figure}
	\centering
	\subfigure[Marginal standard deviations]{
		\includegraphics[width=7cm]{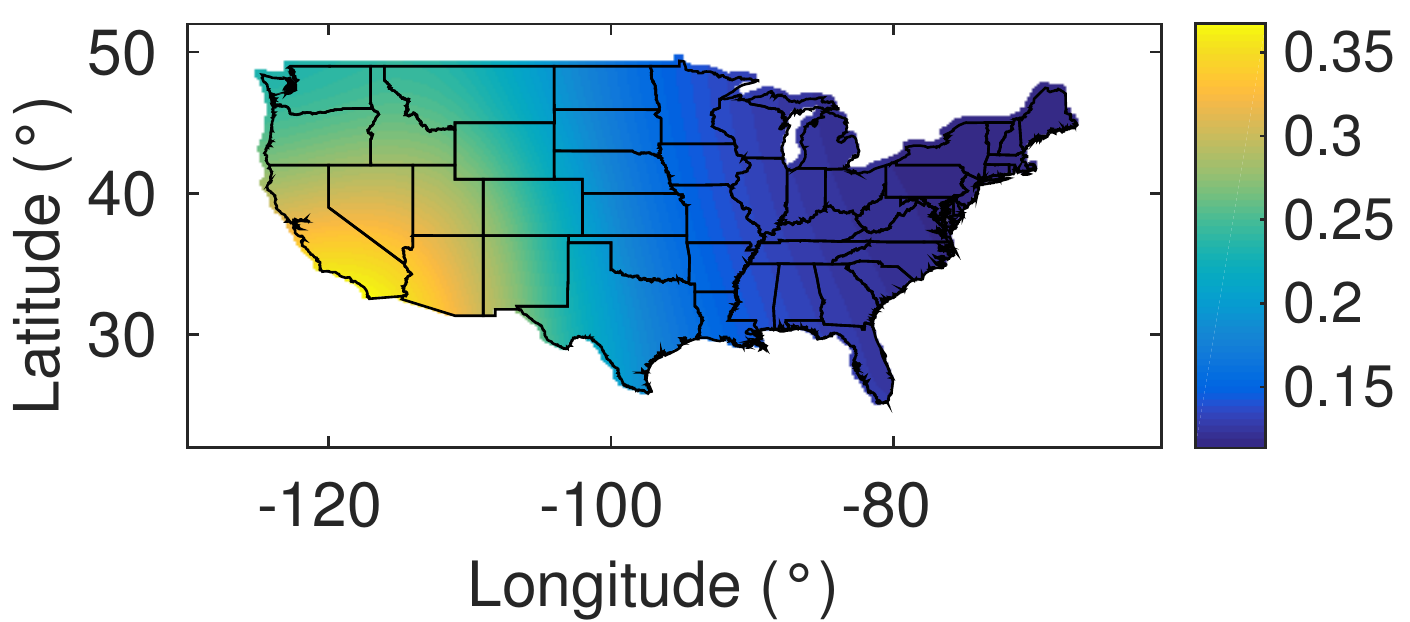}
		\label{fig:sub:deTrendSpatialSTD1NS}
	}
	\subfigure[Estimated correlation structure]{
		\includegraphics[width=7cm]{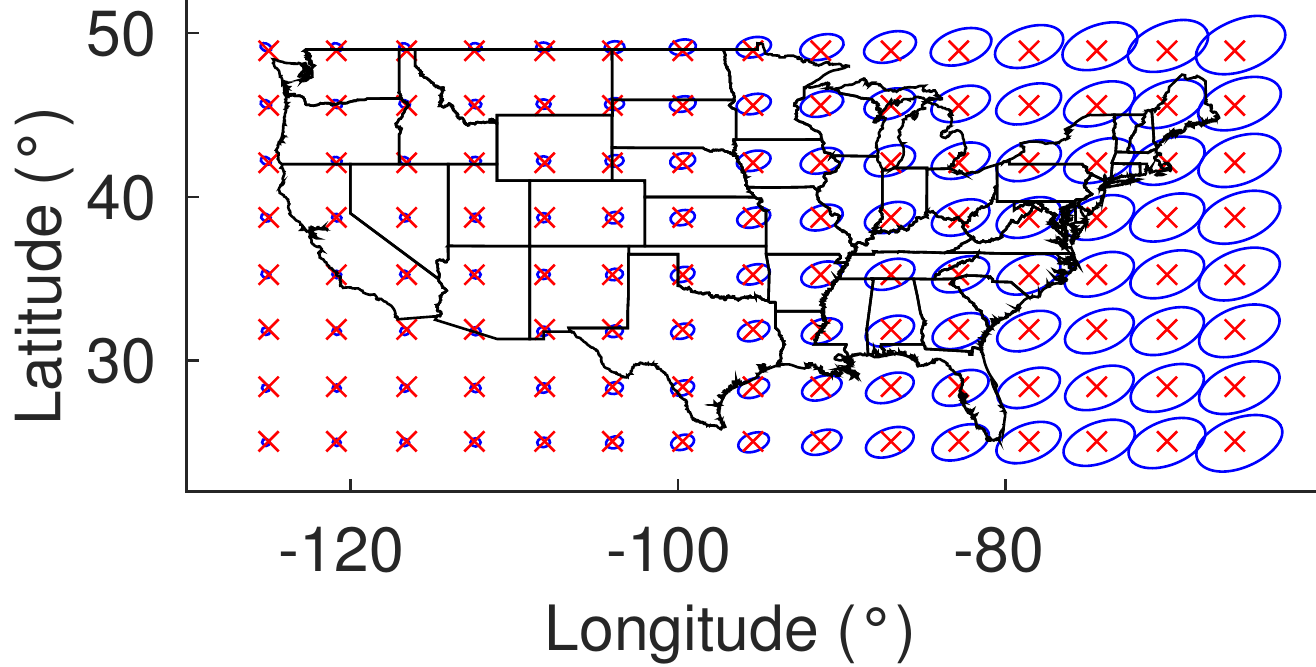}
		\label{fig:sub:deTrendSpatialCor1NS}
	}
	\caption{\subref{fig:sub:deTrendSpatialSTD1NS} Estimated marginal 
			 standard deviations and \subref{fig:sub:deTrendSpatialCor1NS} 
			 estimated 0.7 level contour curves for the correlation 
			 functions with respect to the locations marked with
			 red crosses for the spatial effect in NSTAT1.}
	\label{fig:deTrendSpatial1NS}
\end{figure}

The problem can be seen clearly when looking at the estimated covariance
structure shown in Figure~\ref{fig:deTrendSpatial1NS}. The correlation 
structure in the eastern part looks regular after de-trending the data, but 
the correlation structure in the western region is almost degenerating to
independent noise. This is a problem from a computational perspective, since 
the discretization of the SPDE requires that the range is not too small 
compared to the size of the grid cells, and from a modelling perspective, 
since the parameters are supposed to describe a slowly changing spatial
dependence structure. In the case that the spatial range is that low,
the SPDE models requires a high resolution to properly capture the dependence
between neighbouring grid cells in the discretization, but if the range
is that low, a spatial effect might not be needed. 
Furthermore, Figure~\ref{fig:sub:deTrendSpatialSTD1NS}
shows that the variance of the spatial field is higher in the western region.
This indicates that the nugget effect in the western region needs to be
different from the nugget effect in the eastern region.

\begin{figure}
	\centering
	\subfigure[Log-predictive score]{
		\includegraphics[height=3.6cm]{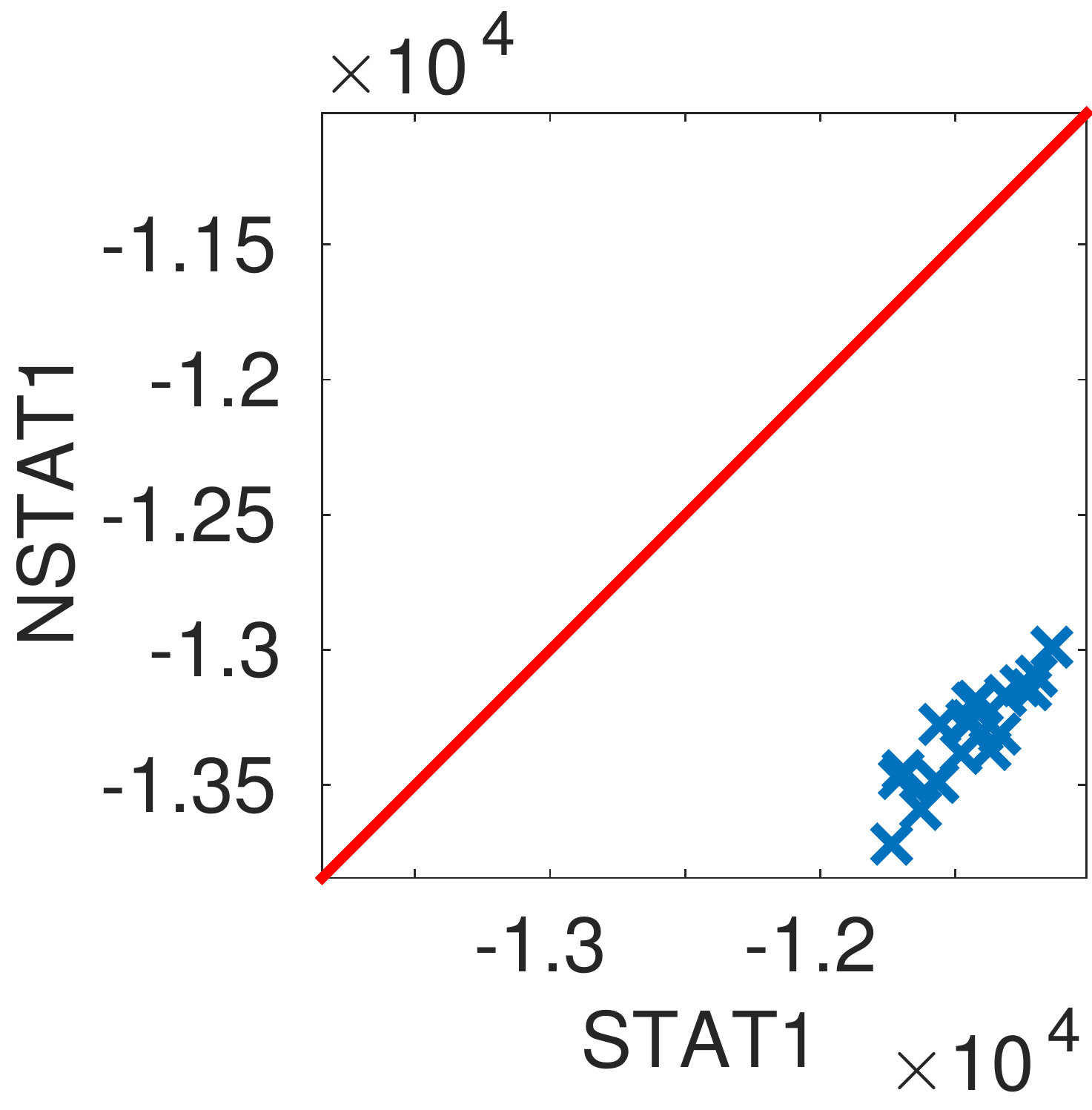}
		\label{fig:sub:deTrendLPD1}
	}
	\subfigure[CRPS]{
		\includegraphics[height=3.3cm]{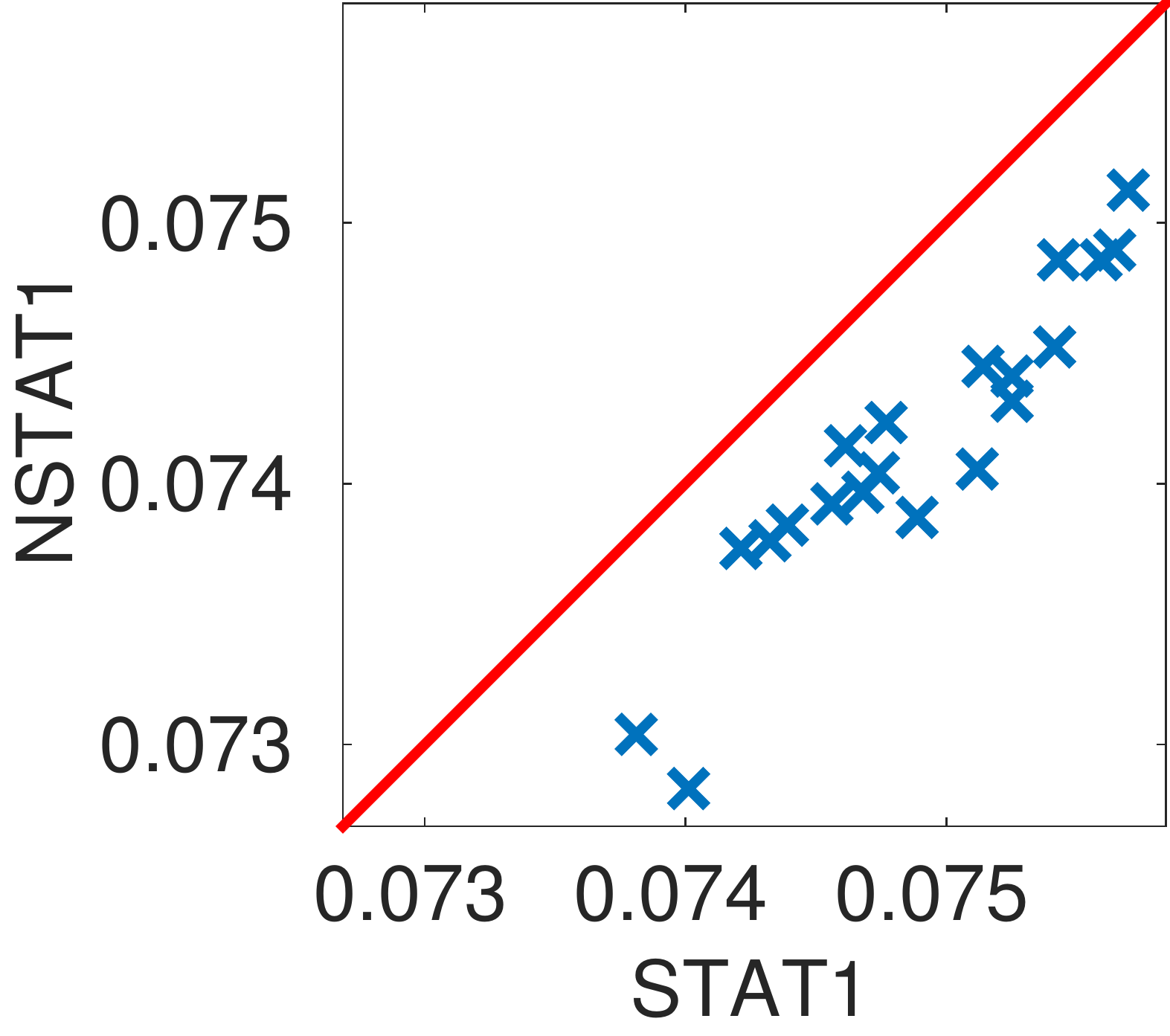}
		\label{fig:sub:deTrendCRPS1}
	}
	\subfigure[RMSE]{
		\includegraphics[height=3.3cm]{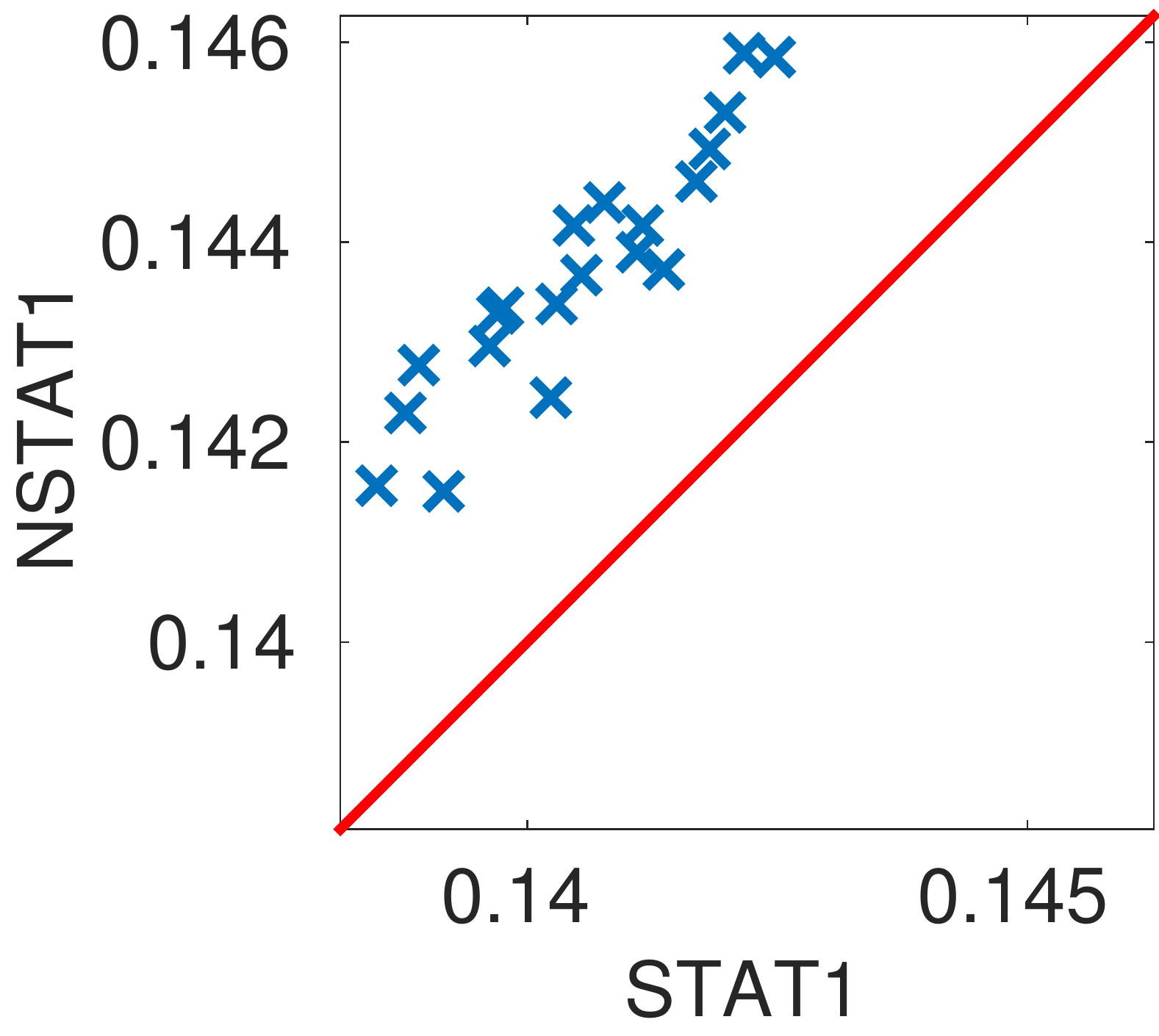}
		\label{fig:sub:deTrendRMSD1}
	}
	\caption{Scatter plots of \subref{fig:sub:deTrendLPD1} Log-predictive 
			 score, \subref{fig:sub:deTrendCRPS1} CRPS and  
			 \subref{fig:sub:deTrendRMSD1} Root mean square error for
			 STAT1 and NSTAT1. The estimates were calculated with hold-out 
			 sets where 20\% of the locations were held-out from each year 
			 as described in Section~\ref{sec:underEast}.
			 }
	\label{fig:deTrendComp1}
\end{figure}

The fits of STAT1 and NSTAT1 are compared with the log-predictive score, 
the CRPS and the RMSE. The scores are calculated by randomly dividing
the data in each year in five parts and then holding out the first part
from each year and do the entire fitting and prediction of this data using only 
the remaining part of the data. Then holding out the second
part of the data in each year and so on, for a total of 5 values. This process
was then repeated three more times for a total of 20 values of the scores. 
Scatter plots comparing the scores for the two models are shown in 
Figure~\ref{fig:deTrendComp1}. 

NSTAT1 has a lower log-predictive score and CRPS than STAT1, but the RMSE
is higher. The conclusions based on the log-predictive score and the CRPS is
the same as for the single realization analysis in 
Section~\ref{sec:ComparisonOneYear}. However, the consistently higher RMSE values
indicate that there is a problem with the model. The problem lies in the western
region where the range is too low, which leads to worse point estimates because
the spatial dependence is not exploited. The flexible non-stationary
model is able to detect that a higher variance is required for the nugget effect
in the western region, but is not able to achieve this in the correct way. 
Even with all the freedom available in the model it is impossible to have 
spatial dependence and different nugget effects because we have put the
non-stationarity in the wrong components of the model. We need to treat
the nugget effects in the western and eastern regions separately.

\subsection{Removing the under-smoothing in the western part}
The results in Section~\ref{sec:underEast} indicate that the nugget effect
is different in the western and the eastern part of the conterminous US. 
Therefore, we fit a stationary model (STAT2) and a non-stationary model (NSTAT2)
with separate nugget effects for locations with longitudes lower than 
\(100\, ^\circ W\) and for locations with longitudes higher than or equal to 
\(100\, ^\circ W\). The placement of the frontier at \(100\, ^\circ W\) 
is motivated by the change from mountainous regions to plains seen in 
Figure~\ref{fig:elevation} and the change from low to high range seen 
in Figure~\ref{fig:sub:deTrendSpatialCor1NS}, but
we do not believe it would be particularly sensitive to the exact
placement as long as it is in the area of transition from mountainous
regions to plains. 
Except for this change,
the models are unchanged, and we use the 
same penalties \(\tau_1\), \(\tau_2\), \(\tau_3\) and \(\tau_4\) for the 
non-stationarity structure. The intention is to see how much the predictions 
and the estimated dependence structure change with different nugget effects, 
but the same penalties.

\begin{figure}
	\centering
	\subfigure[Prediction for STAT2]{
		\includegraphics[width=5.5cm]{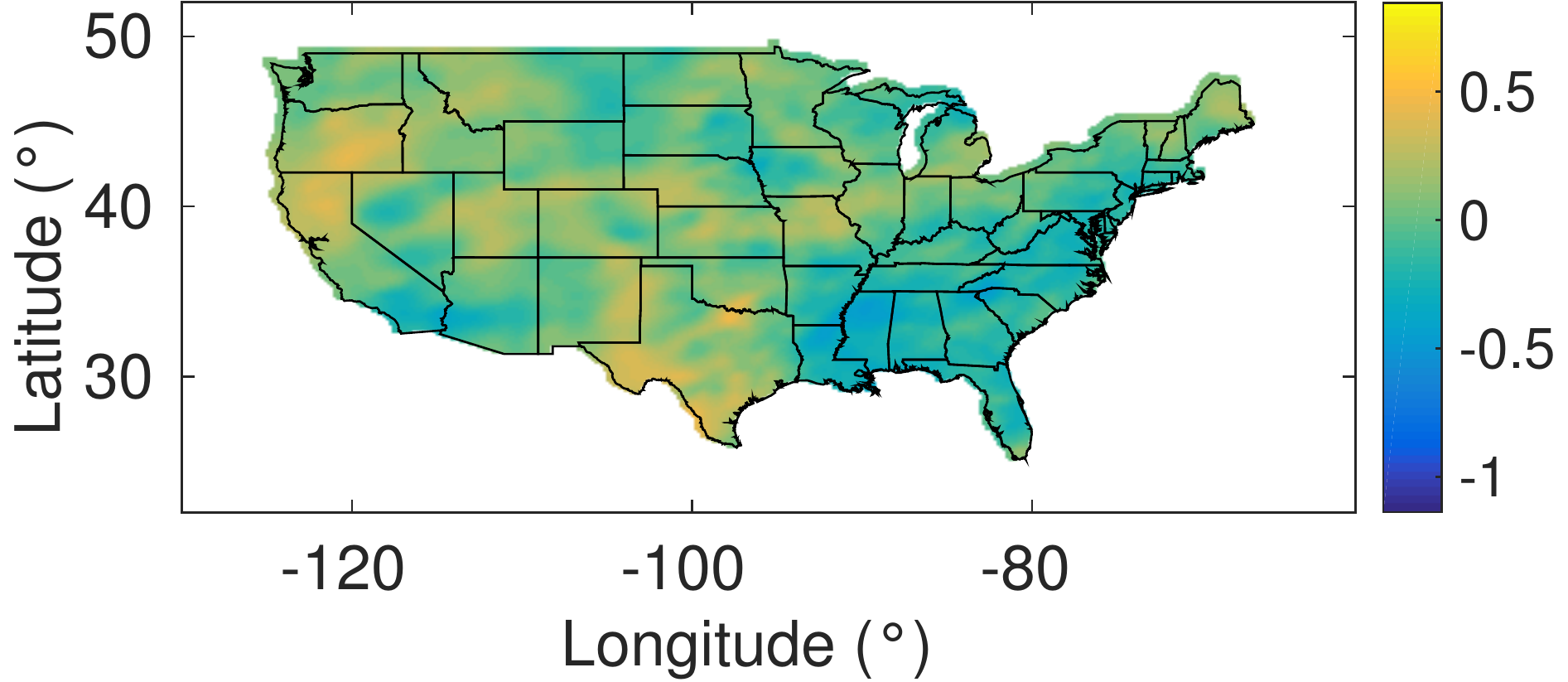}
		\label{fig:sub:deTrendPredRes}
	}
	\subfigure[Prediction for NSTAT2]{
		\includegraphics[width=5.5cm]{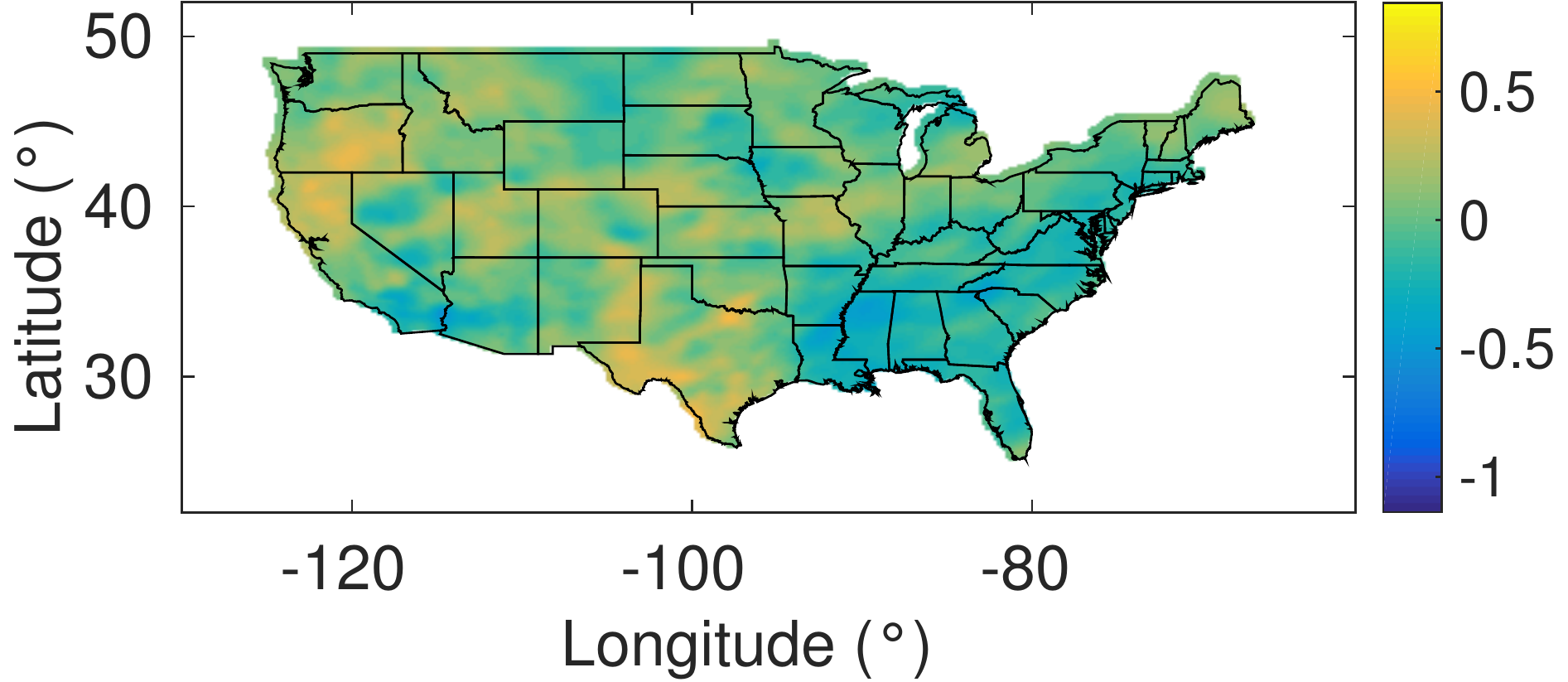}
		\label{fig:sub:deTrendPredResNS}
	}\\
	\subfigure[Prediction standard deviations for STAT2]{
		\includegraphics[width=5.5cm]{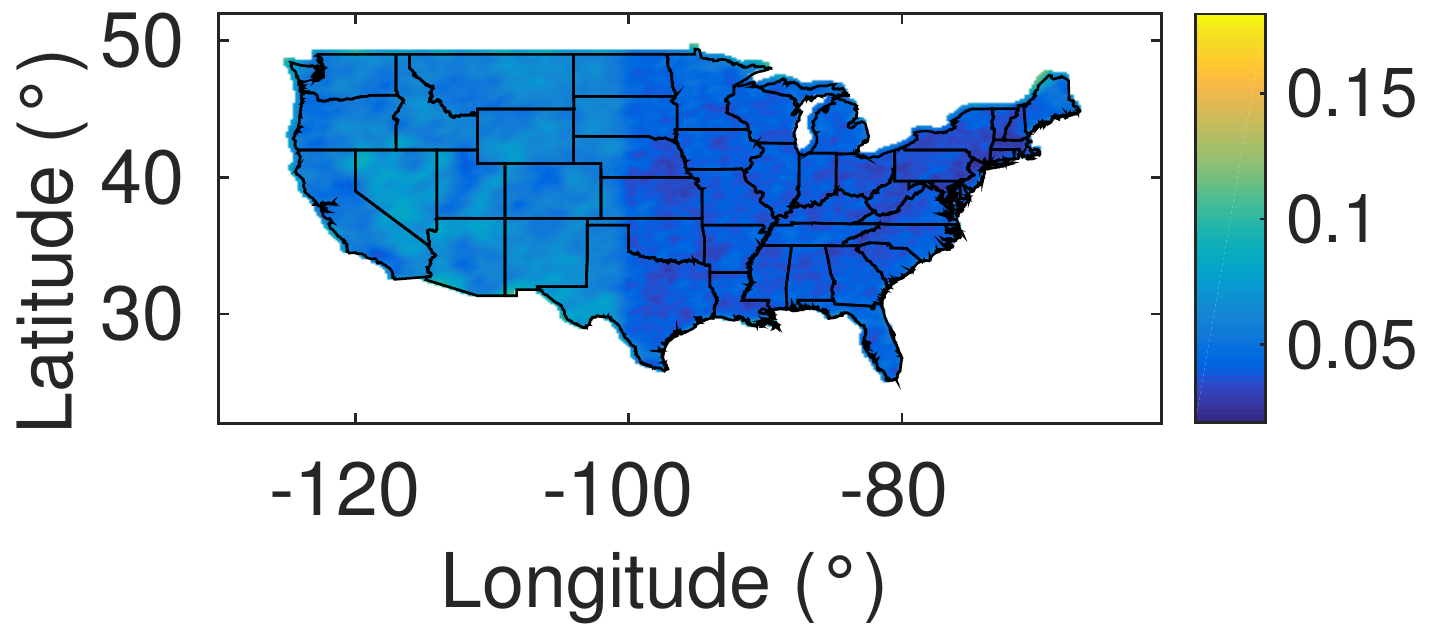}
		\label{fig:sub:deTrendPredStdDev}
	}
	\subfigure[Prediction standard deviations for NSTAT2]{
		\includegraphics[width=5.5cm]{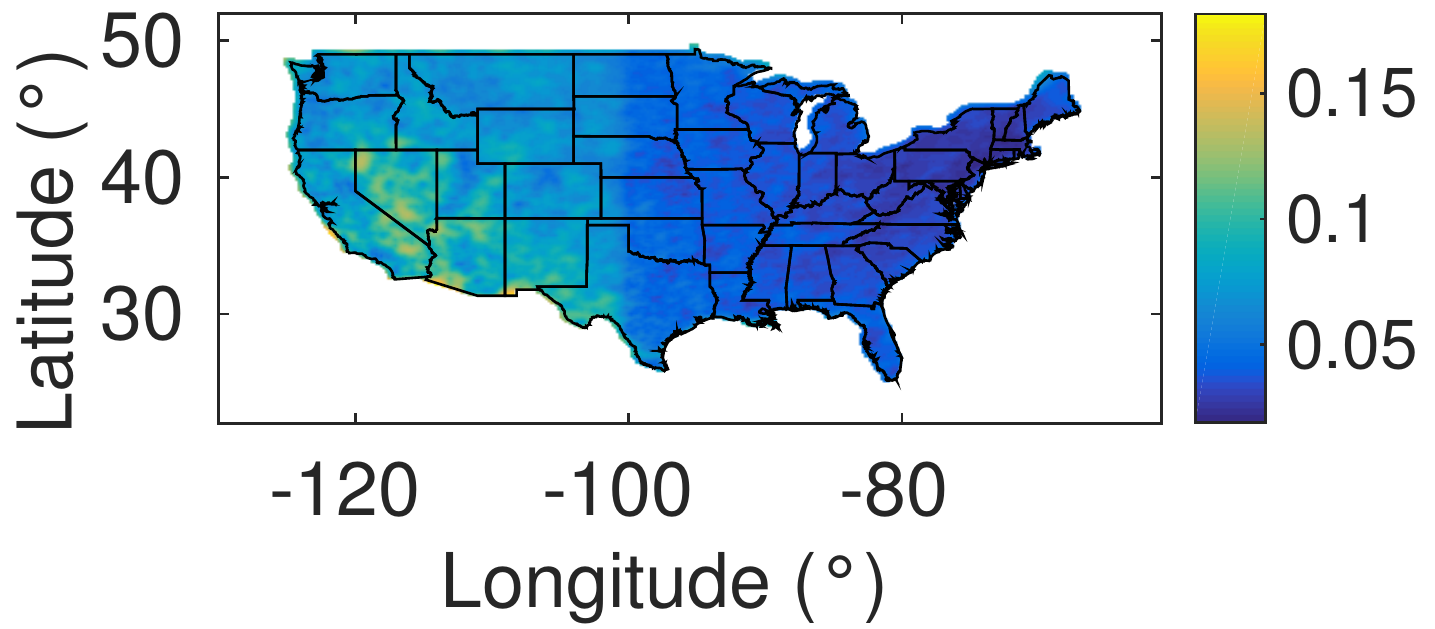}
		\label{fig:sub:deTrendPredStdDevNS}
	}
	\caption{Prediction for de-trended data for year 1981 based
			 on the 15 year period 1971--1985. \subref{fig:sub:deTrendPredRes}
			 shows the prediction for STAT2,
			 \subref{fig:sub:deTrendPredResNS} shows the prediction for NSTAT2, 
			 \subref{fig:sub:deTrendPredStdDev} shows the
			 prediction standard deviations for STAT2 and
			 \subref{fig:sub:deTrendPredStdDevNS} shows the prediction 
			 standard deviations for NSTAT2.}
	\label{fig:deTrendPredNS}
\end{figure}

The predictions and prediction standard deviations are shown in
Figure~\ref{fig:deTrendPredNS}. The prediction standard deviations for
NSTAT2 do not have the strange artifacts in the western region that
are present in Figure~\ref{fig:deTrendPred1} for NSTAT1, but one can 
notice that there is a sharp change in 
prediction standard deviations at longitude \(100\, ^\circ W\). This is by
construction due to the use of different nugget effects for the two 
parts of the conterminous US. STAT2 has an estimated standard deviation
for the nugget effect of \(0.17\) in the western part and of \(0.083\)
in the eastern part and for NSTAT2 the estimated standard
deviation for the nugget effect is \(0.16\) in the western part and is
\(0.083\) in the eastern part.

\begin{figure}
	\centering
	\subfigure[Marginal standard deviations]{
		\includegraphics[width=7cm]{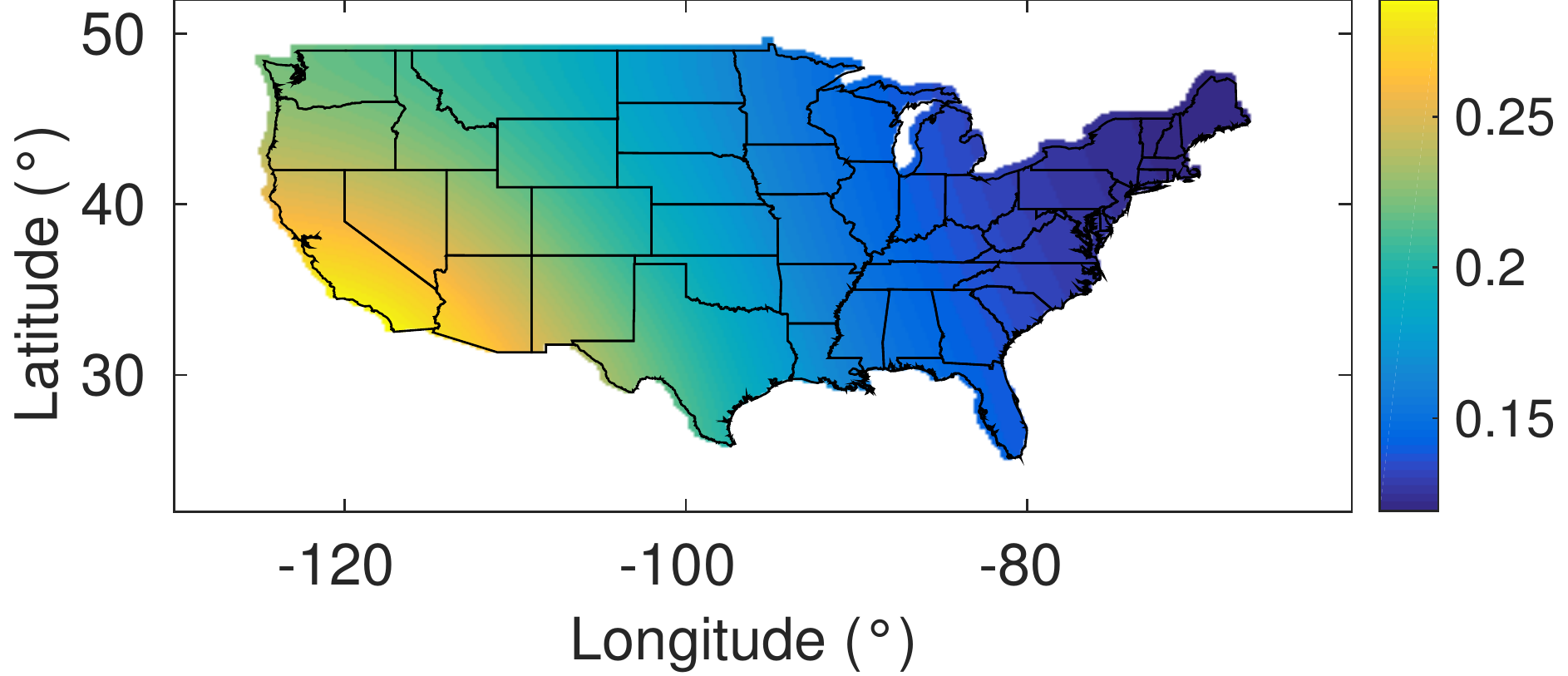}
		\label{fig:sub:deTrendSpatialSTDNS}
	}
	\subfigure[Estimated correlation structure]{
		\includegraphics[width=7cm]{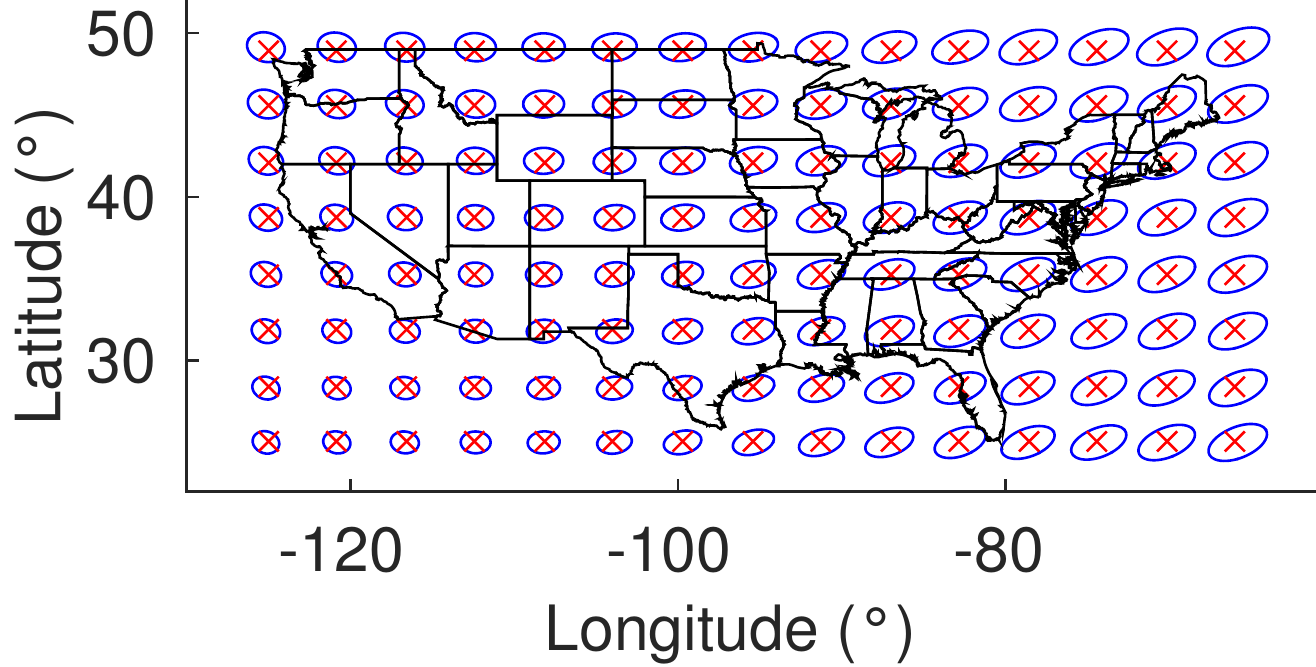}
		\label{fig:sub:deTrendSpatialCorNS}
	}
	\caption{\subref{fig:sub:deTrendSpatialSTDNS} Estimated 
			 marginal standard deviations and
			 \subref{fig:sub:deTrendSpatialCorNS} estimated 0.7 level contour
			 curves for the correlation functions with respect to the locations
			 marked with red crosses for the spatial effect in NSTAT2.}
	\label{fig:deTrendSpatialNS}
\end{figure}

The estimated spatial dependence structure of NSTAT2 is shown in   
Figure~\ref{fig:deTrendSpatialNS}. The clearest change from the dependence
structure of NSTAT1 shown in Figure~\ref{fig:deTrendSpatial1NS} is that the 
non-stationarity in the correlation structure is mostly gone. The appearance 
is much more reasonable than for NSTAT1 since the entire dependence 
structure is changing slowly and there are no areas with unreasonably large 
or small ranges. Some non-stationarity still remains in the marginal standard
deviations, but together these plots indicate that the simple model STAT2, 
which does not use a complex non-stationary spatial field, should fit 
these data well.

\begin{figure}
	\centering
	\subfigure[Log-predictive score]{
		\includegraphics[height=3.5cm]{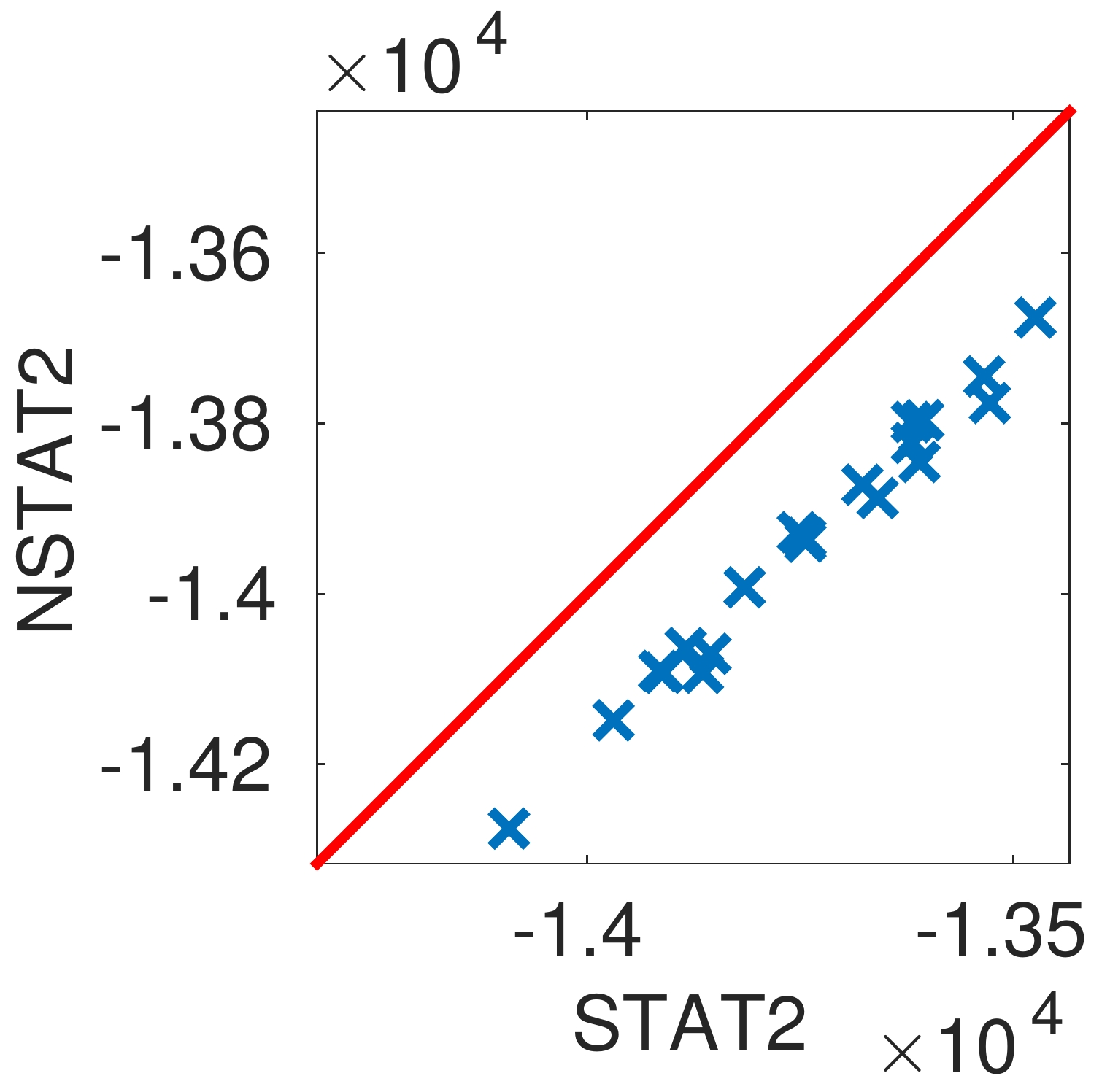}
		\label{fig:sub:deTrendLPD}
	}
	\subfigure[CRPS]{
		\includegraphics[height=3.2cm]{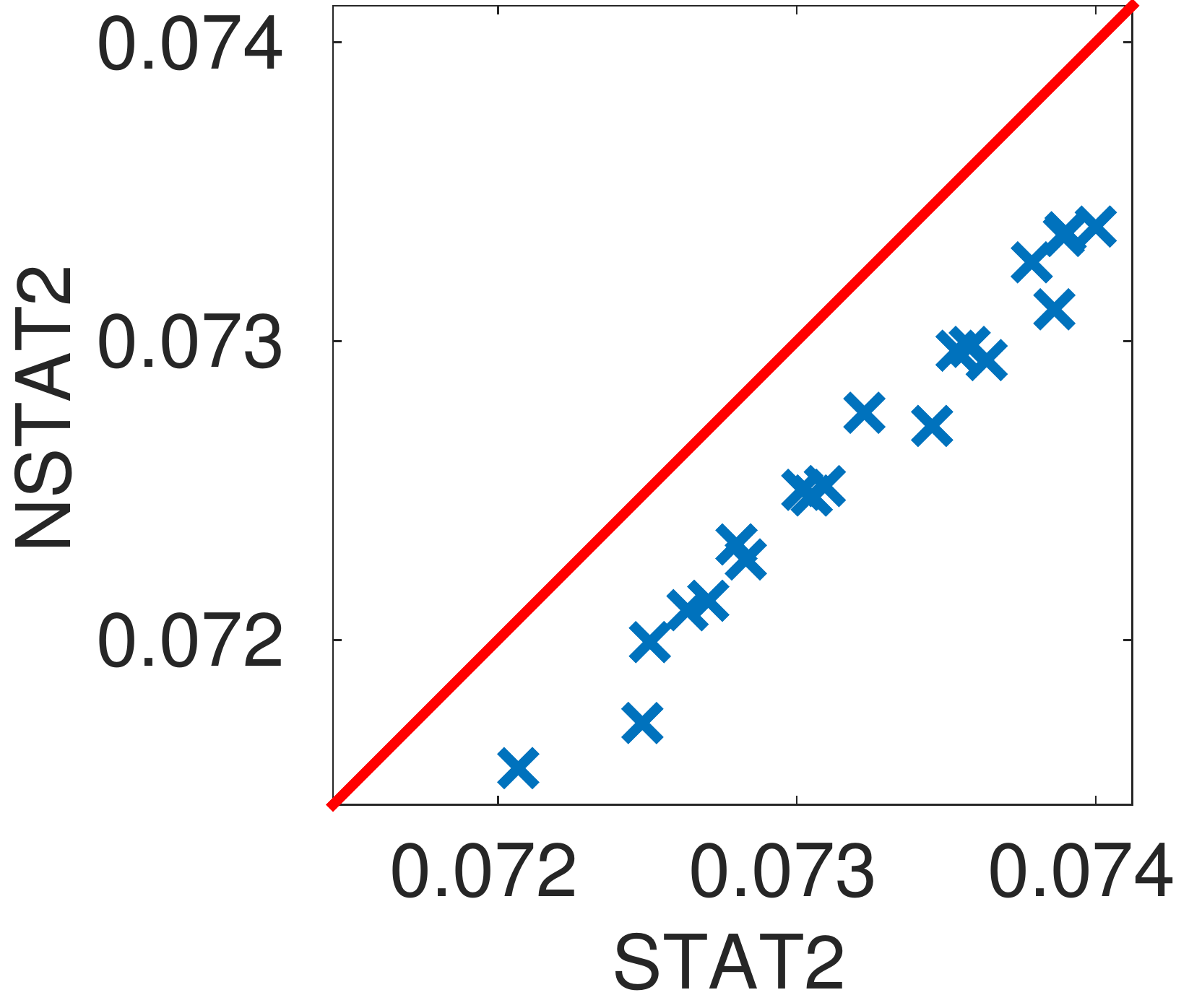}
		\label{fig:sub:deTrendCRPS}
	}
	\subfigure[RMSE]{
		\includegraphics[height=3.2cm]{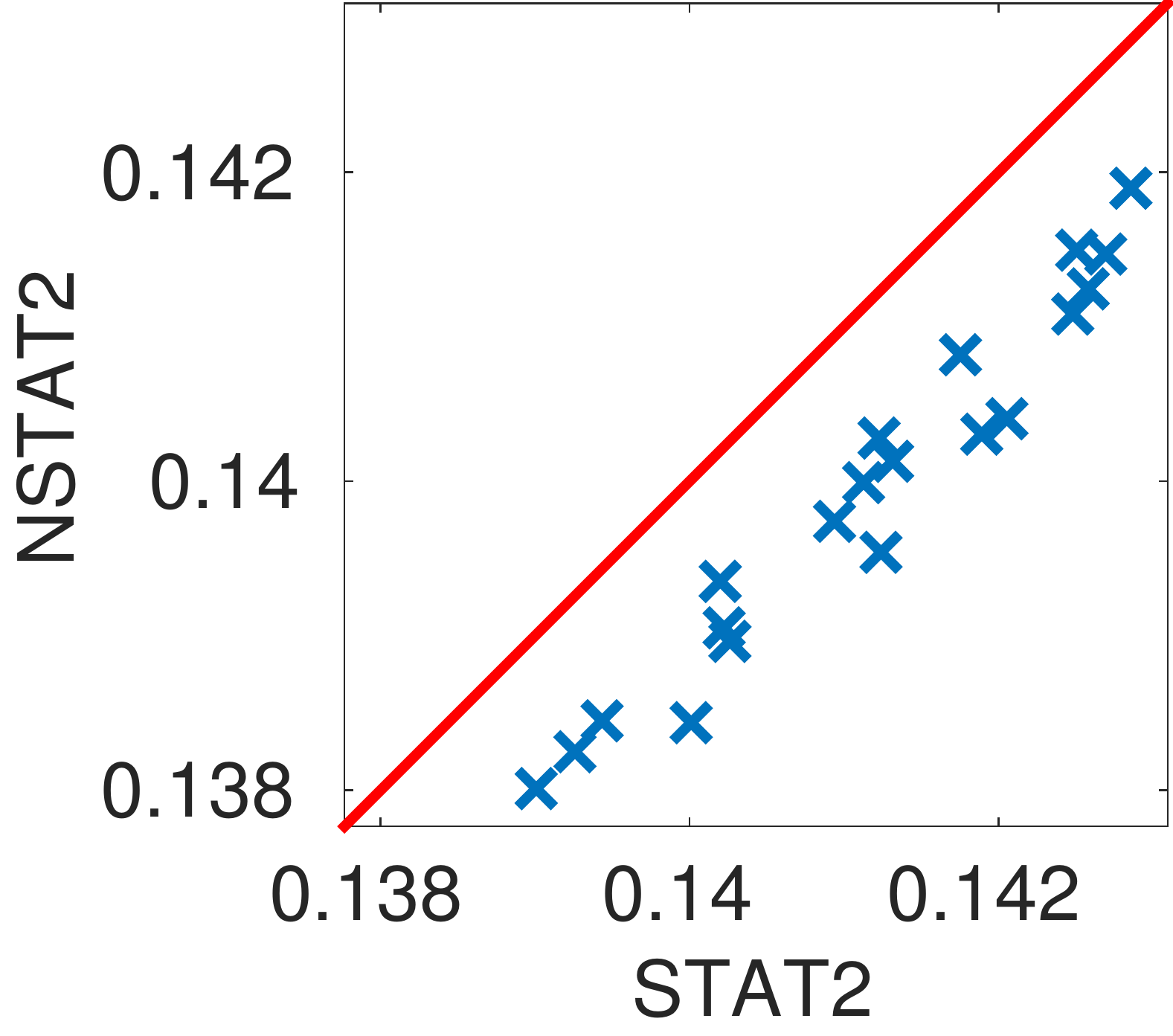}
		\label{fig:sub:deTrendRMSD}
	}
	\caption{Scatter plots of \subref{fig:sub:deTrendLPD} 
			 log-predictive score, \subref{fig:sub:deTrendCRPS} CRPS and 
			 \subref{fig:sub:deTrendRMSD} Root mean square error for STAT2
			 and NSTAT2. The
			 estimates were calculated with hold-out sets where 20\% of
			 the locations were held-out from each year as described in
			 Section~\ref{sec:underEast}.}
	\label{fig:deTrendComp}
\end{figure}

We compare the predictions of STAT2 and NSTAT2 by
the RMSE, the CRPS and the log-predictive score. The results are given in  
Figure~\ref{fig:deTrendComp}. NSTAT2 performs better according to all of 
the scores. The scatter plots of the scores show that NSTAT2 performs better 
for all the hold-out sets, but that the differences in scores are small. 

\subsection{Discussion of models}
\begin{figure}
	\centering
	\subfigure[LPD]{
		\includegraphics[width=3.7cm]{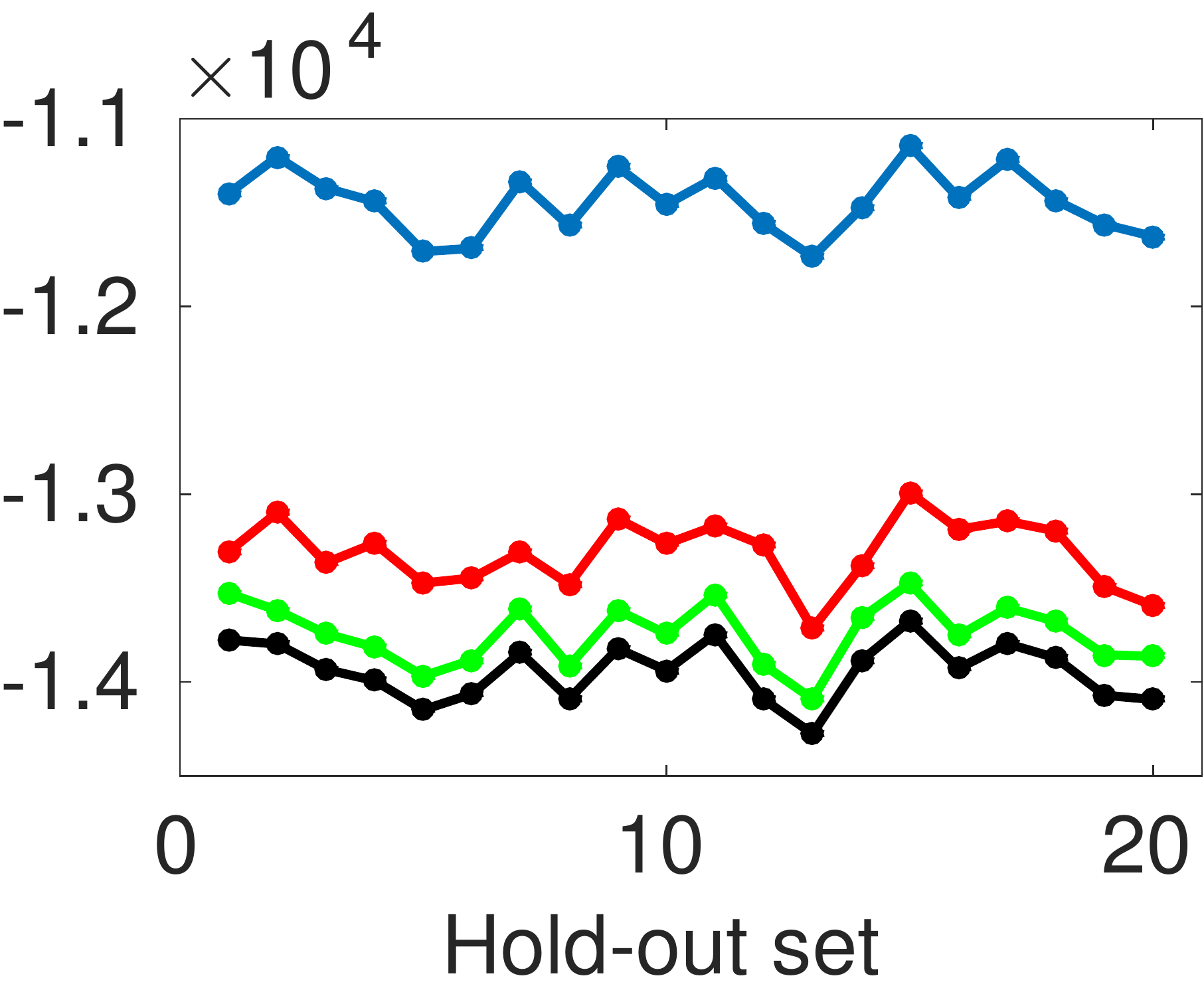}
		\label{fig:sub:deTrendALLLPD}
	}
	\subfigure[CRPS]{
		\includegraphics[width=3.5cm]{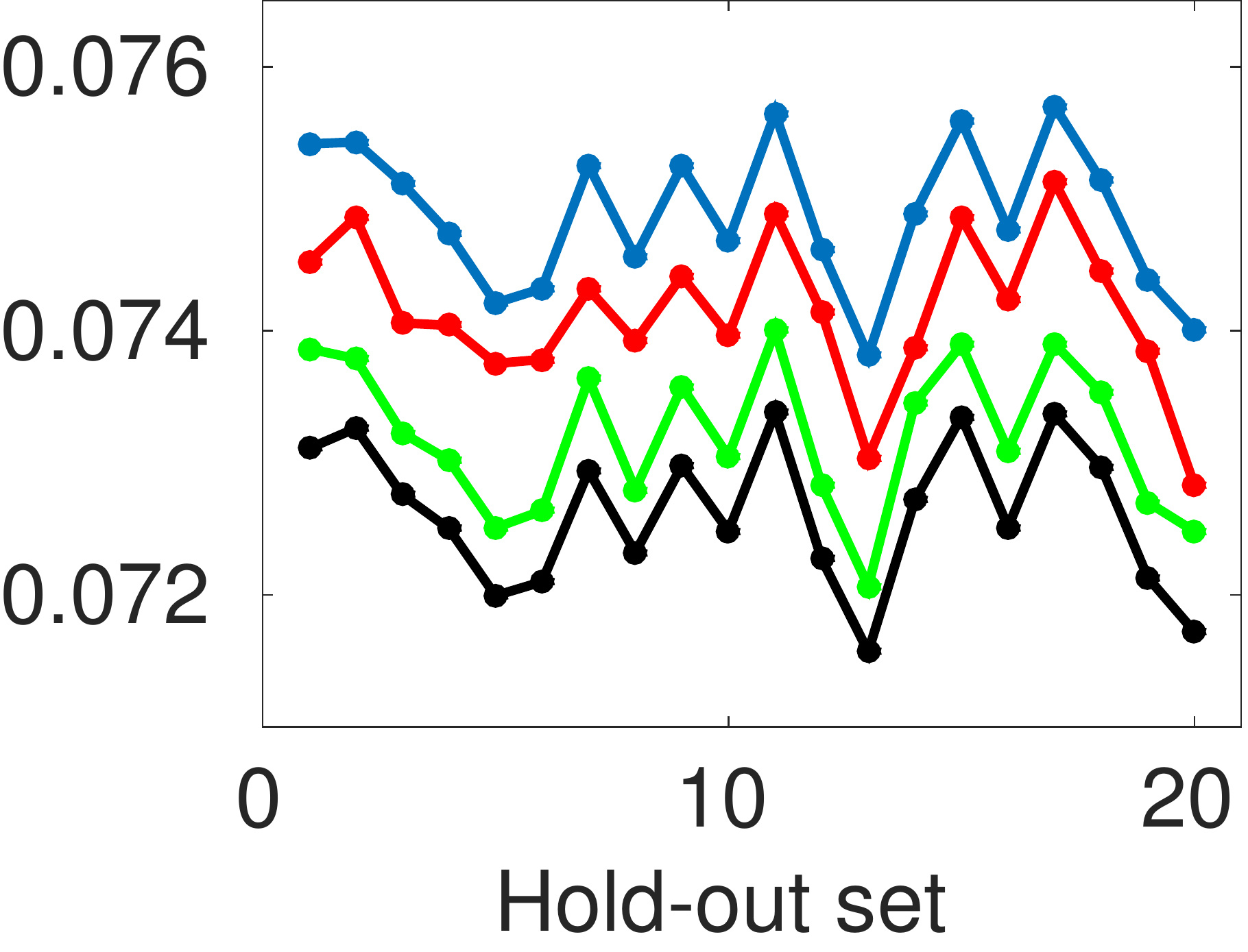}
		\label{fig:sub:deTrendALLCRPS}
	}
	\subfigure[RMSE]{
		\includegraphics[width=3.5cm]{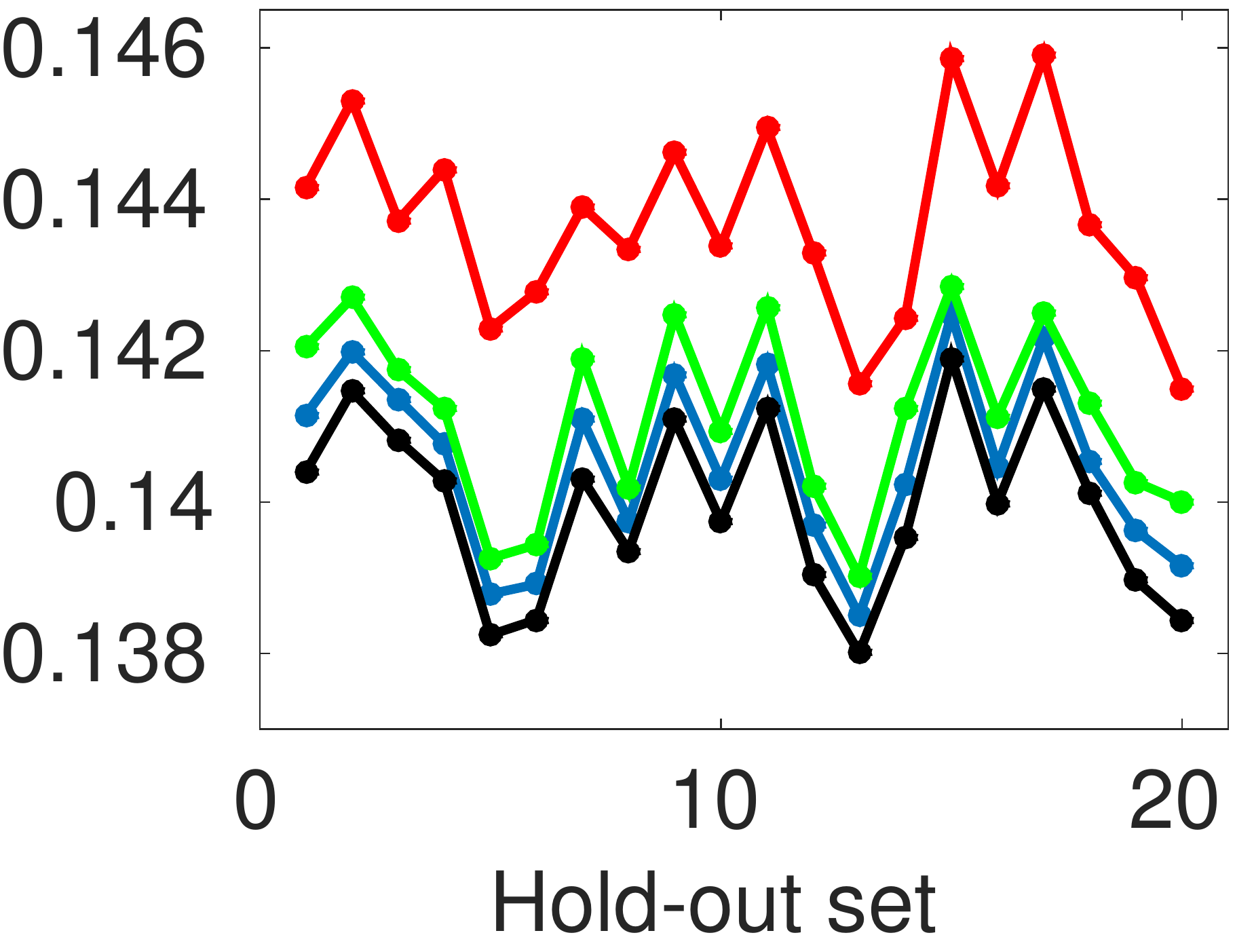}
		\label{fig:sub:deTrendALLRMSD}
	}
	\caption{Comparison of STAT1 (blue), NSTAT1 (red), STAT2 (green) and
			 NSTAT2 (black) based on 
			 \subref{fig:sub:deTrendALLLPD} Log-predictive score,
			 \subref{fig:sub:deTrendALLCRPS} CRPS and
			 \subref{fig:sub:deTrendALLRMSD} RMSE. The
			 estimates were calculated with hold-out sets where 20\% of
			 the locations were held-out from each year as described in
			 Section~\ref{sec:underEast}.}
	\label{fig:deTrendALL}
\end{figure}

The prediction scores for STAT1, NSTAT1, STAT2 and NSTAT2 are shown in 
Figure~\ref{fig:deTrendALL}. The figure shows that the model performing the 
best according to all scores is NSTAT2, but is the extra computation time
worth the effort in this case? The much simpler model STAT2 is performing 
almost as good as NSTAT2 and requires only \emph{one} extra parameter. 
The cost of including one extra parameter is far less than the cost of
introducing the flexible non-stationary model. Additionally, one can see 
that even though the expensive flexible model makes NSTAT1 consistently 
better than STAT1 in the log-predictive score and the CRPS, STAT2 makes
an even greater improvement from STAT1 for the cost of only a single parameter. 

The predictions and prediction standard deviations 
for STAT2 and NSTAT2 in Figure~\ref{fig:deTrendPredNS} are showing less 
extreme differences than the predictions and prediction standard
deviations for STAT1 and NSTAT1 shown in Figure~\ref{fig:deTrendPred1}, 
but there are still some differences
in the prediction standard deviations. Some further gain is possible by
selecting the penalty parameters controlling the non-stationarity
 more carefully. 
We saw some improvement by trying different penalty parameters, but no major 
changes that would change the conclusion. When we take computation time into 
account, STAT2 appears to be the better choice. There is some gain with 
the flexible non-stationary model in NSTAT2, but it comes at a high
computational cost.

The physical cause of the difference in the nugget  effect between the western
region and the eastern region is not known, but
it is unlikely to be caused only by differences in the measurement
equipment. It is more likely that it is caused by differences in the
small-scale behaviour of the process generating the weather 
in the two different regions that is not captured
by the model, but it has not been our intention to find the physical
explanation. The intention has been to demonstrate how such a phenomenon
can affect the estimation of general flexible models for non-stationarity
and the need to carefully evaluate the fitted covariance structures.

\section{Discussion}
\label{sec:Discussion}
The question of whether we need non-stationary spatial models or not, 
is a deeper question than it might seem initially. The first step
of the analysis should be to decide whether it is likely
that non-stationarity is present in the data or not, and in this context
simple data exploration, such as variograms, and formal 
tests~\citep{fuentes2005formal, mikyoung2012, bowman2013inference} 
are useful tools. The second step is to decide which non-stationary
model we want to use and it can be tempting to look for complex models 
that allow for
spatial fields that have large amounts of flexibility in the covariance
structure. We then apply these models with the hope that the high degree of
flexibility means that we will be able to capture any non-stationarity
present in the data, but the analysis of the annual precipitation data
shows that blindly applying such a model might not capture the 
non-stationarity in the correct and best way.

The case study clearly indicates the need to go beyond simply
determining whether or not non-stationarity is present in the data.
We need to determine what type of non-stationarity that is present in the
data. A flexible model will try to adapt to the non-stationarity, but if
the flexibility is available in the wrong parts of the model, the model
might have to do suboptimal things to improve the predictive distributions. For
example, the model might imitate a spatially varying nugget effect by decreasing 
the range and
varying the marginal variances. This adaptation gives 
severe undersmoothing, but simply expanding the model with a smoothly varying 
nugget effect would make the model difficult to identify together with the
rest of the flexibility. Therefore, we should determine 
what is causing the non-stationarity we are seeing before deciding which
non-stationary model to use. 

The first and most obvious source of non-stationarity in a dataset
is the mean structure, and not accounting for this source of 
non-stationarity will confound the non-stationarity in the mean
structure with the non-stationarity in the covariance structure. 
For example, unmeasured covariates can lead to the apparent long range 
dependence and global non-stationarity that we observed in 
the analysis of a single realization.
The method presented in this paper is aimed at modelling local
non-stationarity and is not appropriate for modelling this type of global 
non-stationarity. We handle this apparent structure in the covariances
by de-trending the data, but it is also possible to model jointly
the mean structure and the covariance structure. A simple example
of the latter would be to combine the SPDE models with a small number
of global basis functions to form a hybrid of fixed-rank kriging
and the SPDE models, where the SPDE models captures the short range dependence
and local non-stationarity, and the basis functions capture the long range
dependence and global non-stationarity. Whichever approach is taken, the
paper demonstrates the need to remove the global 
non-stationarity before modelling the local non-stationarity.

After we have removed the global non-stationarity induced by the mean
structure we can model the remaining local non-stationarity, for which
the Markovian structure of the SPDE models offers a computationally
efficient modelling tool.
In the SPDE models we construct a consistent global covariance structure by tying 
together the local behaviour specified by the SPDE at each location,
and the covariance between any two locations will be a combination of the
local behaviour at all locations in the model. We believe that this
approach is a good way to model local non-stationarity that provides
a more flexible, more computationally efficient and easier to parametrize
approach than
the deformation method, while still having a geometric interpretation of varying
the local distance measures.

But modelling local non-stationarity requires information on the small-scale
directional behaviour of the observations, and we would be hesitant to 
estimate flexible non-stationary models for sparser datasets. Methods such 
as the deformation method is routinely applied to much sparser datasets, but there
is no way around the fact that for patches where we do not have observations
we have no idea how the covariances behave. For sparse data it is possible to
imagine multiple covariance structures that could give rise to the observed
empirical covariances and the unobserved structure must be filled by the model 
based on the assumptions and restrictions that we have put into the model.
This can, potentially, lead to highly model dependent estimates since in 
non-stationary modelling the missing covariances do not directly affect 
the observations, and it is important to not allow too much freedom in the 
covariance structure compared to the sparseness of the data, and to realize 
that the features seen in the estimated 
covariance structure will depend on the sparseness of the data.

In an analogous way as for other finite-dimensional methods, there
is a confounding of the nugget effect and the resolution chosen 
for the finite-dimensional approximation. For predictive processes
there exists a solution~\citep{finley2009improving}, but for the
SPDE models it is an active field of research. In a GRF model the nugget 
effect is a combination of the small-scale behaviour and the measurement error, 
where small-scale behaviour is behaviour below the scale which the data
can inform about. The sparser the data is, the more small-scale
variation will be confounded with the nugget effect, but for 
the SPDE models the interpretation of the nugget effect is also tied to the 
discretization and is a combination of measurement
error, small-scale variation and sub-grid variation. The approximation
cannot capture variation within the grid cells and these variations 
increase the nugget variance and decrease the process variances, but
this is only a worry when interpreting these parameters. If the precipitation
data were sparser, the confounding between small-scale variation and
the nugget effect would make it difficult to detect different nugget
effects in the western region and the eastern region, and the approach
might lead to a different conclusion about the nugget effect.

In each of the three cases studied, the flexible non-stationary model performs
better according to the log-predictive score and the CRPS, but 
when we target directly the non-stationarity in the nugget effect, we can 
apply a much simpler model just using two nugget effects. Does this mean 
that the flexible non-stationary model was not useful? No, we were able
to use the flexible non-stationary model to estimate a covariance 
structure that could be used to help determine  
possible sources of the non-stationarity. We could then include these sources 
directly and fit a simpler model performing almost equally well, and we 
could make the same changes to the flexible non-stationary model and fit it
again to become confident that there were no other major uncaptured sources of 
non-stationarity. The idea that the nugget might be the source of 
heterogeneity is not new~\citep{zimmerman1993another}, but the case 
study demonstrates the dangers of putting the heterogeneity in the wrong
components in the model.

If there were knowledge available about what was physically generating
the non-stationarity, it would be possible to make simpler models where 
we reduce the flexibility and control the the covariance structure by 
covariates. The use of two nugget effects is an extreme case of this, 
but covariates in the covariance structure has been a recent direction of
research within all the major families of approaches such as the deformation
method, the process convolution method and the SPDE-based 
method~\citep{Schmidt2011,RSSC:RSSC12027,Rikke2013}. However, even if we 
intend to use covariates, the more general non-stationary models could be 
used to gain intuition about which covariates should be selected and what 
type of non-stationarity they should control. 

The comparison of the different models shows that the scoring rule used to 
evaluate the predictions has a large influence on the conclusion. The use of
a non-stationary model instead of a stationary model mainly affects the 
prediction variances and not the predicted values. Therefore, the 
largest improvements are seen in the log-predictive score and the CRPS, and 
not the RMSE that only evaluates point predictions. However, consistently higher
RMSE values for the flexible non-stationary model compared to the simple
stationary, as observed when fitting the models using a single nugget effect 
to de-trended data, is useful to detect problems with the model such as
undersmoothing.

The spatial structure of the annual precipitation is treated in 
detail in the case study, but the temporal structure is not given the
same amount of focus. The treatment of the yearly data as independent realizations
makes an explicit assumption of independence between years and an
implicit assumption of stationarity in time. The assumption of stationarity 
is necessary to break the fundamental non-separability between the mean structure
and the covariance structure, and without this assumption one would only have a 
\emph{single} realization of the space-time process. If the dataset cannot
be assumed to fulfil these conditions, it is necessary to make a decision
about how to distribute the observed behaviour into mean structure and
covariance structure through a spatio-temporal model. 
An interesting point for future work is whether parts of the patterns in the
estimated covariance structures are caused by different spatial non-stationarity
in different years, and to which degree increasing the spatial resolution would
give additional information and to which degree the complexity of the non-stationarity
varies with time-scale.

One of the major reasons not to use general non-stationary models unless they 
are absolutely needed is that they are computationally expensive. The 
covariate-based approach is less expensive, but requires assumptions about 
how the non-stationarity varies. Another approach would be to estimate the 
model locally in different parts of the domain and then try to piece everything
together for predictions, but looking for the most efficient way to estimate 
the model is not the goal of this paper and the more complex one makes the
model, the more computationally expensive it will be. The point we are trying 
to make is that in applications, time might in many cases be better spent on
considering how to put the non-stationarity into the model than on developing
more complex flexible models and ways to compute them.

Non-stationarity in the covariance structure of spatial models is needed 
even after the non-stationarity in the mean has been removed, but we need to
think carefully about how we handle the non-stationarity. We need to go
beyond determining whether there is non-stationarity or not, and determine what
type of non-stationarity is present and if possible target this non-stationarity
directly instead of using a general flexible model. But in this context the
estimated covariance structure from a general flexible model can in some cases
be a useful tool to determine how to do this.

\section*{Acknowledgements}
We are grateful to the Editor and reviewers for their helpful comments that
have improved the article.

\newpage

\appendix
\section{Computational details for the model}
\label{app:Details}
Let the SPDE
\begin{equation}
	(\kappa^2(\vec{s})-\nabla\cdot\mathbf{H}(\vec{s})\nabla)u(\vec{s}) = \mathcal{W}(\vec{s}), \quad \boldsymbol{s}\in\mathbb{R}^2,
	\label{app:eq:FinalSPDE}
\end{equation}
where \(\mathcal{W}(\cdot)\) is standard Gaussian white noise and 
\(\nabla = (\frac{\partial}{\partial x}, \frac{\partial}{\partial y})\),
describe the desired covariance structure. 
The first thing to notice is that the operator 
in front of \(u\) only contains multiplications with functions and
first order and second order derivatives. All of these operations involve only
the local properties of \(u\) at each location. This means that if \(u\) is 
discretized using a finite-dimensional local basis expansion, the corresponding discretized operators (matrices) should 
only involve variables close to each other. This can be exploited to create
a sparse GMRF which possesses approximately the same covariance structure as \(u\).
The arguments above are not applicable for all smoothnesses, but we are 
constructing a model where the smoothness is fixed to 1 and the range is
allowed to vary spatially (See discussion in~\citet[p. 5]{Fuglstad2014}).  A detailed description of the basis function expansion, the choice of mesh, and the theoretical properties of the methods described in this section in \citet{Lindgren2011, Simpson2011,simpson2011going}.

The first step in creating the GMRF is to restrict 
SPDE~\eqref{app:eq:FinalSPDE} to a bounded domain,
\[
	(\kappa^2(\vec{s})-\nabla\cdot\mathbf{H}(\vec{s})\nabla)u(\vec{s}) = \mathcal{W}(\vec{s}), \qquad \vec{s}\in\mathcal{D} = [A_1,B_1]\times[A_2, B_2]\subset\mathbb{R}^2,
\]
where \(B_1 > A_1\) and \(B_2 > A_2\). This restriction 
necessitates a boundary condition to make the distribution useful and 
proper. For technical reasons the boundary condition chosen is 
zero flux across the boundaries, i.e. at each point of the boundary
the flux \(\mathbf{H}(\vec{s})\nabla_{\vec{n}} u(\vec{s})\), where \(\vec{n}\) is
the normal vector of the boundary at that point, is zero.
 The derivation of a discretized version
of this SPDE on a grid is involved, but for periodic boundary 
conditions the derivation can be found in the supplementary material
to~\citet{Fuglstad2014}. The boundary conditions in this problem involve 
only a slight change in that derivation. 

For a regular \(m \times n\) grid of \(\mathcal{D}\), the end result is the matrix
equation
\[
	\mathbf{A}(\kappa^2, \mathbf{H}) \vec{u} = \frac{1}{\sqrt{V}}\vec{z},
\]
where \(V\) is the area of each cell in the grid, \(\vec{u}\)
corresponds to the values of \(u\) on the cells in the regular grid stacked column-wise, 
\(\vec{z}\sim\mathcal{N}_{mn}(\vec{0}, \mathbf{I}_{mn})\) and
\(\mathbf{A}(\kappa^2, \mathbf{H})\) is a discretized version of 
\((\kappa^2-\nabla\cdot\mathbf{H}\nabla)\). This matrix equation leads to the 
multivariate Gaussian distribution
\begin{equation}
	\vec{u} \sim \mathcal{N}_{mn}(\vec{0}, \mathbf{Q}(\kappa^2, \mathbf{H})^{-1}),
	\label{eq:GMRFeq}
\end{equation}
where \(\mathbf{Q}(\kappa^2, \mathbf{H}) = \mathbf{A}(\kappa^2, \mathbf{H})^\mathrm{T}\mathbf{A}(\kappa^2, \mathbf{H})V\). The precision matrix \(\mathbf{Q}\) is proper and
 has up to 25 non-zero elements in each row,
corresponding to the point itself, its eight closest neighbours and the eight 
closest neighbours of each of the eight closest neighbours. Since the approximation
is constructed from an SPDE, it behaves consistently over different resolution and
converges to a continuously indexed model for small resolutions. Changing the 
resolution changes which features can be represented by the model, but does not 
induce large changes to the covariance structure.

\section{Derivation of the second-order random walk prior}
\label{app:Prior}
Each function, \(f\), is a priori modelled as a Gaussian process described by the 
SPDE
\begin{equation}
	-\Delta f(\vec{s}) = \frac{1}{\sqrt{\tau}}\mathcal{W}(\vec{s}), \qquad \vec{s}\in \mathcal{D} = [A_1, B_1]\times[A_2, B_2], 
	\label{Prior:eq:SPDE}
\end{equation}
where \(A_1 < B_1\), \(A_2 < B_2\) and \(\tau>0\), \(\mathcal{W}\) is standard
Gaussian white noise and 
\(\Delta = \frac{\partial^2}{\partial x^2}+\frac{\partial^2}{\partial y^2}\), with
the Neumann boundary condition of zero normal derivatives at the edges. In practice
this is approximated by representing \(f\) as a linear combination of basis elements
\(\{f_{ij}\}\) weighted by Gaussian distributed weights 
\(\{\alpha_{ij}\}\),
\[
	f(\vec{s}) = \sum_{i=1}^K \sum_{j=1}^L \alpha_{ij} f_{ij}(\vec{s}).
\]
The basis functions are constructed from separate bases \(\{g_i\}\) and \(\{h_j\}\)
for the \(x\)-coordinate and the \(y\)-coordinate, respectively,
\begin{equation}
	f_{ij}(\vec{s}) = g_i(x) h_j(y).
	\label{Prior:eq:fBase}
\end{equation}
For convenience each basis function is assumed to fulfil the boundary condition of
zero normal derivative at the edges.

Let \(\vec{\alpha} = \mathrm{vec}([\alpha_{ij}]_{ij})\), then the task is to
find the best Gaussian distribution for \(\vec{\alpha}\). Where ``best'' is
used in the sense of making the resulting distribution for \(f\) ``close'' to a solution of SPDE~\eqref{Prior:eq:SPDE}. This is done by a least-squares approach 
where the vector created from doing inner products of the left hand side with 
\(-\Delta f_{kl}\)  must be equal in distribution to the vector created from doing 
the same to the right hand side,
\begin{equation}
	\mathrm{vec}\left([\langle -\Delta f, -\Delta f_{kl} \rangle_\mathcal{D}]_{kl}\right) \eqd \mathrm{vec}\left([\langle \mathcal{W}, -\Delta f_{kl} \rangle_\mathcal{D}]_{kl}\right).
	\label{Prior:eq:alpha}
\end{equation}

First, calculate the inner product that is needed
\begin{align*}
	\left\langle -\Delta g_ih_j, -\Delta g_k h_l\right\rangle_\mathcal{D} &= \left\langle \Delta g_i h_j, \Delta g_i h_j\right\rangle_\mathcal{D}\\
	&= \left\langle \left(\frac{\partial^2}{\partial x^2}g_i\right) h_j + g_i \frac{\partial^2}{\partial y^2} h_j, \left(\frac{\partial^2}{\partial x^2}g_{k}\right)h_l + g_k \frac{\partial^2}{\partial y^2} h_{l} \right\rangle_\mathcal{D}.
\end{align*}
The bilinearity of the inner product can be used to expand the expression in a
sum of four innerproducts. Each of these inner products can then be written as a product of two inner products. Due to lack of space this is not done explicitly, but one of these terms is, for example,
\[
	\left\langle\left(\frac{\partial^2}{\partial x^2}g_i\right) h_j, \left(\frac{\partial^2}{\partial x^2}g_k\right)h_l\right\rangle_\mathcal{D} = \left\langle\frac{\partial^2}{\partial x^2}g_i, \frac{\partial^2}{\partial x^2}g_k\right\rangle_{[A_1,B_1]}\left\langle h_j, h_l\right\rangle_{[A_2,B_2]}.
\] 
By inserting Equation~\eqref{Prior:eq:fBase} into Equation~\eqref{Prior:eq:alpha}
and using the above derivations together with integration by parts one can see that
the left hand side becomes
\[
	\mathrm{vec}\left([\langle -\Delta f, -\Delta f_{kl} \rangle_\mathcal{D}]_{kl}\right) = \mathbf{C}\vec{\alpha},
\]
where \(\mathbf{C} = \mathbf{G}_{2}\otimes \mathbf{H}_{0}+2\mathbf{G}_{1}\otimes \mathbf{H}_{1}+\mathbf{G}_{0}\otimes\mathbf{H}_{2}\) with
\[
	\mathbf{G}_{n} = \left[\left\langle \frac{\partial^n}{\partial x^n} g_i, \frac{\partial^n}{\partial x^n} g_j \right\rangle_{[A_1,B_1]}\right]_{i,j}
\]
and
\[
	\mathbf{H}_{n} = \left[\left\langle \frac{\partial^n}{\partial y^n} h_i, \frac{\partial^n}{\partial y^n} h_j \right\rangle_{[A_2, B_2]}\right]_{i,j}	.
\]

The right hand side is a Gaussian random vector where the covariance between the
position corresponding to \(\alpha_{ij}\) and the position corresponding to 
\(\alpha_{kl}\) is given by
\[
	\langle -\Delta f_{ij}, -\Delta f_{kl} \rangle_\mathcal{D}.
\]
Thus the covariance matrix of the right hand side must be \(\mathbf{C}\)
and Equation~\eqref{Prior:eq:alpha} can be written in matrix form as
\[
	\mathbf{C}\vec{\alpha} = \mathbf{C}^{1/2}\vec{z},	
\]
where \(\vec{z}\sim\mathcal{N}_{KL}(\vec{0}, \mathbf{I}_{KL})\).
This means that \(\vec{\alpha}\) should be given the precision matrix
\(\mathbf{Q} = \mathbf{C}\). Note that \(\mathbf{C}\) might be singular
due to invariance to some linear combination of the basis elements.

\section{Conditional distributions}
\label{app:CondDist}
From the hierarchical model
\begin{align*}
	\text{Stage 1: } & \vec{y}|\vec{z}, \vec{\theta} \sim \mathcal{N}_N(\mathbf{S}\vec{z}, \mathbf{I}_N/\tau_{\mathrm{noise}})  \\
	\text{Stage 2: } & \vec{z}|\vec{\theta} \sim \mathcal{N}_{mn+p}(\vec{0}, \mathbf{Q}_z^{-1}),
\end{align*}
the posterior distribution \(\pi(\vec{\theta}|\vec{y})\) can be derived
explicitly. There are three steps involved.

\subsection{Step 1}
Calculate the distribution \(\pi(\vec{z}|\vec{\theta}, \vec{y})\)
up to a constant,
\begin{align*}
	\pi(\vec{z}|\vec{\theta}, \vec{y}) & \propto \pi(\vec{z}, \vec{\theta}, \vec{y}) \\
		& = \pi(\vec{\theta}) \pi(\vec{z}|\vec{\theta}) \pi(\vec{y}|\vec{\theta}, \vec{z}) \\
		& \propto \exp\left(-\frac{1}{2}(\vec{z}-\vec{0})^\mathrm{T}\mathbf{Q}_z(\vec{z}-\vec{0})-\frac{1}{2}(\vec{y}-\mathbf{S}\vec{z})^\mathrm{T}\mathbf{I}_N\cdot\tau_{\mathrm{noise}}(\vec{y}-\mathbf{S}\vec{z})\right)\\
		& \propto \exp\left(-\frac{1}{2}\left(\vec{z}^\mathrm{T}(\mathbf{Q}_z+\tau_{\mathrm{noise}}\mathbf{S}^\mathrm{T}\mathbf{S})\vec{z} - 2\vec{z}^\mathrm{T}\mathbf{S}^\mathrm{T}\vec{y}\cdot\tau_{\mathrm{noise}}\right)\right) \\
		& \propto \exp\left(-\frac{1}{2}(\vec{z}-\vec{\mu}_\mathrm{C})^\mathrm{T}\mathbf{Q}_\mathrm{C}(\vec{z}-\vec{\mu}_\mathrm{C})\right),
\end{align*}
where \(\mathbf{Q}_\mathrm{C} = \mathbf{Q}_z+\mathbf{S}^\mathrm{T}\mathbf{S}\cdot\tau_{\mathrm{noise}}\) and \(\mu_\mathrm{C} = \mathbf{Q}_\mathrm{C}^{-1}\mathbf{S}^\mathrm{T}\vec{y}\cdot \tau_{\mathrm{noise}}\). This is recognised as a Gaussian distribution
\[
	\vec{z}| \vec{\theta}, \vec{y} \sim \mathcal{N}_N(\vec{\mu}_\mathrm{C}, \mathbf{Q}_\mathrm{C}^{-1}).
\]

\subsection{Step 2}
\label{app:CondDist:2}
Integrate out \(\vec{z}\) from the joint distribution of \(\vec{z}\),
\(\vec{\theta}\) and \(\vec{y}\) via the Bayesian rule,

\begin{align*}
	\pi(\vec{\theta},\vec{y}) &= \frac{\pi(\vec{\theta}, \vec{z}, \vec{y})}{\pi(\vec{z}|\vec{\theta}, \vec{y})} \\
		& = \frac{\pi(\vec{\theta})\pi(\vec{z}|\vec{\theta})\pi(\vec{y}| \vec{z}, \vec{\theta})}{\pi(\vec{z}|\vec{\theta}, \vec{y})}.
\end{align*}
The left hand side of the expression does not depend on the value of
\(\vec{z}\), therefore the right hand side may be evaluated
at any desired value of \(\vec{z}\). Evaluating at 
\(\vec{z} = \vec{\mu}_\mathrm{C}\) gives
\begin{align*}
	\pi(\vec{\theta}, \vec{y}) & \propto \frac{\pi(\vec{\theta})\pi(\vec{z} = \vec{\mu}_\mathrm{C})\pi(\vec{y}|\vec{z} = \vec{\mu}_\mathrm{C}, \vec{\theta})}{\pi(\vec{z}=\vec{\mu}_\mathrm{C}|\vec{\theta}, \vec{y})} \\
		& \propto\pi(\vec{\theta}) \frac{|\mathbf{Q}_z|^{1/2}|\mathbf{I}_N\cdot \tau_{\mathrm{noise}}|^{1/2}}{|\mathbf{Q}_\mathrm{C}|^{1/2}}\exp\left(-\frac{1}{2}\vec{\mu}_\mathrm{C}^\mathrm{T}\mathbf{Q}_z\vec{\mu}_\mathrm{C}\right)\times\\
		& \times \exp\left(-\frac{1}{2}(\vec{y}-\mathbf{S}\vec{\mu}_\mathrm{C})^\mathrm{T}\mathbf{I}_N\cdot\tau_{\mathrm{noise}}(\vec{y}-\mathbf{S}\vec{\mu}_\mathrm{C})\right)\times\\
		& \times \exp\left(+\frac{1}{2}(\vec{\mu}_\mathrm{C}-\vec{\mu}_\mathrm{C})^\mathrm{T}\mathbf{Q}_\mathrm{C}(\vec{\mu}_\mathrm{C}-\vec{\mu}_\mathrm{C})\right).
\end{align*}

\subsection{Step 3}
Condition on \(\vec{y}\) to get the desired conditional distribution,
\begin{align}
	\notag \log(\pi(\vec{\theta}|\vec{y})) &= \mathrm{Const} + \log(\pi(\vec{\theta}))+\frac{1}{2}\log(\det(\mathbf{Q}_z))+\frac{N}{2}\log(\tau_\mathrm{noise})+ \\
	& -\frac{1}{2}\log(\det(\mathbf{Q}_\mathrm{C}))-\frac{1}{2}\vec{\mu}_\mathrm{C}^\mathrm{T}\mathbf{Q}_z\vec{\mu}_z-\frac{\tau_\mathrm{noise}}{2}(\vec{y}-\mathbf{S}\vec{\mu}_\mathrm{C})^\mathrm{T}(\vec{y}-\mathbf{S}\vec{\mu}_\mathrm{C}).
	\label{eq:CondDist:post}
\end{align}

\section{Analytic expression for the gradient}
\label{app:Gradient}
This appendix shows the derivation of the derivative of the log-likelihood.
Choose the evaluation point \(\vec{z} = \vec{0}\)
in \ref{app:CondDist:2} to find
\begin{align*}
	\log(\pi(\vec{\theta},\tau_\mathrm{noise}|\vec{y})) &= \mathrm{Const} + \log(\pi(\vec{\theta},\tau_\mathrm{noise}))+\frac{1}{2}\log(\det(\mathbf{Q}_z))+\frac{N}{2}\log(\tau_\mathrm{noise})+ \\
	& -\frac{1}{2}\log(\det(\mathbf{Q}_\mathrm{C}))-\frac{\tau_\mathrm{noise}}{2}\vec{y}^\mathrm{T}\vec{y}+\frac{1}{2}\vec{\mu}_\mathrm{C}^\mathrm{T}\mathbf{Q}_\mathrm{C}\vec{\mu}_\mathrm{C}.
\end{align*}
This is just a rewritten form of Equation~\eqref{eq:CondDist:post} which is more 
convenient for the calculation of the gradient, and which separates the 
\(\tau_\mathrm{noise}\) parameter from the rest of the covariance parameters. First
some preliminary results are presented, then the derivatives are calculated with
respect to \(\theta_i\) and lastly the derivatives are calculated with respect to
\(\log(\tau_\mathrm{noise})\).

Begin with simple preliminary formulas for the derivatives of the conditional 
precision matrix with respect to each of the parameters,
\begin{equation}
	\frac{\partial}{\partial \theta_i} \mathbf{Q}_\mathrm{C} = \frac{\partial}{\partial \theta_i}(\mathbf{Q}+\mathbf{S}^\mathrm{T}\mathbf{S}\cdot\tau_\mathrm{noise}) = \frac{\partial}{\partial \theta_i}\mathbf{Q}
	\label{eq:Gradient:Qc_theta}
\end{equation}
and
\begin{equation}
	\frac{\partial}{\partial \log(\tau_{\mathrm{noise}})} \mathbf{Q}_\mathrm{C} = \frac{\partial}{\partial \log(\tau_{\mathrm{noise}})}(\mathbf{Q}+\mathbf{S}^\mathrm{T}\mathbf{S}\cdot\tau_\mathrm{noise}) = \mathbf{S}^\mathrm{T}\mathbf{S}\cdot\tau_\mathrm{noise}.
	\label{eq:Gradient:Qc_tau}
\end{equation}

\subsection{Derivative with respect to \(\theta_i\)}
First the derivatives of the log-determinants can be handled by an explicit
formula~\citep{petersen2012} 
\begin{align*}
	\frac{\partial}{\partial \theta_i} (\log(\det(\mathbf{Q}))-\log(\det(\mathbf{Q}_\mathrm{C})) &= \mathrm{Tr}(\mathbf{Q}^{-1}\frac{\partial}{\partial \theta_i}\mathbf{Q})-\mathrm{Tr}(\mathbf{Q}_\mathrm{C}^{-1}\frac{\partial}{\partial \theta_i} \mathbf{Q}_\mathrm{C})\\
	&= \mathrm{Tr}\left[(\mathbf{Q}^{-1}-\mathbf{Q}_\mathrm{C}^{-1})\frac{\partial}{\partial \theta_i}\mathbf{Q}\right].
\end{align*}
Then the derivative of the quadratic forms are calculated
\begin{align*}
	\frac{\partial}{\partial \theta_i} \left(-\frac{1}{2}\vec{y}^\mathrm{T}\vec{y}\cdot\tau_\mathrm{noise} + \frac{1}{2}\vec{\mu}_\mathrm{C}\mathbf{Q}_\mathrm{C}\vec{\mu}_\mathrm{C} \right) &= 0 + \frac{\partial}{\partial \theta_i} \left(\frac{1}{2}\vec{y}^\mathrm{T}\tau_\mathrm{noise}\mathbf{S}\mathbf{Q}_\mathrm{C}^{-1}\mathbf{S}^\mathrm{T}\tau_\mathrm{noise}\vec{y}\right)\\
	&= -\frac{1}{2}\vec{y}^\mathrm{T}\tau_\mathrm{noise}\mathbf{S}\mathbf{Q}_\mathrm{C}^{-1}\left(\frac{\partial}{\partial \theta_i}\mathbf{Q}_\mathrm{C}\right)\mathbf{Q}_\mathrm{C}^{-1}\mathbf{S}^\mathrm{T}\tau_\mathrm{noise}\vec{y}  \\
	&= -\frac{1}{2}\vec{\mu}_\mathrm{C}^\mathrm{T}\left(\frac{\partial}{\partial \theta_i}\mathbf{Q}\right)\vec{\mu}_\mathrm{C}.
\end{align*}
Combining these gives
\[
	\frac{\partial}{\partial \theta_i} \log(\pi(\vec{\theta}, \tau_\mathrm{noise}| \vec{y})) = \frac{\partial}{\partial \theta_i} \log(\pi(\vec{\theta},\tau_\mathrm{noise} )) + \frac{1}{2}\mathrm{Tr}\left[(\mathbf{Q}^{-1}-\mathbf{Q}_\mathrm{C}^{-1})\frac{\partial}{\partial \theta_i} \mathbf{Q}\right] - \frac{1}{2}\vec{\mu}_\mathrm{C}^\mathrm{T}\left(\frac{\partial}{\partial \theta_i}\mathbf{Q}\right)\vec{\mu}_\mathrm{C}
\]

\subsection{Derivative with respect to \(\log(\tau_\mathrm{noise})\)}
First calculate the derivative of the log-determinants
\begin{align*}
	\frac{\partial}{\partial \log(\tau_\mathrm{noise})} \left(N\log(\tau_\mathrm{noise}) - \log(\det(\mathbf{Q}_\mathrm{C}))\right) &= N - \mathrm{Tr}\left(\mathbf{Q}_\mathrm{C}^{-1}\frac{\partial}{\partial \log(\tau_\mathrm{noise})} \mathbf{Q}_\mathrm{C}\right)\\
		& = N-\mathrm{Tr}\left(\mathbf{Q}_\mathrm{C}^{-1}\mathbf{S}^\mathrm{T}\mathbf{S}\cdot\tau_\mathrm{noise}\right).
\end{align*}
Then the derivative of the quadratic forms
\begin{align*}
	\frac{\partial\left(-\frac{1}{2}\vec{y}^\mathrm{T}\vec{y}\cdot\tau_\mathrm{noise} + \frac{1}{2} \vec{\mu}_\mathrm{C}\mathbf{Q}_\mathrm{C}\vec{\mu}_\mathrm{C}\right)}{\partial \log(\tau_\mathrm{noise})} &= -\frac{1}{2}\vec{y}^\mathrm{T}\vec{y}\cdot\tau_\mathrm{noise} + \frac{\partial}{\partial \log(\tau_n)} \frac{1}{2}\vec{y}^\mathrm{T}\tau_\mathrm{noise}\mathbf{S}\mathbf{Q}_\mathrm{C}^{-1}\mathbf{S}^\mathrm{T}\tau_\mathrm{noise}\vec{y}\\
	&= -\frac{1}{2}\vec{y}^\mathrm{T}\vec{y}\cdot\tau_\mathrm{noise} +\vec{y}^\mathrm{T}\tau_\mathrm{noise}\mathbf{S}\mathbf{Q}_\mathrm{C}^{-1}\mathbf{S}\left(\frac{\partial \tau_{\mathrm{noise}}}{\partial \log(\tau_{\mathrm{noise}})}\right)\vec{y}+ \\
	&- \frac{1}{2}\vec{y}^\mathrm{T}\tau_\mathrm{noise}\mathbf{S}\mathbf{Q}_\mathrm{C}^{-1}\left(\frac{\partial}{\partial \log(\tau_{\mathrm{noise}})} \mathbf{Q}_\mathrm{C} \right)\mathbf{Q}_\mathrm{C}^{-1}\mathbf{S}^\mathrm{T}\tau_\mathrm{noise}\vec{y}\\
	& = -\frac{1}{2}\vec{y}^\mathrm{T}\vec{y}\cdot\tau_\mathrm{noise} +\vec{\mu}_\mathrm{C}^\mathrm{T}\mathbf{S}^\mathrm{T}\vec{y}\cdot\tau_\mathrm{noise} - \frac{1}{2}\vec{\mu}_\mathrm{C}^\mathrm{T}\mathbf{S}^\mathrm{T} \mathbf{S}\vec{\mu}_\mathrm{C}\cdot\tau_\mathrm{noise}\\
	&= -\frac{1}{2}(\vec{y}-\mathbf{A}\vec{\mu}_\mathrm{C})^\mathrm{T}(\vec{y}-\mathbf{A}\vec{\mu}_\mathrm{C})\cdot\tau_\mathrm{noise}.
\end{align*}
Together these expressions give
\begin{align*}
	\frac{\partial \log(\pi(\vec{\theta}, \tau_\mathrm{noise}| \vec{y}))}{\partial \log(\tau_\mathrm{noise})}  &= \frac{\partial}{\partial \log(\tau_\mathrm{noise})} \log(\pi(\vec{\theta},\tau_\mathrm{noise})) + \frac{N}{2} - \frac{1}{2}\mathrm{Tr}\left[\mathbf{Q}_\mathrm{C}^{-1}\mathbf{S}^\mathrm{T}\mathbf{S}\cdot\tau_\mathrm{noise}\right]+ \\
	&-\frac{1}{2}(\vec{y}-\mathbf{A}\vec{\mu}_\mathrm{C})^\mathrm{T}(\vec{y}-\mathbf{A}\vec{\mu}_\mathrm{C})\cdot\tau_\mathrm{noise}
\end{align*}

\subsection{Implementation}
The derivative $\frac{\partial}{\partial \theta_i} \mathbf{Q}_c$ can be 
calculated quickly since it is a simple functions of $\theta$. The trace of 
the inverse of a matrix \(A\) times the derivative of a matrix \(B\) only requires the
values of the inverse of \(A\) for non-zero elements of \(B\). In the above case
the two matrices have the same type of non-zero structure, but it can happen that
specific elements in the non-zero structure are zero for one of the matrices.
This way of calculating the inverse only at a subset of the locations can be handled as described
in~\citet{col27}.

% Bibliography

%\section*{References}
\renewcommand{\bibfont}{\small}

\bibliographystyle{apalike}
\bibliography{references}

\end{document}

%% file: Tikz/statCovariance.tex
\begin{tikzpicture}
	\usetikzlibrary{decorations.pathreplacing}
	% Axis
	\draw[->, thick] (0,0) -- coordinate (x axis mid) (6cm,0);
    \draw[->, thick] (0,0) -- coordinate (y axis mid) (0,6cm);
    
    % Ticks on axes
    \draw[thick] (0, 1pt) -- (0, -3pt);
    \draw[thick] (2, 1pt) -- (2, -3pt);
    \draw[thick] (4, 1pt) -- (4, -3pt);
    \draw[thick] (1pt, 0) -- (-3pt, 0);
    \draw[thick] (1pt, 2) -- (-3pt, 2);
    \draw[thick] (1pt, 4) -- (-3pt, 4);

    % Labels
    \node[below=0.8cm] at (x axis mid) {$x$};
    \node[rotate=90, above = 0.8cm] at (y axis mid) {$y$};

    % Vectors
    \draw [ultra thick, -latex] (3cm, 3cm) -- (5.121, 5.121) node [midway, above, sloped] (TextNode) {$\boldsymbol{v}_1$};
    \draw [decorate, decoration={brace, amplitude=5pt}, yshift = -2pt, xshift = +2pt] (5.121, 5.121) -- (3,3) node [midway, below, sloped, yshift=-8pt] {$\sqrt{\lambda_1}/\kappa$};
    \draw [ultra thick, -latex] (3cm, 3cm) -- (1.939, 4.061) node [midway, above, sloped] (TextNode2) {$\boldsymbol{v}_2$};
	\draw[rotate around={-45:(3,3)}] (3,3) ellipse (1.5cm and 3cm);
	\draw [decorate, decoration={brace, amplitude=5pt}, yshift = -2pt, xshift = -2pt] (3,3) -- (1.939,4.061) node [midway, below, sloped, yshift=-8pt] {$\sqrt{\lambda_2}/\kappa$};
    \draw [ultra thick, -latex] (3cm, 3cm) -- (1.939, 4.061) node [midway, above, sloped] (TextNode2) {$\boldsymbol{v}_2$};
\end{tikzpicture}